\documentclass[12pt,tightenlines,superscriptaddress]{revtex4PATCH}
\pdfoutput=1

\usepackage{enumitem}

\usepackage{latexsym}
\usepackage{graphicx}
\usepackage{url}
\usepackage{mathrsfs,amssymb,amsmath,amsfonts,amstext,graphics, multirow}
\usepackage{epstopdf}
\usepackage{url}
\usepackage{appendix}
\usepackage{color,slashed}

\usepackage{tikz}
\usepackage{slashed}
\usepackage{cancel}
\usepackage{mciteplus}
\usepackage{verbatim}
\usetikzlibrary{arrows,shapes}
\usetikzlibrary{trees}
\usetikzlibrary{matrix,arrows} 

\usepackage{pdflscape}
\usepackage{xspace}
\usepackage{color}
\usepackage{wrapfig}
\usepackage{rotating}
\usepackage{dcolumn}% Align table columns on decimal point
\usepackage{bm}% bold math
\usepackage{longtable}% Include figure files
\usepackage{changepage} % for expanding margins to accomodate wide table
\usepackage{rotating} % for rotating table 90deg
\usepackage{xcolor}
\usepackage{graphicx}
\usepackage{lipsum}

\usepackage[utf8]{inputenc}											
\usepackage{relsize}

\newcommand{\slac}{SLAC National Accelerator Laboratory, Menlo Park, CA 94025, USA}

\newcommand{\stanford}{Stanford University Physics Department, Stanford, CA 94305, USA}

\newcommand{\jlab}{Jefferson Lab, Newport News, VA 23606, USA}

\newcommand{\perimeter}{Perimeter Institute for Theoretical Physics, Waterloo ON N2L 2Y5, Canada}

\newcommand{\lbl}{Lawrence Berkeley National Laboratory, Berkeley, California 94720, USA}

\newcommand{\irvine}{Department of Physics and Astronomy, University of California, Irvine, California 92697, USA}

\newcommand{\INFNPD}{Istituto Nazionale di Fisica Nucleare, Sezione di Padova,  35131 Padova , Italy}
\newcommand{\INFNCA}{Istituto Nazionale di Fisica Nucleare, Sezione di Cagliari, 09042 Cagliari, Italy}
\newcommand{\INFNGE}{Istituto Nazionale di Fisica Nucleare, Sezione di Genova, 16146 Genova, Italy}
\newcommand{\INFNLNF}{Istituto Nazionale di Fisica Nucleare, Laboratori Nazionali di Frascati, 00044 Frascati, Italy}
\newcommand{\INFNCT}{Istituto Nazionale di Fisica Nucleare, Sezione di Catania, 92125 Catania, Italy}
\newcommand{\INFNRM}{Istituto Nazionale di Fisica Nucleare, Sezione di Roma, 00185 Roma, Italy}
\newcommand{\INFNLE}{Istituto Nazionale di Fisica Nucleare, Sezione di Lecce, 73047, Italy}
\newcommand{\INFNRMtre}{Istituto Nazionale di Fisica Nucleare, Sezione di Roma Tre, 00146 Roma, Italy}
\newcommand{\INFNLNS}{Istituto Nazionale di Fisica Nucleare, Laboratori Nazionali del Sud, 92125 Catania , Italy}
\newcommand{\UNISS}{Universit\`a di Sassari, 07100 Sassari, Italy}
\newcommand{\UNIRM}{Universit\`a di Roma La Sapienza, 00185 Roma, Italy}
\newcommand{\UNIME}{Universit\`a di Messina, 98122 Messina, Italy}
\newcommand{\UNIGE}{Universit\`a di Genova, 16100 Genova, Italy}
\newcommand{\UNISA}{Universit\`a del Salento, 73100 Lecce, Italy}
\newcommand{\UNIPD}{Universit\`a di Padova, 35122 Padova, Italy}
\newcommand{\UNIRMTV}{Universit\`a di Roma Tor Vergata, 00173 Roma, Italy}
\newcommand{\INFNRMtv}{Istituto Nazionale di Fisica Nucleare, Sezione di Roma Tor Vergata, 00173 Roma, Italy}
\newcommand{\UNIMAINZ}{JGU Mainz, Institute for Nuclear Physics and PRISMA Cluster of Excellence, 55128 Mainz, Germany}
\newcommand{\INFNTO}{Istituto Nazionale di Fisica Nucleare, Sezione di Torino, 10125 Torino, Italy}

\newcommand{\orsay}{Institut de Physique Nucl\'eaire, CNRS-IN2P3, Univ. Paris-Sud, Universit\'e Paris-Saclay, 91406 Orsay Cedex, France}

\newcommand{\caltech}{California Institute of Technology, Pasadena, California 91125, USA}

\newcommand{\ethZ}{Institute for Particle Physics, ETH Zurich, 8093 Zurich, Switzerland}

\renewcommand{\mit}{Massachusetts Institute of Technology, Cambridge, MA 02139, USA}

\newcommand{\cernth}{CERN Theory Department, CH-1211 Geneva 23, Switzerland}

\newcommand{\yangstonybrook}{C.N.~Yang Institute for Theoretical Physics, Stony Brook University, Stony Brook, NY 11794, USA}
\newcommand{\stonybrook}{Stony Brook University, Stony Brook, NY 11794, USA}

\newcommand{\riverside}{Department of Physics and Astronomy, University of  California, Riverside, California 92521, USA}
\newcommand{\kentucky}{Department of Physics and Astronomy, University of Kentucky, Lexington, Kentucky 40506, USA}

\newcommand{\calstate}{California State University, Los Angeles, Los Angeles, California 90032, USA}

\newcommand{\yale}{Yale University, New Haven, CT 06520, USA}

\newcommand{\ODU}{Old Dominion University,  Norfolk, VA 23529, USA}

\newcommand{\cornell}{Cornell University,  Ithaca, NY 14853, USA}
\newcommand{\TRIUMF}{TRIUMF,  Vancouver, BC  V6T 2A3, Canada}
\newcommand{\ANL}{Argonne National Laboratory, Lemont, IL 60439, USA}
\newcommand{\FERMILAB}{Fermi National Accelerator Laboratory, Batavia, IL 60510, USA}
\newcommand{\LOSALAMOS}{Los Alamos National  Laboratory, Los Alamos, NM 87545, USA}
\newcommand{\UDAVIS}{University of California, Davis, Davis, CA 95616, USA}
\newcommand{\UBERKELEY}{University of California, Berkeley, Berkeley, CA 94720, USA}
\newcommand{\WM}{College of William and Mary, Williamsburg, VA 23185, USA}
\newcommand{\princeton}{Princeton University, Princeton, NJ 08544, USA}

\newcommand{\uvictoria}{University of Victoria, Victoria, BC V8P 5C2, Canada}
\newcommand{\dubna}{Joint Institute for Nuclear Research, 141980 Dubna, Russia}
\newcommand{\bnl}{Brookhaven National  Laboratory, Upton, NY 11973, USA}
\newcommand{\UCALSB}{University of California, Santa Barbara, Santa Barbara, CA 93106, USA}

\newcommand{\tuebingen}{Institute of Theoretical Physics, University of Tuebingen, D-72076, Tuebingen, Germany}
\newcommand{\tomsk}{Department of Physics, Tomsk State University, 634050 Tomsk, Russia}
\newcommand{\tomskB}{Laboratory of Particle Physics, Mathematical Physics Department, Tomsk Polytechnic University, 634050 Tomsk, Russia}

\newcommand{\gsim}{\lower.7ex\hbox{$\;\stackrel{\textstyle>}{\sim}\;$}}
\newcommand{\lsim}{\lower.7ex\hbox{$\;\stackrel{\textstyle<}{\sim}\;$}}

\usepackage{slashed}

\def\lsim{\mathrel{\rlap{\lower4pt\hbox{\hskip 0.5 pt$\sim$}}
    \raise1pt\hbox{$<$}}}                % less than or approx. symbol
\def\gsim{\mathrel{\rlap{\lower4pt\hbox{\hskip1pt$\sim$}}
    \raise1pt\hbox{$>$}}}

\def\lsim{\mathrel{\rlap{\lower4pt\hbox{\hskip1pt$\sim$}}
    \raise1pt\hbox{$<$}}}
\def\gsim{\mathrel{\rlap{\lower4pt\hbox{\hskip1pt$\sim$}}
    \raise1pt\hbox{$>$}}}

\newcommand{\kev}{{\rm keV}}
\newcommand{\mev}{{\rm MeV}}
\newcommand{\gev}{{\rm GeV}}

\newcommand{\comments}[1]{}

\newcommand{\MeV}{\,\mathrm{MeV}}

\newcommand{\bef}{\begin{figure}[htbp]\begin{center}}
\newcommand{\eef}{\end{center}\end{figure}}

\newcommand{\bea}{\begin{eqnarray}}
\newcommand{\eea}{\end{eqnarray}}

\def\beq{\begin{equation}}
\def\eeq#1{\label{#1}\end{equation}}
\def\eeqn{\end{equation}}
\def\beqa{\begin{eqnarray}}
\def\eeqa#1{\label{#1}\end{eqnarray}}
\def\eeqan{\end{eqnarray}}

\def\leqn#1{(\ref{#1})}
\newcommand{\ee}{\ensuremath{e^+e^-}\xspace}
\newcommand{\elel}{\ensuremath{\ell^{+}\ell^{-}}\xspace}
\newcommand{\mumu}{\ensuremath{\mu^{+}\mu^{-}}\xspace}

\newcommand{\hh}{\ensuremath{h^{+}h^{-}}\xspace}
\newcommand{\pos}{\ensuremath{e^+}\xspace}
\newcommand{\elec}{\ensuremath{e^-}\xspace}
\newcommand{\Ap}{\ensuremath{A^\prime}\xspace}
\newcommand{\dark}{\Ap}
\newcommand{\mdark}{\ensuremath{m_{\Ap}}\xspace}

\newcommand{\abinv}{\ensuremath{\mathrm{ab}^{-1}}\xspace}

\newcommand{\power}[1]{\ensuremath{\times 10^{#1}}}
\newcommand{\eps}{\ensuremath{\varepsilon}\xspace}
\newcommand{\epsq}{\ensuremath{\varepsilon^2}\xspace}
\newcommand{\brem}{\ensuremath{\elec \;{\mathrm Z}\to \elec \; {\mathrm Z} \dark}\xspace}
\newcommand{\pbrem}{\ensuremath{\mathrm{p} \,{\mathrm Z}\to \mathrm{p} \, {\mathrm Z} \dark}\xspace}

\newcommand{\lumiunits}{\ensuremath{\,{\mathrm{cm}}^{-2}{\mathrm{s}}^{-1}}}

\newcommand{\daph}{\ensuremath{\mathrm{DA}\phi\mathrm{NE}}\xspace}
\newcommand{\muA}{\ensuremath{\mu \mathrm{A}}\xspace}
\newcommand{\fbinv}{\ensuremath{\mathrm{fb}^{-1}}\xspace}

\def\babar{\mbox{\slshape B\kern-0.1em{\smaller A}\kern-0.1emB\kern-0.1em{\smaller A\kern-0.2em R}}}

\newcommand{\benum}{\begin{enumerate}}
\newcommand{\eenum}{\end{enumerate}}
\def\lsim{\mathrel{\rlap{\lower4pt\hbox{\hskip 0.5 pt$\sim$}}
    \raise1pt\hbox{$<$}}}                % less than or approx. symbol
\def\gsim{\mathrel{\rlap{\lower4pt\hbox{\hskip1pt$\sim$}}
    \raise1pt\hbox{$>$}}} 
    
    \def\simgt{\mathrel{\lower2.5pt\vbox{\lineskip=0pt\baselineskip=0pt
           \hbox{$>$}\hbox{$\sim$}}}}
\def\simlt{\mathrel{\lower2.5pt\vbox{\lineskip=0pt\baselineskip=0pt
           \hbox{$<$}\hbox{$\sim$}}}}
\begin{document}

\pagestyle{plain}
\title{\Large Dark Sectors 2016 Workshop: Community Report}

% !TEX root = WorkshopReport.tex
\author{Jim Alexander (VDP Convener)}
\affiliation{\cornell}

\author{Marco Battaglieri (DMA Convener)}
\affiliation{\INFNGE}

\author{Bertrand Echenard (RDS Convener)}
\affiliation{\caltech}

\author{Rouven Essig (Organizer)}
\email{rouven.essig@stonybrook.edu}
\affiliation{\yangstonybrook}

\author{Matthew Graham (Organizer)}
\email{mgraham@slac.stanford.edu}
\affiliation{\slac}

\author{Eder Izaguirre (DMA Convener)}
\affiliation{\perimeter}

\author{John Jaros (Organizer)}
\email{john@slac.stanford.edu}
\affiliation{\slac}

\author{Gordan Krnjaic (DMA Convener)}
\affiliation{\FERMILAB}

\author{Jeremy Mardon (DD Convener)}
\affiliation{\stanford}

\author{David Morrissey (RDS Convener)}
\affiliation{\TRIUMF}

% ...

\date{\today}

\maketitle
%%%%%%%%%%%%%%%%%%%%%%%%%%%%%%%%%%%%%%%%%%
\newpage
\tableofcontents

\newpage

% !TEX root = ../WorkshopReport.tex
\section{Executive Summary}

This report, based on the Dark Sectors workshop at SLAC in April 2016, summarizes the scientific importance of searches for dark sector dark matter and forces, the status of this broad international field, the important milestones motivating future exploration, and the promising experimental opportunities to reach these milestones over the next 5-10 years.  

Remarkably, 80$\%$ of the matter in the Universe is an unknown substance --- dark matter --- whose constituents and interactions are quite different from those of ordinary matter.  The 2014 P5 report~\cite{ParticlePhysicsProjectPrioritizationPanel(P5):2014pwa} highlights the vital importance of identifying the physics of dark matter and of making this search as broad as possible --- particularly given the absence of evidence for weakly-interacting massive particles (WIMPs) at the LHC and direct detection experiments. A simple possibility is that dark matter interacts through a new force that is similar in structure to the known forces but couples only indirectly to ordinary matter. While particle physics has traditionally focused on exploring matter at ever-smaller scales through high-energy experiments, testing this dark-sector hypothesis requires innovative low energy experiments that use high-intensity beams and/or high-sensitivity detectors. 

Since 2009, several hundred physicists at over a dozen experiments have mined a wealth of existing data and proven new techniques to search for dark-sector physics, focused primarily on the possible signal of a new force carrier that decays into pairs of charged particles. In parallel, recent experimental and theoretical work has shown the importance, feasibility, and complementarity of searching for dark sectors by hunting for the dark matter itself. 

Building on this progress, proposed dark-sector experiments over the next decade aim to decisively explore simple sub-GeV dark sectors, and cover as much ground as possible for higher-mass or richer dark sectors, by: 
\begin{itemize} 
\item extending the search for visibly decaying force carriers to higher masses and, for force-carriers lighter than about half the proton mass, fully exploring the range of mixing parameters that can arise from simple quantum effects;
\item extending searches for light dark matter production at accelerators, ultimately probing the minimum dark matter coupling required for simple models of sub-GeV thermal dark matter;
\item extending dark matter direct detection searches to lower energy thresholds, to test models of light thermal or freeze-in dark matter and eventually probe the warm dark matter mass limit of 1 keV. 
\end{itemize}
These three goals are complementary, and must be pursued in parallel to fully explore the physics of dark sectors.  Together, they guide the Dark Sectors program in the US and abroad. A robust dark-sector program entails simultaneous investment in several small-scale experiments, support for facilities that can enable these experiments, and judicious use of existing multi-purpose detectors. The discovery of a dark sector would not only shed light on the mystery of dark matter but also open a window on a whole new sector of the Universe.%fundamental particle physics. 

The best ideas of the community for exploring and revealing this new physics are summarized in this report.

\newpage

% !TEX root = ../WorkshopReport.tex
\section{Introduction}

\subsection{Scientific Context and Goals}

Elementary particle physics seeks to discover and understand the most basic constituents of Nature.  Our current knowledge is encompassed in the Standard Model (SM) of particle physics. 
While the SM is phenomenally successful in describing the physics of familiar matter to high precision, in a wide variety of environments, and over a large energy range, it is also known to be incomplete.  In particular, new physics must be responsible for the dark matter, for neutrino masses, and for the matter-antimatter asymmetry in Nature.

The hunt for physics beyond the Standard Model encompasses several distinct directions.  One of these directions is the search for a \emph{dark sector}, which we define to be a collection of particles that are not charged directly under the SM strong, weak, or electromagnetic forces. Such particles are assumed to possess gravitational interactions, and may also interact with familiar matter through several ``portal'' interactions that are constrained by the symmetries of the Standard Model.  

The exciting possibility of a dark sector is motivated in part by the ease with which dark sectors can explain the known gaps in the Standard Model  (dark matter, neutrino masses, and a baryon asymmetry).  The only well-established features of dark matter (DM) are its lack of strong or electromagnetic interactions and its abundance. A dark sector is a very natural scenario to explain this paucity of interactions, and can readily produce  the observed dark matter abundance through thermal freeze-out or freeze-in, as discussed below.  Likewise, sterile neutrinos --- a very simple dark sector --- are the canonical hypothesis for the origin of neutrino masses. The interactions of either sterile neutrinos or a more complex dark sector can readily introduce the ingredients needed to produce a matter-antimatter asymmetry and transmit it to the baryon sector.     A non-trivial dark sector, with new interactions beyond those of the Standard Model, could leave an imprint on other physics in several ways: self-interactions of dark matter may affect the dynamics of galactic structure formation \cite{Carlson:1992fn,Spergel:1999mh}, and portal interactions can affect precision measurements such as the anomalous magnetic moment of the muon \cite{Blum:2013xva} and the proton charge radius \cite{Pohl:2010zza}; indeed, discrepancies between simulation/theory and experiment in all of these measurements have been suggested as possible hints of a dark sector \cite{Ackerman:mha,Feng:2009mn,Tulin:2013teo,Pospelov:2008zw,TuckerSmith:2010ra,Barger:2010aj,Batell:2011qq}. 

More broadly, the possibility of a dark sector is a generic one, with significant implications for our understanding of the Universe.  Despite this, the physics of dark sectors is poorly tested.  The indirect nature of the interactions of a dark sector with ordinary matter means that it could easily evade detection in particle-physics experiments designed to test increasingly higher energies
--- even if the particles of the dark sector are relatively light.  Rather, finding and studying a dark sector requires dedicated searches at high precision and/or intensity.  Although the definition of a dark sector is extremely broad, its physics can be explored effectively and systematically by using the specific portal interactions as a guide.  

It is natural and pragmatic that the greatest focus in the search for dark sectors has been on the most accessible portal, gauge kinetic mixing, and the mass range where high-intensity searches are most feasible (and where conventional searches are least effective), from masses of MeV to a few GeV.  This range (and above) is also a parameter region of great interest for dark-sector DM, where the portal interaction can establish thermal equilibrium between the DM and ordinary matter in the early Universe. Dark sectors in this mass range are particularly relevant to several of the experimental/observational anomalies noted above.  Within this parameter space, there are several natural targets, discussed in \S\ref{targets}, which serve as milestones against which dark-sector experiments can be judged.  

The last decade has seen tremendous progress in the search for MeV-to-GeV mass dark states.  Naturally, the bulk of early results have come from theoretical studies and reanalyses of existing datasets from experiments designed for other purposes.  A growing sequence of dedicated experiments are presently collecting data \cite{Battaglieri:2014hga,Dharmapalan:2012xp,Andreas:2013lya} and in some cases already publishing physics results \cite{Merkel:2011ze,Merkel:2014avp,Abrahamyan:2011gv}.  There is now a tremendous opportunity for rapid progress and new discoveries in this area, with several new and timely proposals to pursue the most important and well-motivated dark-sector targets with new or existing colliding-beam, fixed-target, and direct-detection experiments. A program of experiments, built with the goal of achieving the milestones articulated in this document, has a tremendous potential to revolutionize particle physics through a discovery of a dark sector. 
Even if no discovery is made, this program will have a significant and lasting impact by dramatically narrowing the range of viable dark matter candidates and dark sector scenarios. 

\subsubsection{Dark Sectors 2016 Workshop}

This report summarizes recent developments, important scientific milestones, promising experimental opportunities, and provides an updated discussion of dark-sector theory. The contents are based on the findings of the Dark Sectors 2016 workshop, held at the SLAC National Accelerator Laboratory in April 2016.  The workshop had several plenary sessions and four working groups: Dark Matter at Accelerators (DMA), ``Visible'' Dark Photons (VDP), Direct Detection (DD), and Rich Dark Sectors (RDS).    Each working group had both experimentalist and theorist conveners who coordinated the presentations and discussions, and  a closeout session presented the findings and recommendations of the working groups. Our report is structured in the same way, with corresponding sections assembled by the conveners to reflect the contents and findings of each working group. 
  While this report serves in part as a summary of the current status of the field (see~\cite{Jaeckel:2010ni,Hewett:2012ns,Essig:2013lka} for other recent reviews), the purpose of Dark Sectors 2016 was to discuss the scientific goals of the field, and the best ideas for new experiments that can achieve these goals. No attempt was made to prioritize one experimental approach over another, as many approaches are still in research and development phases, but relative strengths and weaknesses were identified and have been articulated in this report. 

We focus largely on searches for dark photons, decaying either to SM or dark-sector particles, and searches for dark matter coupled to dark photons.  This allows for a comprehensive survey of the wide-ranging phenomenology present in this scenario, and is sufficient to demonstrate the salient features of the most important experimental approaches that are needed.  We comment as well on the theoretical motivations and experimental prospects for more general dark sectors beyond the vector portal.  We will not review in detail developments in searches for axions or axion-like particles, milli-charged particles, or ATLAS and CMS searches for dark sectors, although some of these experiments are discussed in the context of rich dark sectors. 

Below, we summarize the key interactions that dark-sector experiments can probe, guided by fundamental symmetries of the Standard Model. We then describe important theoretical targets in parameter space that serve as concrete goals for the next generation of experiments. 
We discuss the complementarity of different approaches, emphasizing what is already well known regarding searches for weak-scale physics:~only a comprehensive and multi-faceted approach can probe the range of dark-sector possibilities.  The working group summary sections follow. 

\subsubsection{The ``Portal'' Interactions}

Dark sectors typically include one or more mediator particles coupled to the SM via a {\it portal}. The portal relevant for dark sector-SM interactions depends on the mediator spin and parity:~it can be a scalar $\phi$, a pseudoscalar $a$, a fermion $N$, or a vector $A'$.  The gauge and Lorentz symmetries of the SM greatly restrict the ways in which the mediator can couple to the SM. The dominant interactions between the SM and these mediators are therefore the following SM gauge singlet operators: 
\bea
\mathcal{L} \supset 
\begin{cases}
-\dfrac{\epsilon}{2\cos\theta_W}\, B_{\mu\nu} F'^{\mu\nu}\,, & {\textrm{ vector portal}} \\ 
(\mu \phi + \lambda \phi^2) H^\dagger H \,,&  {\textrm{ Higgs portal}} \\ 
y_n L H N \,, &  {\textrm{ neutrino portal}}\\
\frac{a}{f_a} F_{\mu\nu} \widetilde{F}^{\mu\nu}\,, &  {\textrm{ axion portal.}}
\end{cases}
\eea
Here, $H$ is the SM Higgs doublet with charge assignment $(1, 2, +\frac{1}{2})$ under the SM gauge group $SU(3)_c \times SU(2)_L \times U(1)_Y$, $L$ is a lepton doublet of any generation transforming as $(1,2, -\frac{1}{2})$, $B_{\mu \nu} \equiv \partial_\mu B_\nu -\partial_\nu B_\mu$ is the hypercharge field strength tensor, $F_{\mu\nu}$ ($\widetilde{F}_{\mu\nu}$) is the (dual) field-strength tensor of the SM photon field, $\theta_W$ is the weak mixing angle, and $F_{\mu\nu}'\equiv \partial_\mu A'_\nu -\partial_\nu A'_\mu$ is the field strength of a dark $U(1)_D$ vector boson. 
The first three operators are renormalizable (dimension-4), while the axion portal is dimension-5 and suppressed by some (high) mass scale $f_a$.  These four portals are arguably the most important ones to consider when discussing dark sectors. At the nonrenormalizable level, additional portals can arise from dimension-6 operators
involving a light (or even massless) vector mediator and SM fermions \cite{Dobrescu:2004wz}. Dimension-4 couplings to the SM quarks can also be sizable for a leptophobic vector mediator of mass at the GeV scale \cite{Dobrescu:2014ita}. 

Our focus will be on the vector portal, but we briefly comment on the other portals.  If the mediator is a scalar, it can interact via the Higgs portal. This is probed in various ways, including exotic Higgs decays at high-energy colliders such as the LHC \cite{Curtin:2013fra} (more detailed references are provided in \S\ref{rds:appendix}).  If we require the scalar $\phi$ to be sub-GeV, then various constraints already exist~\cite{Bird:2004ts,Bird:2006jd,Pospelov:2007mp,Schmidt-Hoberg:2013hba,Krnjaic:2015mbs}.  Fermionic mediators $N$ play the role of a right handed neutrino with a Yukawa coupling $y_{\nu}$.  $N$ can itself be a viable, cosmologically metastable (non-thermal) DM candidate in a narrow mass range, $m_N\sim$~keV~\cite{Dodelson:1993je}.  For $m_N$ in the MeV-to-GeV range, there are strong constraints from beam dumps, rare meson decays, and Big Bang Nucleosynthesis~\cite{Deppisch:2015qwa}.
%\re{improve} 
For pseudoscalar mediators, an extensive literature exists, see \textit{e.g.}~\cite{Agashe:2014kda} and references therein. 
 
In what follows, we focus on the vector portal as it is the most viable for thermal models of light DM (LDM).  If the mediator is a vector boson from an additional $U(1)_D$ gauge group under which  LDM is charged, the ``kinetic mixing'' interaction $\epsilon/\cos\theta_W B^{\mu \nu} F^\prime_{\mu \nu}/2$ is  invariant under gauge transformations of both $U(1)_D$ and $U(1)_Y$ \cite{Okun:1982xi,holdom:1985ag}. Here $\epsilon$ is {\it a priori} a free parameter, though it often arises from loops of heavy states charged under both groups, so it is generically expected to be small, $\epsilon\sim 10^{-3}$ or smaller \cite{holdom:1985ag}.  Additionally, its phenomenology is representative of a broader class of well-motivated models, such as the scenarios where the mediator couples preferentially to baryonic, or leptonic, or $(B-L)$ currents. Finally, simple models of the vector portal admit various couplings of DM to the $U(1)_D$ gauge boson. The DM could, for example, have Majorana couplings to the mediator that affect the phenomenology (see~\cite{Izaguirre:2015zva} for an example), or it could live in a rich sector ({\it e.g.,} see~\cite{Morrissey:2014yma}). 

\subsection{Important Milestones}\label{subsec:intro-milestones}

To discuss important vector-portal milestones, we introduce a simple and minimal dark-sector model.  This model can easily be expanded to describe more complicated theories.

\subsubsection{Models}
A starting point for our discussion is the \textbf{minimal kinetically mixed dark photon}, which couples through the vector portal discussed above. The dark photon is a vector field ${A^\prime}_\mu$ with Lagrangian
\bea
{\cal L}_{\rm A'} = -\frac{1}{4}  {F^\prime}^{\mu\nu} F^\prime_{\mu\nu} + \frac{1}{2} \frac{\epsilon}{\cos\theta_W} B^{\mu\nu} F^\prime_{\mu\nu} - \frac{1}{2} m_{A^\prime}^2 {A^\prime}^{\mu} {A^\prime}_{\mu}
\eea
where ${F^\prime}_{\mu\nu} \equiv \partial_\mu {A^\prime}_\nu - \partial_\nu {A^\prime}_\mu$ is the dark photon field strength and $B_{\mu\nu} \equiv \partial_\mu {B}_\nu - \partial_\nu {B}_\mu$ is the Standard Model hypercharge field strength.  This model, parameterized by the dark photon mass $m_{A^\prime}$ and kinetic mixing parameter $\epsilon$, comprises one of the simplest possible dark sectors in its own right, and it can also represent the mediator portion of a larger dark sector. For MeV--GeV-mass dark photons, the dominant effect of this kinetic mixing, after electroweak symmetry breaking, is an analogous mixing $\frac{1}{2} \epsilon F^\prime_{\mu\nu} F^{\mu\nu}$ with the Standard Model electromagnetic field strength $F^{\mu\nu}$.  In the basis with diagonal and canonically normalized kinetic terms, the result of the kinetic mixing is that the dark photon acquires a coupling of strength $e\epsilon$ to the electromagnetic current.

Since the existence of DM  is among the main motivations for a dark sector, it is natural to consider an extension of the minimal dark-photon model that includes a DM candidate. This DM can be a fermion, $\chi$, or a scalar boson, $\phi$, that couples to the dark photon through dark-sector gauge interactions 
\bea
{\cal L}_{DM(f)} & =  &\bar \chi ( i \displaystyle{\not}{D}- m_\chi) \chi,\quad \rm{or} \\\label{dmF}
{\cal L}_{DM(s)} & =  &(D^\mu \phi)^* (D_\mu \phi) -m_\phi^2 |\phi|^2,  \label{dmS}
\eea
where $D_\mu \equiv (\partial_\mu - i g_D A_\mu)$ and $g_D$ is the dark-sector coupling. The dark photon mediates interactions between DM and the SM electromagnetic current that could be observed in
various laboratory experiments. 
 
Because the $A^\prime$ vector boson is massive, the dark gauge symmetry may be spontaneously broken by a dark Higgs boson. In this case, Majorana mass terms may also be allowed,
\bea
\Delta{\cal L}_{DM(f)} & =  &- \frac{\delta}{2} \bar\chi^c \chi+\mathrm{h.c.},\quad \rm{or} \\\label{dmFdelta}
\Delta{\cal L}_{DM(s)} & =  &-\frac{\delta^2}{2}\phi\phi+\mathrm{h.c.}  \label{dmSdelta	}
\eea 
Because these mass terms must vanish with the restoration of the gauge symmetry,  they are naturally small ($\delta$ naturally $\ll m_\chi$).  The effect of such ``inelastic splittings'' on dark matter phenomenology is significant --- they split a Dirac dark matter fermion into two Majorana states (or a complex scalar into two real ones), and the leading dark gauge interaction mediates a transition from the light state to the heavier one, or vice versa \cite{TuckerSmith:2001hy}. The DM present in the halo is typically the lighter Majorana state; if the DM kinetic energy in the halo is insufficient to up-scatter into the heavier state, then signals in direct and indirect detection experiments are absent, while accelerator constraints on DM are largely unchanged relative to the case with unbroken gauge symmetry.

\subsubsection{Targets in Parameter Space}
\label{targets}
While the model above has a broad parameter space, several specific regions are particularly important targets -- either because they are theoretically well-motivated, or because exploring these regions decisively tests outstanding anomalies ({\it e.g.,} $(g-2)_\mu$ \cite{Pospelov:2008zw}), or because they represent a challenging new experimental frontier. 
These targets are:
\begin{itemize}
\item{\bf{Thermal DM freeze-out}}:~Thermal freeze-out of dark matter (DM) annihilations into ordinary matter is a simple and predictive explanation for the origin of the DM abundance measured today.  It is the basis for the ``WIMP  Miracle'' in TeV-scale weakly-interacting DM models but applies equally to dark-sector DM, and predicts a thermally-averaged DM annihilation cross-section of $\approx 3\times 10^{-26}\rm{cm}^3/s$ with only mild dependence on the DM mass. 

For all mediators types and LDM candidates, $\chi$, there is an important distinction between ``secluded'' annihilation to pairs of  mediators (via $\chi \chi$ $\to$ $A^\prime A^\prime$  for  $m_\chi > m_{A'}$) followed by mediator decays to SM particles \cite{Pospelov:2007mp}, and ``direct'' annihilation to SM final states (via virtual mediator exchange  in the $s$-channel, $ \chi \chi \to$ $A^{\prime *} \to$  SM SM for $m_{\chi} < m_{A'}$) without an intermediate step. 

For the secluded process, the annihilation rate scales as 
\bea 
\hspace{-1cm } ({\rm ``secluded" ~annihilation}) ~~~~~~ \langle \sigma v \rangle \sim \frac{  g_D^4 }{  m_\chi^2  } ~~ ~,~~
\eea
where $g_D$ is the coupling between the mediator and the LDM, and there is no dependence on the SM-mediator coupling $g_{\rm SM} = \epsilon e$. Since extremely
small values of  $g_{\rm SM}$ can be compatible with thermal LDM in this regime, the secluded scenario does not lend itself to decisive laboratory tests. For secluded annihilations to be kinematically allowed, $m_{A'} < m_\chi$, which is suggestive of visible dark photon decays.
  
The situation is markedly different for  the direct annihilation regime in which $m_{\chi} < m_{A'}$ where the annihilation rate scales as  
\bea
\hspace{1cm}({\rm ``direct" ~annihilation})~~~~\langle \sigma v \rangle   \sim       \frac{      g_{D}^2 \,  g_{\rm SM}^2  \,  m_{\chi}^2 }{   m_{A'}^4  \!\!}.
\eea
This regime offers a clear, predictive target for discovery or falsifiability, since the dark coupling $g_{D}$  and mass ratio $m_{{\chi}}/m_{A'}$ are 
at most ${\cal O}(1)$, so there is a minimum SM-mediator coupling compatible with a thermal history (larger values of $g_D$ require non-perturbative dynamics in the mediator-SM coupling or intricate model building).
 This mixing target, at the level of 
$\epsilon \sim 10^{-7} m_{A^\prime}^2/(m_\chi \MeV \sqrt{\alpha_D})$ with $\alpha_D = g_D^2/4\pi$, is an important benchmark for both dark photon and dark matter searches.

\item{\bf Direct-detection down to the warm DM mass limit of $\sim$keV}:  
In general, DM in thermal equilibrium with ordinary matter in the early Universe can decouple while it is relativistic if its mass is low enough.  DM that remains relativistic until after matter-radiation equality (\textit{i.e.}~for temperatures $\lesssim 1$~eV), so-called ``hot'' DM, is well excluded by cosmological observations. However, even if DM is heavier and becomes non-relativistic before matter-radiation equality, it can still wash out small-scale structure.  The current lower bound is about 3.3~keV for thermally produced DM and comes from Lyman-$\alpha$ forest measurements \cite{Viel:2013apy}.  Dark matter above this mass is consistent with the known structure formation history of the Universe.  Therefore, reaching the keV-mass threshold with direct detection and other experiments presents an interesting target.  Current direct detection experiments have excellent sensitivity down to DM masses of about a GeV, and a number of planned and proposed experiments promise to extend this range to near a MeV.  Going beyond this, to directly test DM masses in the keV-MeV range, presents a motivated but significant challenge.

\item{{\bf Freeze-in}}:~An alternative to the thermal DM scenario is one where the DM abundance is established via ``freeze-in'' \cite{Hall:2009bx}. In freeze-in scenarios, the dark sector is never in thermal equilibrium with the SM, but out-of-equilibrium scattering gradually populates the DM. Because thermal equilibrium is not established between the dark and visible sectors, the couplings in freeze-in scenarios are typically very small.

Consider a simple example with a kinetically mixed dark photon, $A^\prime$, which can decay to two DM particles, $\chi$. The dark photon is gradually produced through out-of-equilibrium scatterings, and the dark photons then decay to establish the DM abundance. To obtain the observed DM abundance, the couplings must satisfy \cite{Hall:2009bx}
\bea
\frac{\epsilon^2\,m_\chi}{m_{A'}} \sim 10^{-26}\,g_*^{3/2},
\eea  
where $g_*$ is the number of degrees of freedom in the SM plasma at the temperature $T\sim m_{A'}$. 

A variation of this model is to take $m_{A'}$ to be tiny, $m_{A'}\ll$eV.  In this case, the DM abundance is also built up slowly over time through SM particles annihilating to DM particles through an off-shell $A'$.  However, the final DM abundance is independent of $m_{A'}$, and there is only a mild dependence on $m_\chi$ above $\sim 1$~MeV.  For example, for $m_\chi = 100$~MeV, the couplings needed to obtain the correct relic abundance are given 
by~\cite{Essig:2011nj,Chu:2011be}
\bea
\alpha_D \epsilon^2  \sim 3\times 10^{-24}\,.
\eea

Such small couplings are incompatible with a thermal origin of DM, but are motivated by the freeze-in scenario. They are impossible to test at colliders or fixed-target experiments, but novel low-threshold direct detection techniques have an opportunity to probe them in the near future.  It is important to note that such small couplings are only accessible in models with very light mediators, and so this approach is complementary to the accelerator-based searches for mediators in the MeV-GeV range.

\item{\bf{The 2-loop region for Kinetic Mixing ($\epsilon \gtrsim 10^{-6}$), from MeV to GeV energies:}} While the strength of kinetic mixing $\epsilon$ is arbitrary, certain ranges of this coupling are motivated because they readily arise from the quantum effects of heavier particles.   In particular, loops of a heavy particle that carries both hypercharge and $U(1)_D$ charge typically generate $\epsilon \sim 10^{-4}-10^{-2}$ \cite{holdom:1985ag}.  Moreover, the particle content and gauge couplings of the SM raise suggestive hints that, at high energies, hypercharge may be unified with other SM gauge forces into a larger non-Abelian gauge symmetry (\textit{i.e.} a Grand Unified Theory or GUT) \cite{Dimopoulos:1981zb}.   In this case, the enhanced symmetry of the GUT guarantees that $\epsilon$ must vanish at tree-level and that the 1-loop contributions to $\epsilon$ from heavy particles also vanish.  In fact, the leading contribution to $\epsilon$ arises at 2-loop order,  with characteristic size $\epsilon \sim 10^{-6}-10^{-3}$ \cite{delAguila:1988jz}.

Fully exploring the range of kinetic mixing strength that can arise from 1- or 2-loop effects is therefore an important milestone for dark-sector searches (although it should be noted that much smaller $\epsilon$ are also compatible with GUTs, and can arise for example if the mixing is non-perturbative or arises at 3-loop level).  At present, there are plausible strategies to explore the full 2-loop mixing parameter space only for dark photon masses $\lesssim$ 1 GeV --- but this upper mass limit is imposed simply by experimental feasibility, not theoretical motivation.

\item{{\bf Closing the partially visible muon $g-2$ region}}:~The anomalous magnetic dipole moment of the muon, $(g-2)_\mu$, is a property of the muon that is precisely measured and is more sensitive than $(g-2)_e$ to new interactions with GeV-scale mediators. There is a long-standing discrepancy between the SM prediction for $(g-2)_\mu$ and its experimentally measured value \cite{Bennett:2006fi,2012RvMP...84.1527M}. While improved theoretical calculations and experimental precision should shed light on this discrepancy in the future, it is worth investigating how new interactions could account for the anomaly. In particular, dark sector states such as dark photons can account for the discrepancy with $\epsilon\gtrsim 10^{-3}$~\cite{Gninenko:2001hx,Pospelov:2008zw}. The $(g-2)_\mu$ anomaly provides a concrete target that can be probed by dark-sector experiments. 

In the simplest example of a visibly decaying, kinetically mixed dark photon, the entire $(g-2)_\mu$ preferred region is now excluded \cite{Batley:2015lha}. However, a dark photon can decay predominantly into dark-sector states, in which case there is still a region of parameters that can account for $(g-2)_\mu$ and is compatible with other experiments. For example, if $A'$ decays invisibly, or decays partially visibly ($A'\rightarrow \chi_2 \chi_1,\,\chi_2\rightarrow \chi_1+\mathrm{SM}$ for invisible $\chi_1$) as in models of inelastic DM, then $A'$ can explain the $(g-2)_\mu$ discrepancy. Similarly, new vector bosons with suppressed couplings to electrons and light-flavor quarks can evade current constraints and still be compatible with the excess in $(g-2)_\mu$, necessitating dedicated searches for vector bosons with enhanced couplings to muons and taus. 
  
\item{\bf{Direct Dark Photon Searches above a GeV}}: Searches for dark photons above 1 GeV in mass are more challenging for several reasons --- these dark photons have lower production cross-sections and (in the case of visible decays) shorter decay lengths for a given $\epsilon$.  As such, while dedicated fixed-target experiments have a crucial role to play in low-mass dark sector searches, the range $m_{A\prime}\gtrsim 1$ GeV is best explored through searches at multi-purpose colliders --- including high-luminosity B-factories (Belle-II) and high-energy pp collider experiments (ATLAS, CMS and LHCb) \cite{Essig:2013vha,Hoenig:2014dsa,Curtin:2014cca,ilten:2015hya, ilten:2016tkc}.  Given the unique mass reach of these experiments, it is essential to develop techniques and upgrades that can improve their sensitivity to dark sectors and exploit them fully (\textit{e.g.} dedicated trigger).
  
\end{itemize}

\subsection{Complementarity Between Experimental Approaches}

The discussion above shows that, even in a relatively minimal model of DM with a dark photon, a range of distinct phenomenology is expected. An experimental program to test GeV-scale hidden sectors should therefore encompass a wide variety of signatures. Below, we comment on a few cases in which different experiments probe dark sectors in complementary ways, with many more details provided in the body of the report.

In the well-motivated scenario of an invisibly-decaying mediator, there are two main approaches to observing a dark-sector signature:~either a dark photon is produced at a beam dump and promptly decays to invisible states, and the invisible states scatter off of material in a detector placed downstream from the beam dump, or the presence of the invisible states is inferred from missing momentum and/or missing mass (either at a high-energy collider or fixed-target experiment).  The beam-dump method allows for the direct detection of the invisible states, but suffers from a small scattering rate in the detector.  In contrast, the strength of missing-momentum techniques is that their sensitivity scales only with the production cross section, and no additional rate penalty is required to re-scatter states downstream. The technique is also independent of the dark-sector coupling so long as any visible component of the dark-photon decay is not vetoed by the experimental search strategy.  

There also exists a complementarity between direct detection experiments and accelerator-based probes. In the \mev-\gev~mass range, both direct detection and accelerator-based techniques can probe elastic scattering DM scenarios, but only accelerator based experiments are able to broadly explore scenarios such as inelastic DM. Beam-dump and missing-momentum experiments are also readily capable of probing meta-stable states unrelated to DM, or any sub-dominant components of the total DM density.  On the other hand, the prospects for direct detection look better than for accelerator techniques in the case of ultra low-mass (sub-\kev) mediators, where the elastic cross section is enhanced by the small momentum transfer in scattering. If the dark photon itself is the DM, direct detection experiments can have excellent sensitivity to DM absorption (and similar absorption signals can arise for scalars or pseudoscalars). Perhaps most exciting, there are several scenarios discussed above in which different experimental approaches can see a signal for the same model, allowing for different aspects of the dark-sector scenario to be studied comprehensively.  

A variety of experimental approaches are also motivated by rich dark sectors. A dark sector that possesses a large number of particles and interactions can modify some of the relationships used in defining the target parameter spaces listed above, and this can lead to new signatures in all types of dark-sector experiments. To ensure broad coverage of such signatures, we advocate for the development of new avenues of communication between theorists and experimentalists to ensure that, where possible, existing experiments adopt strategies that broaden, rather than narrow, their sensitivities to new particles and interactions. It is also important to develop new computational tools that facilitate experimental studies of rich dark sectors and provide a link between experimental signatures, and astrophysical and cosmological properties of dark-sector models.

\newpage

\section{Visible Dark Photons}\label{sec:VDP}
\vspace{-0.3cm}
\begin{flushleft}
\textit{Conveners:~Jim Alexander, Maxim Perelstein.~Organizer Contact: Tim Nelson}
\end{flushleft}

\subsection{Theory Summary}

The Standard Model (SM) of particle physics describes strong, weak, and electromagnetic interactions in terms of a gauge theory based on the $SU(3)\times SU(2)\times U(1)_Y$ symmetry group. While phenomenologically successful, the model does not provide insight into the origin of this symmetry. It is quite possible that a more complete theory of nature will include additional gauge interactions. Additional gauge groups appear in many theoretical extensions of the SM, such as supersymmetric models or string theory. In addition, the existence of dark matter motivates extending the SM to include a ``dark sector", consisting of fields with no SM gauge charges. The dark sector may well include additional gauge symmetries. In fact, as discussed in the Introduction, an Abelian gauge boson of the dark sector can provide a natural ``portal" coupling between the dark sector and the SM. This motivates experimental searches for non-SM gauge bosons associated with such extended symmetry structures, and this section will discuss such experimental searches. 

Our focus will be on accelerator experiments looking for gauge bosons with masses roughly between 1 MeV and 10 GeV. The lower bound of this range is defined primarily by the existing bounds from accelerator experiments, cosmology, and astrophysics.
%, which imply that lighter gauge bosons are only allowed if their interactions with the SM matter are extremely weak. 
The upper bound is dictated by the kinematic reach of the high-intensity accelerator facilities considered here. Of course, these searches are complemented by the experiments at energy-frontier facilities such as the LHC, which are sensitive to extra gauge bosons with higher masses, up to a few TeV, albeit with lower sensitivity to the portal couplings. 

The production of non-SM gauge bosons in collider experiments  relies on the couplings of the new vector bosons to SM particles, primarily electrons and quarks. In the simplest scenario, such couplings arise from the ``kinetic mixing" interaction, which mixes the gauge boson of a non-SM ``dark" gauge group $U(1)_D$ with the SM photon:
\beq
{\cal L}_{\rm kin.mix.} = \frac{1}{2} \epsilon F^{\mu\nu} F^\prime_{\mu\nu}.
\eeq{kinmix}
Here $F$ and $F^\prime$ are field strength tensors of the SM $U(1)_{\rm em}$ and the dark $U(1)_D$, respectively, and $\epsilon$ is a dimensionless parameter~\footnote{Since the energies of the experiments considered here are well below the electroweak scale, it is appropriate to  consider only the $U(1)_{\rm em}$ gauge group of the SM. In a more fundamental theory above the weak scale, the kinetic mixing would involve the $U(1)_Y$ group instead; such a theory would always reduce to~\leqn{kinmix} below the weak scale as discussed in the Introduction.}. This coupling generically arises in theories that include new fields charged under both $U(1)_D$ and $U(1)_{\rm em}$. If the kinetic mixing appears at the one-loop level, $\epsilon$ can be estimated to be in the range $\sim 10^{-4}-10^{-2}$. In some cases, the one-loop contribution to the kinetic mixing may vanish; for example, this occurs if the heavy states that induce it appear in multiplets of an $SU(5)$ or a larger Grand Unified Theory (GUT) group. In this case, the leading contribution is at two loops, and $\epsilon\sim 10^{-6}-10^{-3}$, with values as low as $10^{-7}$ possible if both $U(1)$'s are in unified groups. Notice that since kinetic mixing is a marinal operator, these estimates are independent of the masses of the heavy particles that give rise to it. 

\begin{figure}[t]
\begin{center}
\includegraphics[width=150mm]{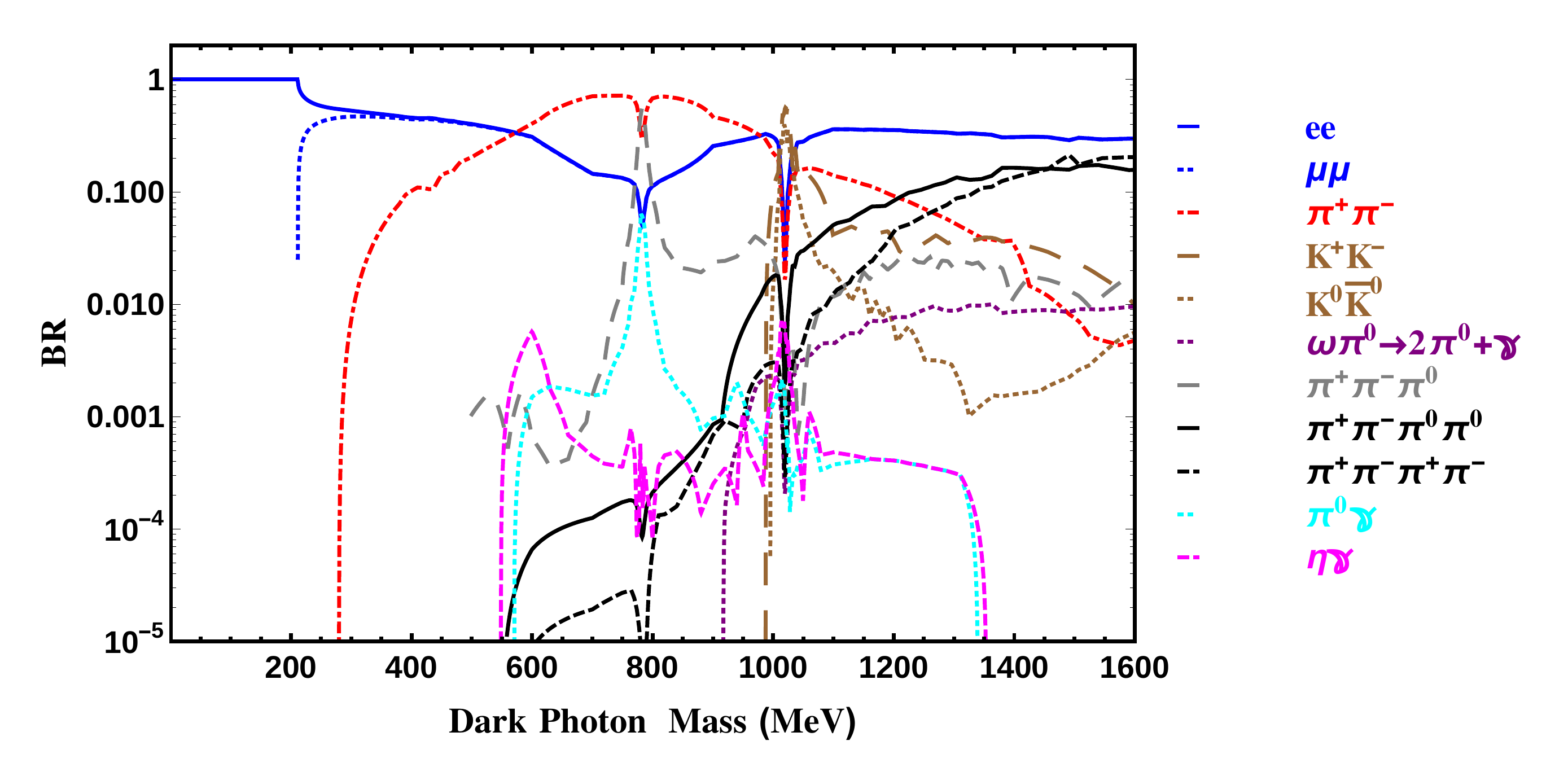}
\caption{Visible dark photon decay branching ratios (figure from Ref.~\cite{Liu:2014cma}).}
\label{fig:BRs}
\end{center}
\end{figure}

The physical consequences of the kinetic mixing are best understood in the basis where the kinetic terms are canonical. In this basis, the theory contains two gauge bosons, the ordinary photon $A$ and the dark photon $A^\prime$. The interactions between the dark photon and SM particles are described by
\beq
{\cal L}_{\rm int} = \epsilon\, e A^\prime_\mu J_{\rm EM}^\mu\,,
\eeq{Aprime_coupling}    
where $J^\mu_{\rm EM}$ is the usual electromagnetic current, so that the $A^\prime$ couplings to SM particles are proportional to their electric charges. The interaction in~\leqn{Aprime_coupling} is responsible both for the production of dark photon in SM particle collisions, and also for its decays into SM states. The simple structure of the interaction leads to a highly predictive theory:~for example, the predicted branching ratios of $A^\prime$ decays are shown in Fig.~\ref{fig:BRs}. 

In addition, the dark photon may couple to other non-SM particles in the dark sector:~for example, there may be new matter states charged under $U(1)_D$, which may include particles that constitute dark matter. If decays of the dark photon to the dark-sector states are kinematically forbidden, such couplings are irrelevant to the phenomenology of the experiments discussed here, and the branching ratios of Fig.~\ref{fig:BRs} hold. This case is referred to as the {\it ``visible dark photon"} model, and is the focus of this section of the report.
If, on the other hand, the dark photon can decay into dark-sector states, the branching ratios into the SM would be (uniformly) reduced. In the simplest case, the dark sector decays of the $A^\prime$ would not be seen by the standard particle detectors, and are referred to as ``invisible" (it is possible that dedicated downstream detectors may be sensitive to long-lived dark sector states produced in $A^\prime$ decays; this will be discussed in the Dark Matter at Accelerators section of this Report). Such invisible decays can nevertheless be detected by using missing-mass or missing-momentum techniques, as discussed below, and thus are included in this section. Depending on the model of the dark sector, $A^\prime$ decays into mixed final states containing both SM and dark sector particles are also possible. The large variety of possible final states puts a premium on search approaches that are insensitive to the specific decay channel, such as the missing-mass technique.

\subsection{Strategies for Dark Photon Searches}

Current and planned dark photon searches can be characterized by their strategies for production and detection of the dark photon.  The main production channels include:
\begin{itemize}
	\item  {\bf Bremsstrahlung:} \brem, for electrons incident on a nuclear target of charge $Z$. In a fixed-target configuration the \dark is produced very forward, carrying most of the beam energy (for 
$E_\mathrm{beam}\gg\mdark$) while the electron emerges at a larger angle.  Acceptance for the forward-moving \dark can be nearly complete; high-resolution spectrometers with lower acceptance require higher-current beams. Mass reach extends nominally up to beam energy but falls rapidly with mass. Proton-beam fixed-target experiments also exploit bremsstrahlung production, \pbrem. 
	\item {\bf Annihilation:} $\ee\to\gamma\dark$.  This production process is favored for searches that emphasize invisible \dark decay modes, in which the unseen \dark is reconstructed as a missing-mass; visible modes can also contribute. Annihilation channels are pursued in both fixed-target experiments with \pos beams, and \ee collider experiments; a proposal has been made for a very asymmetric collider that would cover a more extended range of masses~\cite{bw}. The accessible \dark mass in all cases is limited by $\sqrt{s}$. 

	\item {\bf Meson decay:} Dalitz decays, $\pi^{0}/\eta/\eta^{\prime} \to \gamma\dark$, and rare meson decays such as $K\to\pi\dark$, $\phi\to\eta\dark$, and $D^{\ast}\to D^{0}\dark$, may produce low-mass dark photons if their coupling to quarks is nonzero.  Hadronic environments, either in colliders or fixed-target setups, offer copious meson production and make this a favored production channel.  The rare meson decay mechanism plays a role in \ee colliders, {\it e.g.} $\ee\to\phi$ (KLOE, KLOE-2). %\cite{Aubert:2008as}
The \dark mass reach is limited by the parent meson mass.
	\item {\bf Drell-Yan:}  $q\bar q \to \dark \to (\elel~ \text{or}~ \hh)$. This process is useful in hadron colliders and proton fixed-target experiments.

\end{itemize}
The methods of \dark detection may be broadly summarized as follows:
\begin{itemize}
	\item {\bf Bump hunt in visible final-state invariant mass:} $\dark\to\ell^{+}\ell^{-}$ or $\dark\to h^{+}h^{-}$ against high background. Firm control of statistical and systematic issues is necessary to achieve reliable results when $S/{B}$ may be in the $\sim 10^{-6}-10^{-4}$ range. 
	\item {\bf Bump hunt in missing-mass:} In $\ee\to\gamma\dark$ or meson decay production channels, invisible \dark decay may be detected indirectly as a bump in a missing-mass distribution. 
The visible SM part of the final state is reconstructed; the initial state must be known. The same challenges as noted above apply to missing-mass bump hunts.
	\item {\bf Vertex detection} in $\dark\to\ell^{+}\ell^{-}$. The \dark decay length scales with $(\epsq m_{\dark})^{-1}$, implying that searches for displaced vertices in visible decay modes probe the very low-\eps regions of parameter space. Experiments with vertex reconstruction also do visible invariant mass bump hunts.
\end{itemize} 

The three detection strategies listed above correlate with regions of the $(\epsq,\mdark)$ parameter plane. This is illustrated in Figure 1, which shows a cartoon of the sensitivity regions for the three generic experimental approaches. The horizontal axis maps the available kinematic reach; the vertical axis is determined by integrated luminosity (increasing downwards);  and the diagonal direction corresponds to increasing decay length. 
The gap between regions A and B, which has come to be called ``Mont's Gap'' after JLAB Director Hugh Montgomery's observation that HPS coverage in coupling strength was incomplete, highlights the challenge to fill in the transition region between bump hunts and displaced vertex searches by either increased luminosity (for bump hunts) or improved vertex resolution for short decay lengths, or both.

\begin{figure}[t!]
\begin{center}
\includegraphics[scale=0.5]{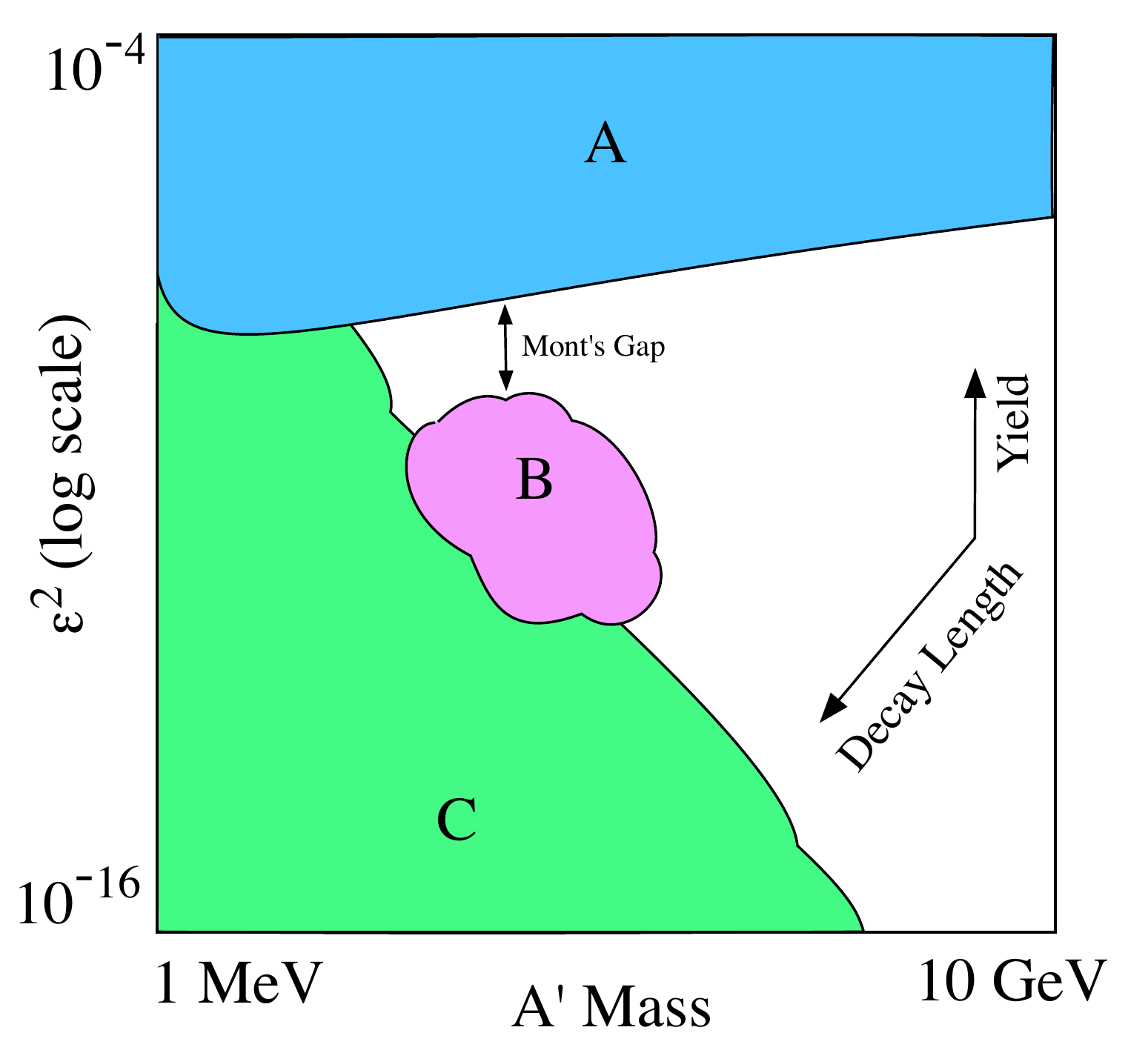}
\caption{Cartoon of \epsq vs. \dark mass parameter plane. Region A: bump hunts,  visible or invisible modes. Region B: displaced vertex searches, short decay lengths; Region C: displaced vertex searches, long decay lengths.}
\label{default}
\end{center}
\end{figure} 

\subsection{Brief Summary of Existing Constraints}

In the case of the visible dark photon model, the current experimental situation is summarized in Fig.~\ref{f:visdarkpho}. Experiments searching for a bump in $\ell^+\ell^-$ invariant mass distribution rule out values of  $\epsilon$ above $\sim 10^{-3}$ in the 10 MeV-10 GeV mass range, with the strongest bounds coming from NA-48/2~\cite{Batley:2015lha}, A1~\cite{Merkel:2014avp} and BaBar~\cite{Lees:2014xha} experiments. These experiments employ a variety of dark photon production mechanisms, including meson decays (NA-48/2), bremsstrahlung (A1), and annihilation (BaBar). They are complemented by beam dump experiments, such as E141~\cite{Riordan:1987aw} and E137~\cite{Bjorken:1988as} at SLAC, E774~\cite{Bross:1989mp} at Fermilab, and others, which place {\it upper} bounds on $\epsilon$ as explained above. There is also a constraint from the measurement of the anomalous magnetic moment of the electron $a_e$ ~\cite{Pospelov:2008zw}. Together, the existing constraints already rule out the possibility that the visible dark photon model can explain the observed deviation of the anomalous magnetic moment of the muon from the SM prediction. Dark photon masses below 10 MeV are also essentially ruled out. However, large part of the parameter space remain unexplored, including the region suggested by the ``2-loop target" (see Introduction).

The situation is significantly less constrained in the case of dark photons with a significant decay branching fraction to dark-sector (``invisible") final states, see Fig.~\ref{f:invisdarkpho}. In this case, the strongest bounds come from the E787~\cite{Adler:2004hp} and E949~\cite{Artamonov:2008qb} kaon decay experiments at BNL, as well as BaBar~\cite{Aubert:2008as}. The bound from the $a_e$ measurement also applies, since it relies on virtual dark photon contribution and as such is insensitive to the $A^\prime$ decay mode. Note that a large part of the parameter space where the dark photon could explain the $a_\mu$ anomaly is still allowed in this case.    
   
\subsection{Thumbnail summaries of ongoing and proposed experiments}

We summarize briefly the main features of the current and proposed experiments searching for dark photons by any or all of the techniques reviewed above.\\
\par

{\bf Electron Beams:}
\begin{itemize}

\item {\bf APEX (JLab):} Production is by electron bremsstrahlung in Jefferson Lab's Hall A, with 120 $\mu$A beam current.  The experiment uses a high-resolution pair spectrometer with low acceptance, $\sigma_{M}/M\sim0.5\%$, covering the \dark mass range $65<\mdark<600$ MeV.  Detection strategy is the \mdark bump hunt. First publication was in 2010, with 2nd run scheduled for 2018.   
References:~\cite{Abrahamyan:2011gv,Essig:2010xa}.

\item {\bf A1 (Mainz):} Production by electron bremsstrahlung, in the Microtron beam at Mainz (180-855 MeV; 100 $\mu$A). Using a high resolution spectrometer, the experiment conducts a bump hunt in visible decays $\dark\to\ee$.  The parameter coverage is $40\,\mev < \mdark < 300\,\mev$, with \epsq reaching to $8\power{-7}$. First publication was in 2011. Future plans include a possible upgrade to include displaced-vertex reconstruction. References:~\cite{Merkel:2011ze,Merkel:2014avp}

\item {\bf HPS (JLab):} Production by electron bremsstrahlung, $20-200$ MeV \mdark reach, $1-2$ MeV \mdark resolution. The experiment is performed in the CEBAF \elec beam with $50-500$ nA current in the range $1-6$ GeV. The experiment has high acceptance and vertex detection by silicon trackers, with vertex resolution in the range $1-5$\,mm, depending on \mdark. Data taking has started, with first publications expected in 2016-2017; upgrades are envisioned for 2018, together with higher beam energies. %; 10\% of the allocated run time has been used. 
Two search strategies are employed:~\mdark bump hunt, and displaced vertex detection, leading to two exclusion islands in the $(\epsq,\mdark)$ parameter plane. Anticipated upgrades aim to close Mont's Gap with improved vertexing and increased luminosity. References:~\cite{Battaglieri:2014hga}.

\item {\bf DarkLight (JLab):} Production by electron bremsstrahlung in Jefferson Lab's Low-Energy Recirculator Facility with an electron beam of 10 mA at 100 MeV. The experiment uses a windowless gas target in the recirculating beam, searching for the visible $\dark\to\ee$ final state. A silicon layer detects the recoil proton, implying possible sensitivity to invisible decays (\emph{i.e.}, fully-known initial state). Initial engineering run (2012) demonstrated the use of the internal gas target; the next stage (2016) will include operation of the solenoid, with future runs adding detector elements. References:~\cite{Balewski:2013oza,Balewski:2014pxa}.

\item {\bf MAGIX (Mainz):} Production by electron bremsstrahlung in the future ERL beam MESA, designed for 1 mA current at 155 MeV, with a projected luminosity of ${\cal L}\sim 1\power{35}$. The experiment will use a windowless gas-jet or cluster-jet target and deploy a high resolution spectrometer to reconstruct \ee final states.  It is expected to probe a mass range $10\,\mev < \mdark < 60\,\mev$ with \epsq reaching to $3\power{-9}$ at lower masses.  As with DarkLight, the possibility to observe the recoil proton from the initial bremsstrahlung event opens the door to a missing-mass approach and sensitivity to invisible decay modes.  Earliest operation expected in $\sim2019+$. Reference: \cite{Denig:2016dqo}.

\item {\bf NA64 (CERN):} 100 GeV secondary electrons at CERN's SPS. Range of currents: up to  $2\times 10^6$ e$^-$/spill. Pulse structure: 2-4 spills of 4.8 s per minute. The electron beam absorption in an upstream calorimeter (ECAL1) is accompanied by the emission of bremsstrahlung \dark in the reaction $eZ\rightarrow eZ\dark$. A part of the primary beam energy is deposited in the ECAL1, while the rest of the energy is transmitted by the \dark through the ECAL1 and deposited in another downstream calorimeter ECAL2 by the $e^+e^-$ pair in $\dark\rightarrow e^+e^-$. The \dark production signature is an excess of events with the two-tracks in a tracker and two-shower signature in the ECAL1 and ECAL2.  Timeline: request for 2 weeks test beam in 2017, and 6 weeks physics run in 2018. Expected to collect a few $10^{11}$ EOT. After CERN long shutdown (2019-2020) expected to collect more than $10^{12}$ EOT.

\item{\bf ``SuperHPS'' (SLAC):} Production by electron bremsstrahlung using the proposed DASEL (``{\bf DA}rk {\bf S}ector {\bf E}xperiments at {\bf L}CLS-II'') electron beam at SLAC.  SuperHPS is similar to HPS (see above), with improved acceptance, mass resolution, and rate capability.
Reference:~\cite{superhps-a,superhps-b}.

\item{\bf ``TBD'' (Cornell):} Production by electron bremsstrahlung using the Cornell-BNL FFAG-ERL Test Accelerator (CBETA) high-intensity electron ring under construction at Cornell.  Expected machine parameters are I=100mA at 76 MeV, I=80 mA at 146 MeV, I=40 mA at 286 MeV. First beams expected $\sim$2019; experimental proposals are welcomed. Reference:~\cite{cbeta}.

%\item {\bf A2:} High intensity photoproduction of pseudoscalar mesons in the Mainz Microtron could be used in the future for \dark searches via meson decay mode. 

\end{itemize}

{\bf Positron Beams:}
\begin{itemize}
\item {\bf VEPP3 (BINP):} \ee annihilation production in 500 MeV circulating \pos beam with internal hydrogen gas target providing ${\cal L}=10^{33}\lumiunits$. The \mdark range is $5-22$ MeV, with mass resolution $\sim 1\,\mev$. Missing-mass mode reconstruction allows inclusive detection of invisible \dark decays, with visible decays also possible; detection is by missing-mass bump hunt.  First run is anticipated for 2019-2020.   References:~\cite{Wojtsekhowski:2009vz,Wojtsekhowski:2012zq}

\item {\bf PADME (Frascati):}  \ee annihilation production by 550 MeV \pos beam incident on thin, active diamond target. The experiment is sensitive to invisible decays, detected by bump hunt in the \dark missing-mass distribution; visible modes are also detected and explicitly reconstructed. The \dark mass reach is up to 24 MeV. Rich final states such as $\ee\to h^{\prime}\dark\to\dark\dark\dark$ will also be searched for. First run is anticipated for end 2018, with a possible future upgrade to 1 GeV beam energy. References:~\cite{Raggi:2014zpa}

\item {\bf MMAPS (Cornell):}  \ee annihilation production  with 6.0 GeV \pos beam of 2 nA incident on thick beryllium target. Expected luminosity is ${\cal L}=10^{34}\lumiunits$.  As with VEPP3 and PADME, detection is via the missing-mass mode, providing sensitivity to both invisible and visible final states.  \dark mass reach is $20-78$ MeV, with \mdark resolution $10-1$ MeV over the same range. No first-run date is set yet. Reference:~\cite{cornell}.
\end{itemize}

{\bf $\mathbf{e^{+}e^{-}}$ Colliders:}
\begin{itemize}
\item {\bf Belle-II (KEK):} \ee annihilation at $\sqrt{s}\sim10$ GeV with sensitivity to visible ($\gamma\elel$) and invisible (mono-photon) modes.  Belle-II will also search for rich final states, such as $\ee\to h^{\prime}\dark\to\dark\dark\dark$.  Trigger strategies are under active development for monophoton modes, taking advantage of Belle-II's no-projective-cracks design.  The range of sensitivity in \dark mass is  $20\,\mev-10\,\gev$.  The first Belle-II physics run is anticipated in 2018.
References:~\cite{TheBelle:2015mwa}.

\item {\bf KLOE2 (Frascati):} Production modes include meson decay ($\phi\to\eta\dark$), annihilation ($\ee\to\dark\gamma$), and dark-higgsstrahlung ($\ee\to\dark h^{\prime}$).  
The \daph-2 upgrade triples previous luminosity, and KLOE2 upgrades include tracking GEM vertex detector and forward calorimetry.
 Searches include both visible (\dark$\to$ \ee, \mumu, $\pi\pi$) and invisible \dark decay modes.  Parameter reach in KLOE is  $\epsq<1\power{-5}\sim1\power{-7}$, while KLOE2/\daph2 will improve limits by a factor of two. References:~\cite{Archilli:2011zc,Babusci:2012cr,Babusci:2014sta,::2016lwm,kloe-ee,Babusci:2015zda}.
\end{itemize}

{\bf Hadron Beams:}
\begin{itemize}
\item {\bf NA62 (CERN):} Using a high intensity kaon source, the primary goal is to measure $Br(K^{+} \to \pi^{+}\nu \bar{\nu})$ with 10\% accuracy and with 10:1 signal to background ratio; the result is then
also used to derive an upper bound for 
$Br(K^+ \to \pi^+\dark), ~\dark\to \chi \chi$ process. 
Additionally, decay products of $\pi^0$, $\eta$, and $D$ mesons produced at the target are expected to be used for \dark searches.
400 GeV protons in slow extraction from the CERN SPS strike a beryllium target to produce a 750MHz hadron beam at 75 GeV, from which $K^{+}$ are selected. Projected accumulation by 2018 is $10^{19}$ protons on target.

\item {\bf SeaQuest (FNAL):} 120 GeV protons on target from FNAL Main Injector. Dark photon production is by Drell-Yan, meson decay, and proton-bremsstrahlung. The \dark search will look for muon pairs emerging from a beam dump; the primary signal is a bump in the dimuon mass distribution, with vertex requirements helping to suppress backgrounds. A parasitic run in 2017 with slightly augmented triggering and
electromagnetic calorimetry will establish baseline performance. Further running with more extensive upgrades for particle ID could follow. The expected range of sensitivity is $200\,\mev < \mdark < 10\,\gev$ for the bump-hunt mode, and up to 2 GeV for the displaced vertex mode. Alternatively, by analyzing existing and anticipated 
dimuon data from FNAL E906/SeaQuest, a 95\% CL limit on a dark photon mass from 215-5600 MeV should be possible. References: \cite{gardner:2015wea}

\item {\bf SHiP (CERN):} An ambitious, broad-spectrum search for hidden particles, using the 400 GeV, proton beam from the CERN SPS. For dark photons, production can be accomplished by proton Bremsstrahlung, Drell-Yan, QCD Compton scattering, and meson decay. SHIP is sensitive to visible final states with long decay lengths ($\sim$10's m) producing a displaced-vertex signature. Dark photon parameter sensitivity is in the range $10^{-18}<\epsq<10^{-8}$ and $\mdark<10\,\gev$, covering an extensive zone in the displaced vertex lobe. The projected running period is 2026-2031, with an integrated accumulation of $2\power{20}$ protons on target. Reference: \cite{Anelli:2015pba}.
\end{itemize}

{\bf Proton-proton collisions:}
\begin{itemize}
\item {\bf LHCb (CERN):} Notable production mechanisms include exclusive rare heavy quark decay modes such as $D^{\ast}\to D^{0}\dark(\to\ee)$ and $B\to K^{\ast}\dark(\to\mumu)$ for low mass coverage, $\mdark < 140\,\mev$. Inclusive visible decays $\dark\to\mumu$, with and without displaced-vertex reconstruction provide sensitivity to tens of \gev, punctuated by exclusion zones to remove known resonances. Triggerless operation foreseen in the upgrade for LHC Run 3 (2021-2023) and beyond is expected to increase sensitivity reach substantially. 
References: \cite{ilten:2016tkc,aaij:2015tna,ilten:2015hya}.
\end{itemize}

{\bf Stopped Muons:}
\begin{itemize}
\item {\bf Mu3e (PSI):} The experiment
searches primarily for the charged lepton number violating (cLFNV) decay
$\mu^+ \rightarrow e^+\, e^+e^-$. The primary beam is 2.3 mA protons at 590
MeV; particle tracking with silicon will achieve 0.3\gev\ resolution, 
and timing with scintillating fibers will provide 100ps resolution. 
Experimental sensitivity in  $Br(\mu \rightarrow 3e)$ is expected to
reach $10^{-15}$ in 2018, and $10^{-16}$ after beam intensity upgrades in 2020.
A bump hunt search will be carried out in the $e^+ e^-$ invariant mass 
spectrum, up to the muon rest mass.  
References:~\cite{Blondel:2013ia,Echenard:2014lma}.
\end{itemize}

\subsection{Projections for future experiments}

\begin{figure}[htbp]
\begin{center}
\includegraphics[scale=0.4]{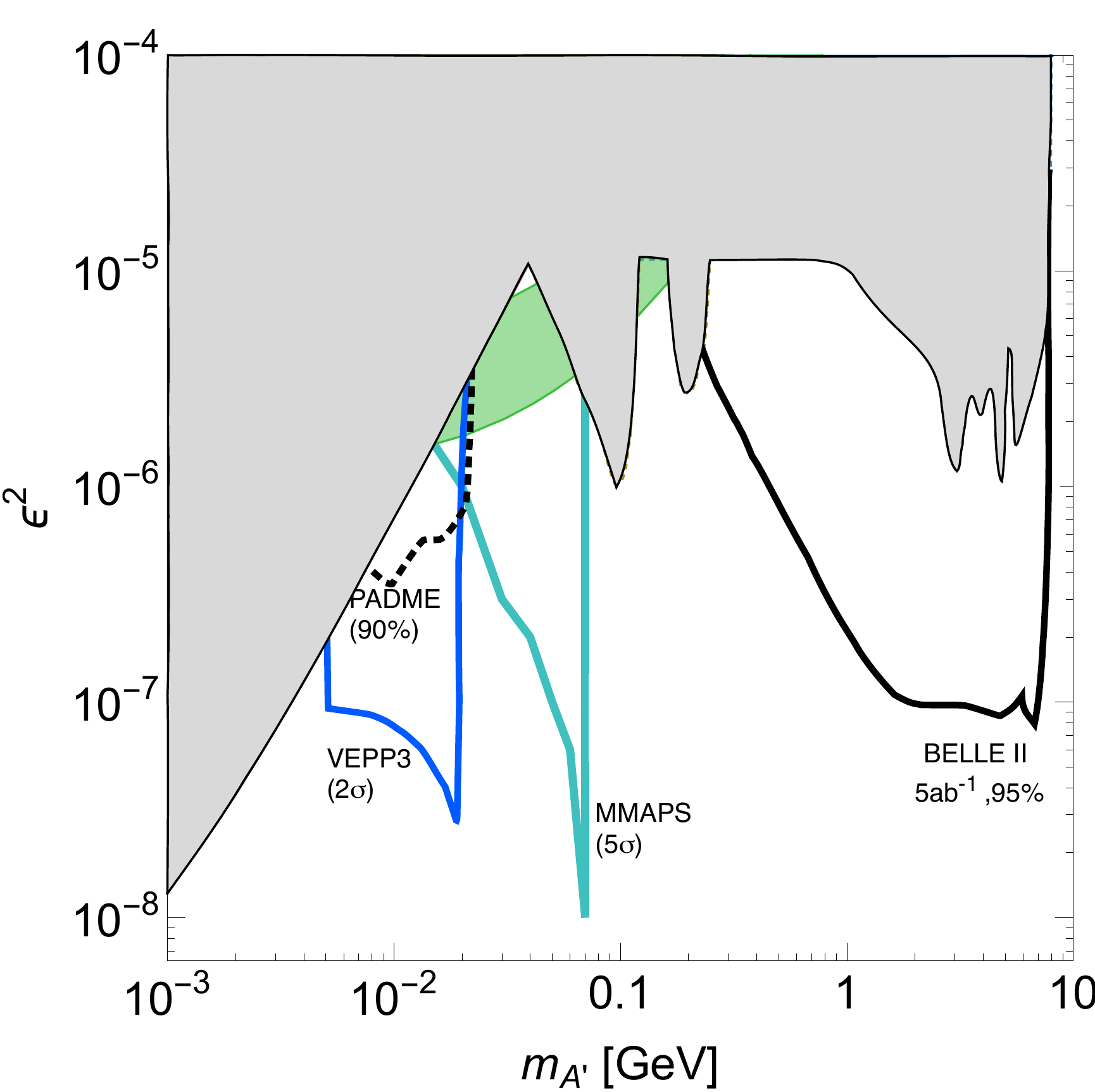}
\caption{\dark sensitivity for missing-mass experiments, allowing invisible decay modes. Existing exclusions, shown in gray, have been smoothed. }
\label{f:invisdarkpho}
\end{center}
\end{figure}

\begin{figure}[t!]
\begin{center}
\includegraphics[scale=0.45]{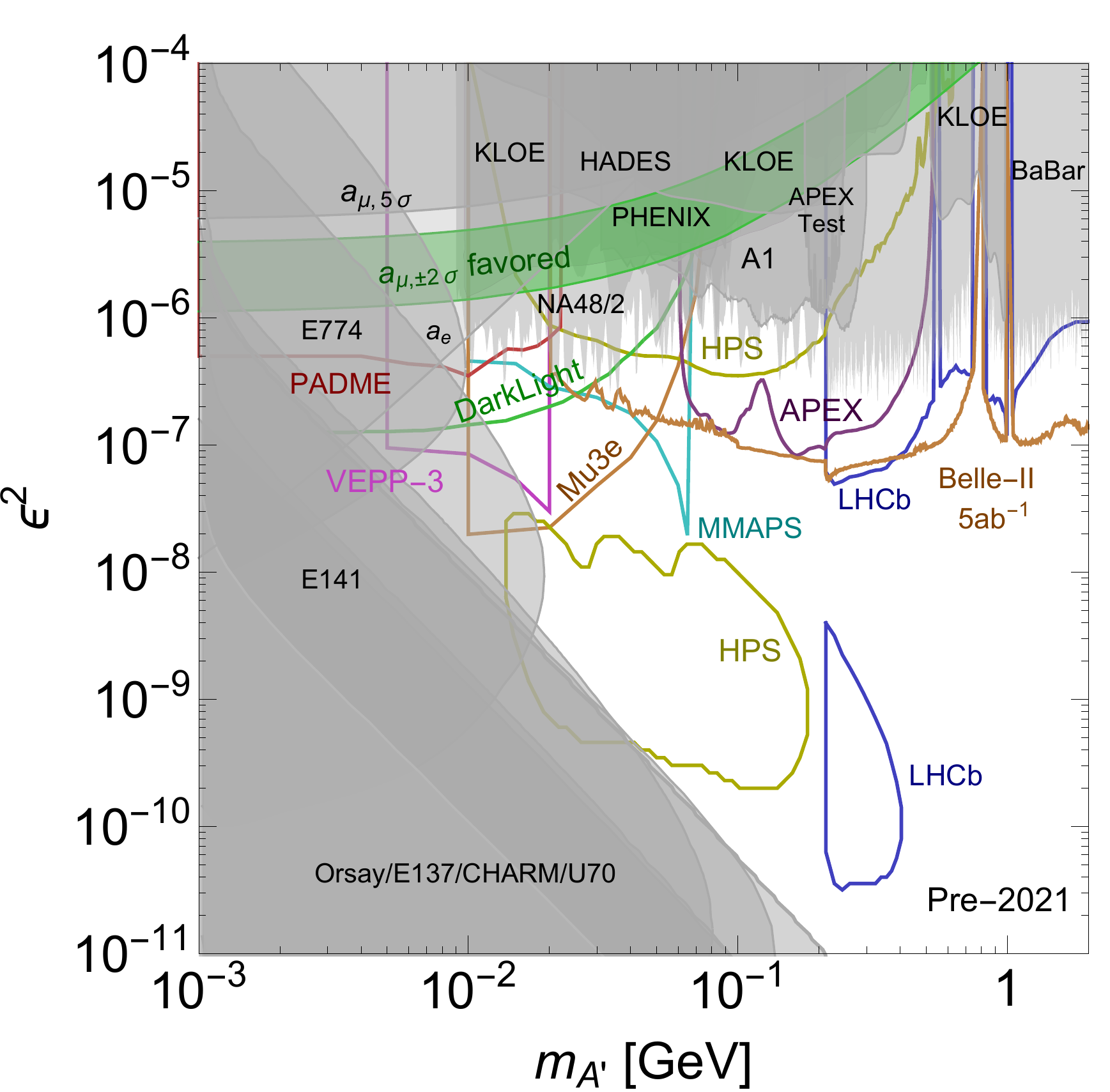}
~~\includegraphics[scale=0.45]{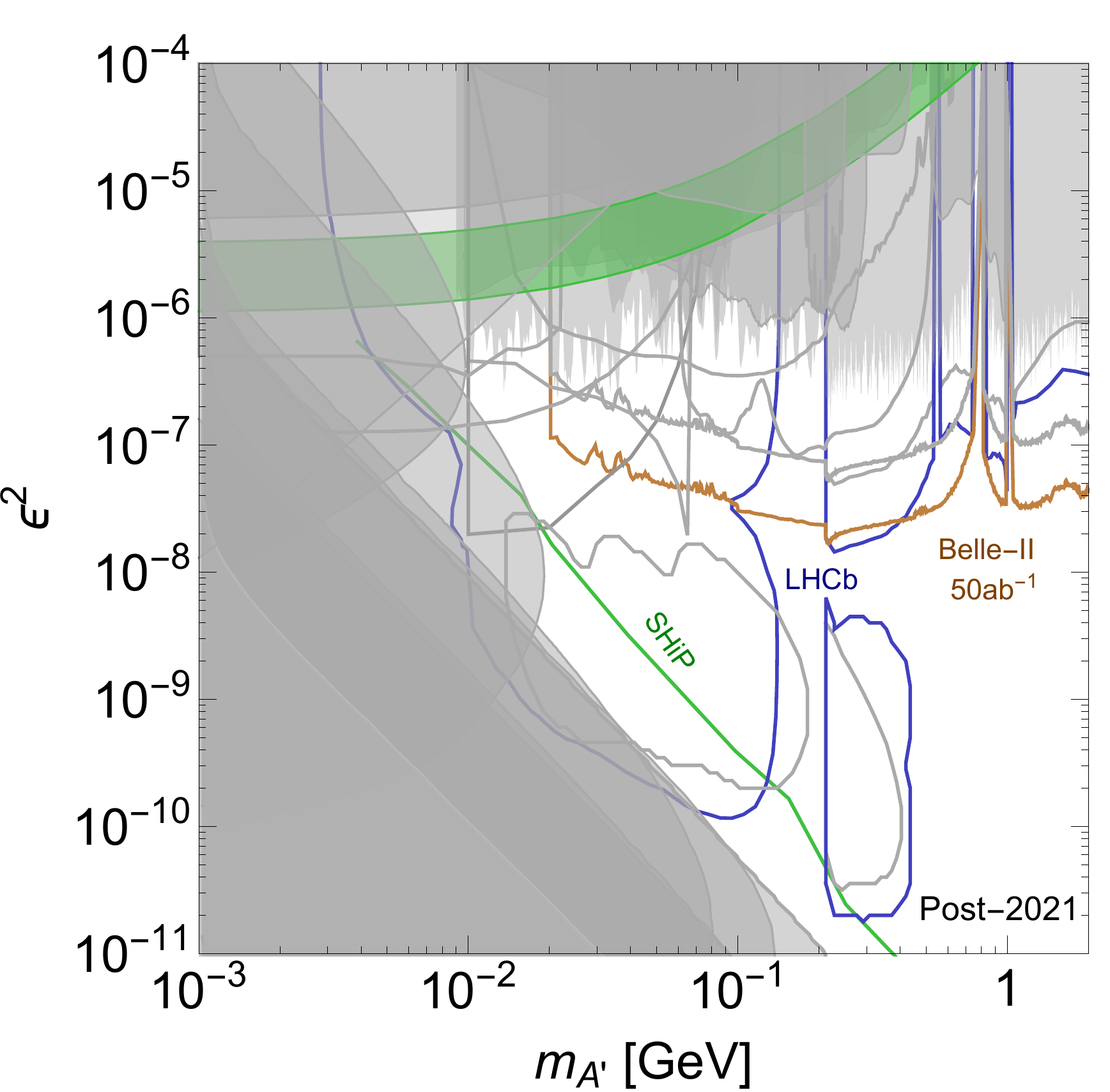}
\caption{Sensitivity to \dark for exclusive experiments seeking visible decay modes $\dark\to\elel$. 
{\bf Left:} Experiments capable of delivering results over the next 5 years to 2021. 
Shaded regions show existing bounds.  Green band shows $2\sigma$ region in which an \dark can explain the 
discrepancy between the calculated and measured value for the muon $g-2$.  
{\bf Right:} Longer term prospects beyond 2021 for experimental sensitivity.  All projections on left plot are repeated in gray here.  
Note that LHCb and Belle-II can probe to higher masses than 2 GeV and SHIP can probe to lower values of $\epsilon$ than indicated. 
%Mu3e reference:~\cite{Echenard:2014lma}.
}
\label{f:visdarkpho}
\end{center}
\end{figure}

Figures \ref{f:invisdarkpho} and \ref{f:visdarkpho} illustrate the dark-photon parameter plane, \epsq versus \mdark, with existing exclusion zones indicated in gray, and anticipated 
%$5\sigma$ 
exclusion reaches of planned experiments indicated by colored curves.  Table \ref{tab:BigTable} summarizes actual and/or projected performance and characteristics of dark photon experiments.

\newpage
\newcommand{\ang}{80}
\begin{sidewaystable}[htp]
\label{tab:BigTable}
\begin{adjustwidth}{-0cm}{+0cm} % to accomodate wide table
\begin{center}
\renewcommand{\arraystretch}{1.5}
\caption{Summary of dark photon experiments. }
\small
\begin{tabular}{|l|c|c|c|c|c|c|c|c|c|c|c|c|}
\hline
          \rotatebox{0}{Experiment} 
        & \rotatebox{\ang}{Lab} 
        & \rotatebox{\ang}{Production}  
        & \rotatebox{\ang}{Detection} 
        & \rotatebox{\ang}{Vertex}  
        & \rotatebox{\ang}{Mass(MeV)}  
        & \rotatebox{\ang}{Mass Res. (MeV)}
        & \rotatebox{\ang}{Beam}
        & \rotatebox{\ang}{Ebeam (GeV)}
        & \rotatebox{\ang}{Ibeam or Lumi}
        & \rotatebox{\ang}{Machine}
        & \rotatebox{\ang}{1st Run}
        & \rotatebox{\ang}{Next Run}  
        \\ 
\hline
\hline
        APEX %\rotatebox{90}{Experiment}      Numbers confirmed by Bogdan 
        & JLab %\rotatebox{90}{Lab}
        & e-brem%\rotatebox{90}{Production}  
        & \elel%\rotatebox{90}{Detection} 
        & no%\rotatebox{90}{Vertex}  
        & $65-600$%\rotatebox{90}{Mass}  
        & 0.5\%%\rotatebox{90}{Mass Resolution}
        & \elec %\rotatebox{\ang}{Beam}
        & 1.1--4.5%\rotatebox{90}{Ebeam}
        & 150 \muA %\rotatebox{90}{Ibeam}
        & CEBAF(A)%\rotatebox{90}{Machine}
        & 2010%\rotatebox{90}{1st Run}
        & 2018%\rotatebox{90}{2nd Run or Upgrade}  
        \\ 
\hline
        A1 %\rotatebox{90}{Experiment}  
        & Mainz %\rotatebox{90}{Lab}
        & e-brem%\rotatebox{90}{Production}  
        & \ee%\rotatebox{90}{Detection} 
        & no%\rotatebox{90}{Vertex}  
        & $40-300$%\rotatebox{90}{Mass}  
        & ?%\rotatebox{90}{Mass Resolution}
        & \elec %\rotatebox{\ang}{Beam}
        & 0.2--0.9%\rotatebox{90}{Ebeam}
        & 140 \muA%\rotatebox{90}{Ibeam}
        & MAMI%\rotatebox{90}{Machine}
        & 2011%,14%\rotatebox{90}{1st Run}
        & --%\rotatebox{90}{2nd Run or Upgrade}  
        \\
\hline
         HPS %\rotatebox{90}{Experiment}   % numbers confirmed by Tim Nelson
        & JLab %\rotatebox{90}{Lab}
        & e-brem%\rotatebox{90}{Production}  
        & \ee%\rotatebox{90}{Detection} 
        & yes%\rotatebox{90}{Vertex}  
        & $20-200$%\rotatebox{90}{Mass}  
        & 1--2%\rotatebox{90}{Mass Resolution}
        & \elec %\rotatebox{\ang}{Beam}
        & 1--6%\rotatebox{90}{Ebeam}
        & 50--500 nA%\rotatebox{90}{Ibeam}
        & CEBAF(B)%\rotatebox{90}{Machine}
        & 2015%\rotatebox{90}{1st Run}
        & 2018%\rotatebox{90}{2nd Run or Upgrade}  
        \\
\hline
        DarkLight %\rotatebox{90}{Experiment}  % numbers confirmed by Ross Corliss
        & JLab %\rotatebox{90}{Lab} 
        & e-brem%\rotatebox{90}{Production}  
        & \ee%\rotatebox{90}{Detection} 
        & no%\rotatebox{90}{Vertex}  
        & $<80$%\rotatebox{90}{Mass}  
        & ?%\rotatebox{90}{Mass Resolution}
        & \elec %\rotatebox{\ang}{Beam}
        & 0.1%\rotatebox{90}{Ebeam}
        & 10 mA%\rotatebox{90}{Ibeam}
        & LERF%\rotatebox{90}{Machine}
        & 2016%\rotatebox{90}{1st Run}
        & 2018%\rotatebox{90}{2nd Run or Upgrade}  
        \\
\hline
        MAGIX %\rotatebox{90}{Experiment}  
        & Mainz %\rotatebox{90}{Lab}
        & e-brem%\rotatebox{90}{Production}  
        & \ee%\rotatebox{90}{Detection} 
        & no%\rotatebox{90}{Vertex}  
        & $10-60$%\rotatebox{90}{Mass}  
        & ?%\rotatebox{90}{Mass Resolution}
        & \elec %\rotatebox{\ang}{Beam}
        & 0.155%\rotatebox{90}{Ebeam}
        & 1 mA%\rotatebox{90}{Ibeam}
        & MESA%\rotatebox{90}{Machine}
        & 2020%\rotatebox{90}{1st Run}
        & --%\rotatebox{90}{2nd Run or Upgrade}  
        \\
\hline
        NA64 %\rotatebox{90}{Experiment}  
        & CERN %\rotatebox{90}{Lab}
        & e-brem%\rotatebox{90}{Production}  
        & \ee%\rotatebox{90}{Detection} 
        & no%\rotatebox{90}{Vertex}  
        & $1-50$%\rotatebox{90}{Mass}  
        & ?%\rotatebox{90}{Mass Resolution}
        & \elec %\rotatebox{\ang}{Beam}
        & 100 %\rotatebox{90}{Ebeam}
        & $2\times 10^{11}$ EOT/yr %\rotatebox{90}{Ibeam}
        & SPS%\rotatebox{90}{Machine}
        & 2017%\rotatebox{90}{1st Run}
        & 2022%\rotatebox{90}{2nd Run or Upgrade}  
        \\
\hline
        Super-HPS  %\rotatebox{90}{Experiment}  
        & SLAC %\rotatebox{90}{Lab}
        & e-brem%\rotatebox{90}{Production}  
        & vis%\rotatebox{90}{Detection} 
        & yes%\rotatebox{90}{Vertex}  
        & $<500$%\rotatebox{90}{Mass}  
        & ?%\rotatebox{90}{Mass Resolution}
        & \elec %\rotatebox{\ang}{Beam}
        & $4-8$%\rotatebox{90}{Ebeam}
        & 1 \muA%\rotatebox{90}{Ibeam}
        & DASEL%\rotatebox{90}{Machine}
        & ?%\rotatebox{90}{1st Run}
        & ?%\rotatebox{90}{2nd Run or Upgrade}  
		\\
\hline
        {\em (TBD)} %\rotatebox{90}{Experiment} 
        & Cornell %\rotatebox{90}{Lab} 
        & e-brem%\rotatebox{90}{Production}  
        & \ee%\rotatebox{90}{Detection} 
        & ?%\rotatebox{90}{Vertex}  
        & $<100$%\rotatebox{90}{Mass}  
        & ?%\rotatebox{90}{Mass Resolution}
        & \elec %\rotatebox{\ang}{Beam}
        & 0.1-0.3%\rotatebox{90}{Ebeam}
        & 100 mA%\rotatebox{90}{Ibeam}
        & CBETA%\rotatebox{90}{Machine}
        & ?%\rotatebox{90}{1st Run}
        & ?%\rotatebox{90}{2nd Run or Upgrade}  
        \\
%\hline
%        A2 %\rotatebox{90}{Experiment}  
%        & Mainz %\rotatebox{90}{Lab}
%        & meson%\rotatebox{90}{Production}  
%        & \ee%\rotatebox{90}{Detection} 
%        & ?%\rotatebox{90}{Vertex}  
%        & $<200$%\rotatebox{90}{Mass}  
%        & ?%\rotatebox{90}{Mass Resolution}
%        & \elec %\rotatebox{\ang}{Beam}
%        & 0.2--1.6%\rotatebox{90}{Ebeam}
%        & 140 \muA%\rotatebox{90}{Ibeam}
%        & MAMI%\rotatebox{90}{Machine}
%        & ?%\rotatebox{90}{1st Run}
%        & --%\rotatebox{90}{2nd Run or Upgrade}  
%        \\
\hline\hline
        VEPP3 %\rotatebox{90}{Experiment}  
        & Budker %\rotatebox{90}{Lab}
        & annih%\rotatebox{90}{Production}  
        & invis%\rotatebox{90}{Detection} 
        & no%\rotatebox{90}{Vertex}  
        & $5-22$%\rotatebox{90}{Mass}  
        & 1%\rotatebox{90}{Mass Resolution}
        & \pos %\rotatebox{\ang}{Beam}
        & 0.500%\rotatebox{90}{Ebeam}
        & $10^{33}\lumiunits$%\rotatebox{90}{Ibeam}
        & VEPP3%\rotatebox{90}{Machine}
        & 2019%\rotatebox{90}{1st Run}
        & ?%\rotatebox{90}{2nd Run or Upgrade}  
        \\
\hline
        PADME %\rotatebox{90}{Experiment}    Numbers confirmed by mauro.
        & Frascati  %\rotatebox{90}{Lab}
        & annih %\rotatebox{90}{Production}  
        &  invis%\rotatebox{90}{Detection} 
        &  no%\rotatebox{90}{Vertex}  
        &  $1-24$%\rotatebox{90}{Mass}  
        &  $2-5$%\rotatebox{90}{Mass Resolution}
        & \pos %\rotatebox{\ang}{Beam}
        &  0.550%\rotatebox{90}{Ebeam}
        &  $\le10^{14}\,\pos$OT/y%\rotatebox{90}{Ibeam}
        &  Linac%\rotatebox{90}{Machine}
        &  2018%\rotatebox{90}{1st Run}
        &  ?%\rotatebox{90}{2nd Run or Upgrade}  
        \\
\hline
        MMAPS %\rotatebox{90}{Experiment}  
        & Cornell %\rotatebox{90}{Lab}
        & annih %\rotatebox{90}{Production}  
        &  invis%\rotatebox{90}{Detection} 
        &  no%\rotatebox{90}{Vertex}  
        &  $20-78$%\rotatebox{90}{Mass}  
        &  $1-6$%\rotatebox{90}{Mass Resolution}
        & \pos %\rotatebox{\ang}{Beam}
        &  6.0%\rotatebox{90}{Ebeam}
        &  $10^{34}\lumiunits$%\rotatebox{90}{Ibeam}
        &  Synchr%\rotatebox{90}{Machine}
        &  ?%\rotatebox{90}{1st Run}
        &  ?%\rotatebox{90}{2nd Run or Upgrade}  
        \\
\hline
\hline
        KLOE 2 %\rotatebox{90}{Experiment}      % numbers confirmed by Enrico Granziani
        & Frascati %\rotatebox{90}{Lab}
        & several  %\rotatebox{90}{Production}  
        &  vis/invis%\rotatebox{90}{Detection} 
        &  no%\rotatebox{90}{Vertex}  
        &  $<1.1\,\gev$%\rotatebox{90}{Mass}  
        &  1.5%\rotatebox{90}{Mass Resolution}
        & \ee %\rotatebox{\ang}{Beam}
        &  0.51%\rotatebox{90}{Ebeam}
        &  $2\times 10^{32}\lumiunits$%\rotatebox{90}{Ibeam} Lumi=2e32
        &  DA$\phi$NE%\rotatebox{90}{Machine}
        &  2014%\rotatebox{90}{1st Run}
        &  -%\rotatebox{90}{2nd Run or Upgrade}  
        \\
\hline
        Belle II %\rotatebox{90}{Experiment}   % numbers confirmed by Chris Hearty
        & KEK %\rotatebox{90}{Lab}
        & several  %\rotatebox{90}{Production}  
        &  vis/invis%\rotatebox{90}{Detection} 
        &  no%\rotatebox{90}{Vertex}  
        &  $\lsim10\,\gev$%\rotatebox{90}{Mass}  
        &  $1-5$%\rotatebox{90}{Mass Resolution}
        & \ee %\rotatebox{\ang}{Beam}
        &  $4\times 7$%\rotatebox{90}{Ebeam}
        &  $1\sim10~\abinv$/y%\rotatebox{90}{Ibeam}  1-10 abinv/yr
        &  Super-KEKB%\rotatebox{90}{Machine}
        &  2018%\rotatebox{90}{1st Run}
        &  -%\rotatebox{90}{2nd Run or Upgrade}  
        \\
\hline
\hline
        SeaQuest %\rotatebox{90}{Experiment}  %numbers confirmed by Ming Liu
        & FNAL %\rotatebox{90}{Lab}
        & several %\rotatebox{90}{Production}  
        &  \mumu%\rotatebox{90}{Detection} 
        &  yes%\rotatebox{90}{Vertex}  
        &  $\lsim10\,\gev$%\rotatebox{90}{Mass}  
        &  $3-6\%$%\rotatebox{90}{Mass Resolution}
        &  p %\rotatebox{\ang}{Beam}
        &  120%\rotatebox{90}{Ebeam}
        &  $10^{18}$~POT/y%\rotatebox{90}{Ibeam}
        &  MI%\rotatebox{90}{Machine}
        &  2017%\rotatebox{90}{1st Run}
        &  2020%\rotatebox{90}{2nd Run or Upgrade}  
        \\
\hline
        SHIP %\rotatebox{90}{Experiment}  % confirmed by Gaia Lanfranchi
        & CERN %\rotatebox{90}{Lab}
        &  several%\rotatebox{90}{Production}  
        &  vis%\rotatebox{90}{Detection} 
        &  yes%\rotatebox{90}{Vertex}  
        &  $\lsim10\,\gev$%\rotatebox{90}{Mass}  
        &  $1-2$%\rotatebox{90}{Mass Resolution}
        &  p %\rotatebox{\ang}{Beam}
        &  400 %\rotatebox{90}{Ebeam}
        &  $2\times10^{20}$~POT/5y%\rotatebox{90}{Ibeam}
        &  SPS%\rotatebox{90}{Machine}
        &  2026%\rotatebox{90}{1st Run}
        &  -%\rotatebox{90}{2nd Run or Upgrade}  
        \\
\hline
        LHCb %\rotatebox{90}{Experiment}  % confirmed by Mike Williams
        & CERN %\rotatebox{90}{Lab}
        &  several %\rotatebox{90}{Production}  
        &  \elel%\rotatebox{90}{Detection} 
        &  yes%\rotatebox{90}{Vertex}  
        &  $\lsim40\,\gev$%\rotatebox{90}{Mass}  
        &  $\sim 4$%\rotatebox{90}{Mass Resolution}
        &  pp %\rotatebox{\ang}{Beam}
        &  6500%\rotatebox{90}{Ebeam}
        &  $\sim10\,\fbinv$/y%\rotatebox{90}{Ibeam}
        &  LHC%\rotatebox{90}{Machine}
        &  2010%\rotatebox{90}{1st Run}
        &  2015%\rotatebox{90}{2nd Run or Upgrade}  
        \\
\hline
\end{tabular}
\renewcommand{\arraystretch}{1.0}
\end{center}
\end{adjustwidth}
\end{sidewaystable}

\subsection{Summary of ongoing and proposed experiments}
The experimental community for dedicated dark sector searches has grown substantially in the last eight years and as the list above illustrates, the experiments, whether ongoing or proposed, have expanded to cover a wide range of production modes and detection strategies. Experiments like APEX, A1, HPS, and DarkLight, that take advantage of explicit final state reconstruction, push deep into the \epsq parameter range, with sensitivity in \mdark up to a few hundred MeV. In the coming years, experiments like VEPP3, PADME, and MMAPS will address a more limited parameter range, but as missing mass experiments, eliminating aspects of model dependence by being fully agnostic as to the final state. Collider experiments allow probes to much higher masses than can be reached in fixed-target experiments.  Some, like Belle-II and LHCb, will have trigger schemes specifically optimized for dark sector searches.  Taken together, the set of existing and planned experiments form a suite of balanced and complementary approaches, well-suited to the search for new phenomena whose physical characteristics and potential manifestations cannot be predicted in detail ahead of time.

\newpage

\section{Dark Matter at Accelerators}\label{sec:DMA}
\vspace{-0.3cm}
\begin{flushleft}
\textit{Conveners:~Marco Battaglieri, Eder Izaguirre, Gordan Krnjaic, Adam Ritz, Richard G Van de Water.~Organizer Contact: Philip Schuster}
\end{flushleft}

\subsection{Introduction}

A premier strategy for detecting and measuring dark matter interactions with familiar matter is to use dedicated experiments that can produce dark matter in the lab, typically with the aid of high-intensity beams and/or high precision detectors. 
Consequently, numerous accelerator techniques to probe dark matter have been developed and proven over several decades, and these provide a solid basis on which future progress with Intensity and Energy Frontier experiments can be mapped out. 
To that end, this section provides a self-contained introduction to the broad class of light dark matter models coupled to the Standard Model (SM) through a new light mediator, with a focus on simple vector mediator models that can accommodate all existing data. A key scientific goal for the field is to probe thermal models of GeV-scale light dark matter to a decisive level of sensitivity. This defines a series of parameter space targets -- summarized below -- that experiments should aim to reach. After providing a brief snapshot of existing constraints, we summarize exciting ongoing efforts and near-term opportunities to make significant experimental progress with existing and upcoming accelerators and several different detector approaches. 

\subsection{Theory Summary}

In many models of dark matter, the dark matter particles do not have direct couplings to the Standard Model. Instead, they interact with the SM through a ``mediator'', a particle that couples to both the SM and the DM. While the mass range over which this can occur extends into the weak-scale, we focus here on the less explored sub-GeV mass range for both the DM particle and the mediator. The gauge and Lorentz symmetries of the SM greatly restrict the ways in which the mediator can couple to the SM. One expects the dominant interactions to be the so-called renormalizable portals: those interactions consisting of SM gauge singlet operators with mass dimension less than or equal to $4$:
\bea
\hat {\cal O}_{\rm portal} ~ = ~~~H^\dagger H~~~,~~~ LH ~~~,~~~ B_{\mu \nu}~~~,
\eea
and a new SM-neutral degree of freedom, which can be a scalar $\phi$, a fermion $N$, or a vector $A'$.
Here $H$ is the SM Higgs doublet with charge assignment $(1, 2, +\frac{1}{2})$ under the SM gauge group $SU(3)_c \times SU(2)_L \times U(1)_Y$, $L$ is a lepton doublet of any generation
transforming as $(1,2, -\frac{1}{2})$, and $B_{\mu \nu} \equiv \partial_\mu B_\nu -\partial_\nu B_\mu$ is the hypercharge field strength tensor. 
%Although there could also be higher dimension effective operators to connect to the mediators, direct searches for the states that resolve such operators require suppression scales in excess of the electroweak scale, which generically would lead to over-abundance of dark matter if these were the predominant interactions that set the DM relic abundance.

If the mediator is a scalar particle $\phi$, the only allowed renormalizable interactions are through the Higgs portal via $\phi H^\dagger H$  and $\phi^2 H^\dagger H$, which induce mass mixing  between $\phi$ and the SM Higgs boson after electroweak symmetry breaking. The simplest example in this scenario, however, is already sharply constrained by existing experiments \cite{Krnjaic:2015mbs}. If the mediator is a fermion $N$, its interaction with the SM proceeds through the neutrino portal $\sim y_\nu LHN$ and it plays the role of a right-handed neutrino with a Yukawa coupling $y_{\nu}$. If DM is {\it not}  thermal in origin, $N$ can itself be a viable, cosmologically metastable DM candidate in a narrow mass range \cite{Dodelson:1993je}. Since $N$ is stipulated to be sub-GeV, obtaining the observed neutrino masses (without additional field content) requires Yukawa couplings of order $y_\nu \lesssim 10^{-7}$, which are too small to allow thermalization to take place at early times \cite{Feng:2010gw}. We note the possibility that $N$ is the mediator, but not the DM; in this case additional particle content in the dark sector would be needed to accommodate DM.
 
 We therefore focus on the third scenario, as it is the most viable for models of light DM, and also spans a range of accessible phenomenology. If the mediator is a vector force carrier (referred to as a ``dark photon'') from an additional $U(1)_D$ gauge group under which  LDM is charged, the ``kinetic mixing" interaction $\epsilon_Y B^{\mu \nu} F^\prime_{\mu \nu}$ is gauge invariant under both $U(1)_D$ and $U(1)_Y$. Electroweak symmetry breaking induces a mixing with the SM photon $\epsilon_Y B^{\mu \nu} F^\prime_{\mu \nu} \to \epsilon F^{\mu \nu} F^\prime_{\mu \nu}$, which is responsible for the relevant phenomenology considered in light DM models. Here $\epsilon$ is, {\it a priori} a free parameter, though it often arises in UV complete models after heavy states charged under both groups are integrated out at a high scale, so it is generically expected to be small, $\epsilon \sim 10^{-3}$ or smaller \cite{holdom:1985ag}.  Additionally, important variations, such as the scenarios where the mediator couples preferentially to baryonic, leptonic, or $(B-L)$ currents each have a phenomenology that is similar to the kinetic mixing scenario. 
 
In these scenarios, dark matter consists of new particles that couple to the SM via the dark photon. The nature of the DM particle content is important in establishing constraints on the model, as well as the parts of parameter space favored by cosmological and astrophysical observations of DM. For example, the phenomenology of DM depends on the spin of the particles and the mediator, since this determines whether DM scattering proceeds through the $s$- or $p$-wave, and consequently certain processes are velocity suppressed in the halo today. Additionally, the phenomenology can change based on how many DM particles exist in the dark sector, and whether they have diagonal or off-diagonal couplings to the mediator (see Refs.~\cite{Morrissey:2014yma,Izaguirre:2015zva}, and references therein for examples).  In what follows we consider the two scenarios of scalar/fermion DM communicating with the SM through a vector mediator.
  
\subsection{Defining Thermal Targets}

\noindent For all mediators and LDM candidates $\chi$, there is a basic distinction between ``secluded" annihilation to pairs of  mediators (via $\chi \chi$ $\to$ $A^\prime A^\prime$ for $m_\chi > m_{A^\prime}$) followed by 
mediator decays to SM particles \cite{Pospelov:2007mp}, and ``direct" annihilation to SM final states (via virtual mediator exchange  in the $s$-channel, $ \chi \chi \to$ $A^{\prime*}$ $\to$  SM SM for $m_{\chi} < m_{A^\prime}$) without an intermediate step. For concreteness, the discussion below leading up to the definition of a thermal target is made for fermionic DM. A similar discussion is applicable for scalar DM, up to the difference between p-wave and s-wave annihilation.

For the secluded process, the annihilation rate scales as 
\bea 
\hspace{-1cm } ({\rm ``secluded" ~annihilation}) ~~~~~~ \langle \sigma v \rangle \sim \frac{  g_D^4 }{  m_\chi^2  } ~~ ~,~~
\eea
where $g_{D}$ is the coupling between the mediator and the LDM, and there is no dependence on the SM-mediator coupling $g_{\rm SM}$. Since arbitrarily small values of  $g_{\rm SM}$ can be compatible with thermal LDM in this regime, the secluded scenario does not lend itself to decisive laboratory tests, though severe indirect constraints from CMB data do exist for this case. 
  
The situation is markedly different for  the direct annihilation regime in which $m_{\chi} < m_{A^\prime}$. Here the annihilation rate scales as  
\bea
\hspace{1cm}({\rm ``direct" ~annihilation})~~~~\langle \sigma v \rangle   \sim       \frac{      g_{D}^2 \,  \alpha \epsilon^2  \,  m_{\chi}^2 }{   m_{A^\prime}^4  \!\!}~~ ~~~~~~~~~  ~,~~
\eea
where $\alpha$ is the QED fine structure constant, and $\epsilon=\epsilon_Y\cos{\theta_w}$.
This offers a clear, predictive target for discovery or falsifiability since the dark coupling $g_{D}$  and mass ratio $m_{{\chi}}/m_{A^\prime}$ are 
at most ${\cal O}(1)$ in this $m_{A^\prime} > m_{\chi}$ regime. Thus, there is a minimum SM-mediator coupling $\epsilon$ compatible with a thermal history; larger values of $g_D$ require non-perturbative dynamics in the mediator-SM coupling or intricate model building.

In the direct annihilation regime, the minimum annihilation rate requirement translates into a minimum value of the dimensionless combination 
\bea\label{eq:generic-thermal-target}
\frac{   \alpha g_D^2 \, \epsilon^2  }{4\pi} \, \left(\frac{m_\chi}{m_{A^\prime}}\right)^4 \gtrsim  \langle \sigma v \rangle_{\rm relic}   \, m_\chi^2~,~~
\eea
which, up to order one factors, is valid for every DM/mediator variation provided that $m_{\chi} < m_{A^\prime}$.

\subsection{Brief Summary of Existing Constraints}

DM searches at accelerators can be divided into two broad categories: fixed-target and collider experiments. We summarize the different strategies in Table~\ref{tab:detectionstrategy}. 

Existing constraints have primarily been deduced by recasting a number of prior experimental searches. Before moving to future opportunities, we provide a brief summary of these limits. For lower mass DM, the strongest constraints follow from limits on anomalous scattering at proton and electron fixed target experiments such as LSND \cite{deNiverville:2011it,Kahn:2014sra,lsnd2001} and E137 \cite{e137,e137_exp}. For specific mass ranges, limits on invisible pion \cite{pi0invis}, kaon \cite{Pospelov:2008zw,Artamonov:2009sz} and $J/\Psi$ \cite{Ablikim:2007ek} decays are significant, while monophoton searches at \babar\ \cite{babar1,babar2} are stringent at higher masses. Monojet searches \cite{monojet,monojet2} are generally less constraining, but relevant for leptophobic mediators. Finally, vector mediator exchange induces corrections to $g-2$ of the electron and muon, which impose constraints at low mass \cite{g2_1,g2_2,g2_3,Pospelov:2008zw} and, in the case of the muon $g-2$ anomaly, have identified a region of interest in parameter space. A number of these existing limits are shown in the figures below.

In what follows, we summarize current and future opportunities. The summary bullets only provide key information, such as beam type and energy, detector type and detection strategy, and schedule -- additional references are provided for details. \\

\subsection{Proton Beam-Dump Experiments}
 
 \begin{figure*}[th]
\centerline{ 
\includegraphics[width=0.75\textwidth]{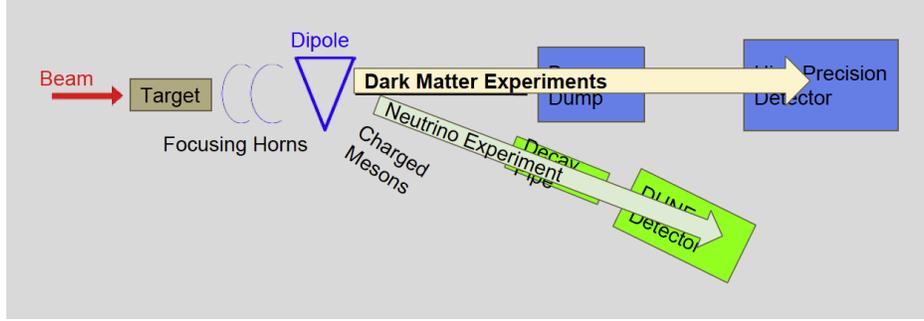}\hspace*{0.3cm}}
\caption{\footnotesize Schematic of the dual purpose LBNF beam-line
  that can 
 simultaneously produce both a charged and neutral beam.  The charged
beam decays into neutrinos while the neutral beam can couple to LDM. }
 \label{fig:LBNFsetup}
\end{figure*}

\begin{table}[t] 
\begin{center}
\begin{tabular*}{1.0\textwidth}{@{\extracolsep{\fill}}|c|c|c|}
\hline
%\\[-7pt]
 {\bf  Experiment Class}  & {\bf Production Modes} & {\bf Detection} \\
\hline
B-factory & $e^+e^-\rightarrow \gamma A'$ & missing mass \\
Electron fixed-target  & $e^-Z\rightarrow e^-Z A'$ & DM scatter or missing energy/mass \\
Hadron collider & $pp\rightarrow (\rm{jet}/\gamma) A'$ & missing energy \\
Positron fixed-target  & $e^+e^-\rightarrow \gamma A'$ & missing mass\\
Proton fixed-target  & $\pi^0/\eta/\eta'\rightarrow \gamma A'$,~$q\bar{q}\rightarrow A'$,~$p Z\rightarrow p Z A'$ & DM scatter downstream \\
 \hline
 \end{tabular*}
 \caption{Catalogue of complementary experimental strategies to search for light DM.}
 \label{tab:detectionstrategy}
\end{center}
\end{table}

\begin{figure*}
 \includegraphics[width=0.6\textwidth]{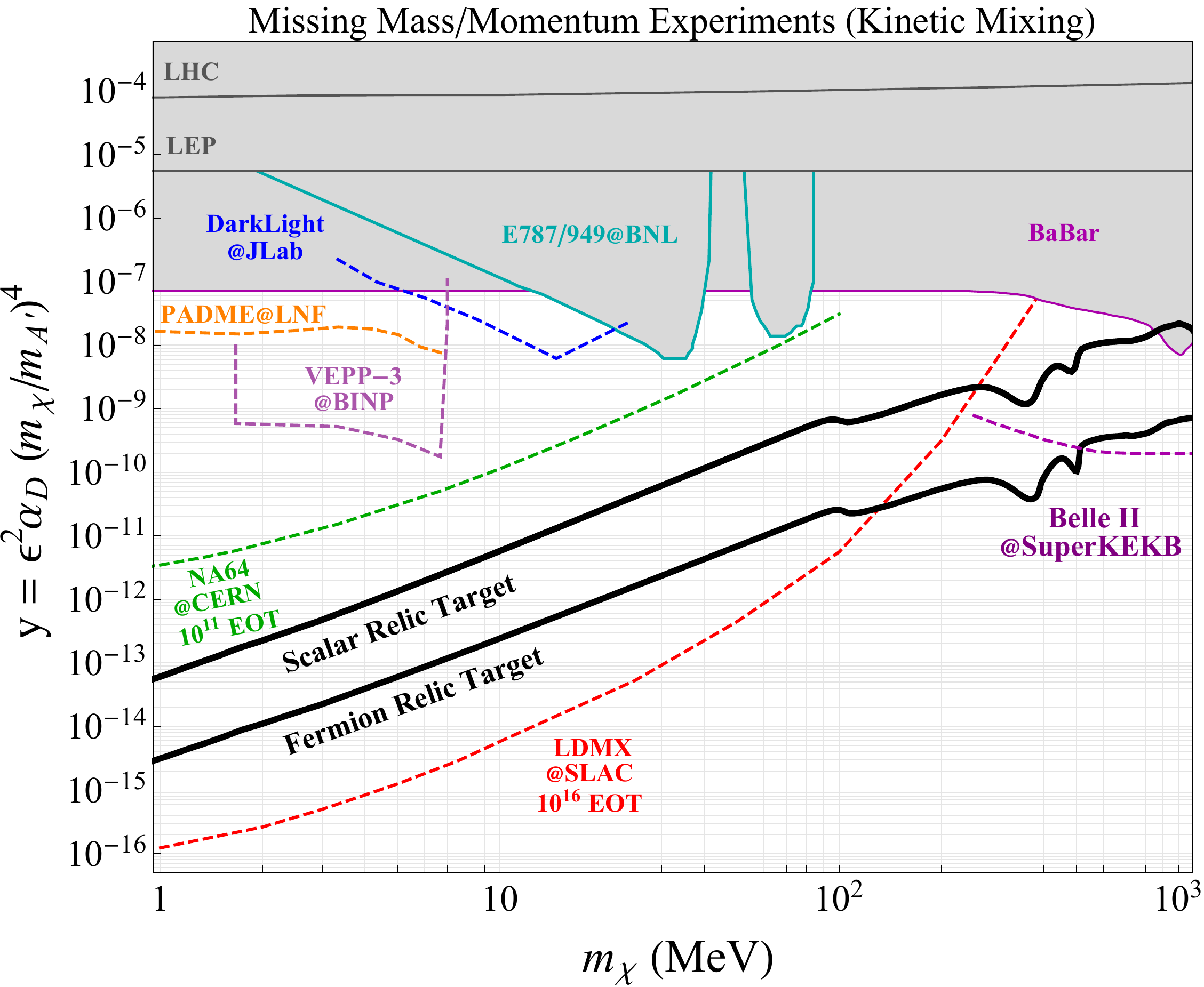}       \\
\caption{\footnotesize  
Yield projections for various proposed DM search strategies involving missing mass/momentum plotted
alongside constraints based on the same experimental technique and the thermal target for scalar and fermion DM candidates.
Here all bounds and projections
conservatively assume $m_\chi = 3\, m_{A^\prime}$, but the thermal targets are invariant as this ratio changes (see \cite{Izaguirre:2015zva} for a discussion). For larger ratios, 
the experimental curves shift downward to cover more parameter space; for small ratios $ m_\chi >  m_{A^\prime}$, there
is no thermal target as the DM annihilation proceeds trough $\chi \chi \to A^\prime A^\prime$, which is independent of the SM coupling $\epsilon$. This plot serves to compare proposed missing momentum based searches against similar constraints; bounds not based on 
missing momentum techniques (e.g. direct detection or beam dump searches) are omitted. Here, the shaded regions represent excluded parameter space, dashed projections are based on signal-yield estimates and solid curves represent sensitivity estimates based on background studies.   }
 \label{fig:MMPlot}
\end{figure*}

\begin{itemize}
\item {\bf MiniBooNE at FNAL}: 8 GeV Booster Neutrino Beamline (BNB) protons. Can run in target mode (Be), and off-target mode (Fe). Mineral oil Cherenkov detector, 450 ton fiducial mass, situated 540 m downstream. Main production mode via $\pi^0/\eta\eta' \rightarrow \gamma (A'\rightarrow \chi\bar\chi)$ and $q\bar{q}\rightarrow A'\rightarrow \chi\bar\chi$. Detection via $\chi e \rightarrow \chi e$, or $\chi N \rightarrow \chi N$ elastic scattering, or via inelastic such as $\chi N\rightarrow \chi (\Delta \rightarrow N \pi^0)$. Completed running in off-target mode with $1.86\times10^{20}$ POT, analysis ongoing. See Ref.~\cite{Batell:2009di,deNiverville:2011it,deNiverville:2012ij,Dharmapalan:2012xp,Batell:2014yra,Morrissey:2014yma,Kahn:2014sra,Gorbunov:2014wqa,Blumlein:2013cua} for more details. 

\item {\bf T2K at J-PARC}: 30 GeV protons. The near and far detectors are two degrees off-axis, and the timing structure of the bunches in each spill can be used to cleanly separate beam-related backgrounds at the far-detector, Super-Kamiokande, a 50 kiloton water Cerenkov detector 295 km from the target. The production modes are as for MiniBooNE, but the high degree of background reduction can compensate for the reduced angular acceptance in utilizing the far detector. Initial analysis will focus on de-excitation gammas from the neutral current quasielastic~(NCQE) interaction on oxygen (see \cite{Abe:2014dyd} for related studies of neutrino scattering), again testing the underlying $\chi N \rightarrow \chi N$ process. The final dataset is expected to be $7.8 \times 10^{21}$ POT by 2021.

\item {\bf SBN at FNAL}: 8 GeV BNB protons. Three Liquid Argon TPC detectors (LArTPC), 112 ton, 89 ton, and 476 ton fiducial mass situated 110 m, 470 m, and 600 m respectively downstream. Production and detection channels as in MiniBooNE. Current plan to collect $6\times10^{20}$ POT, beginning in
2018, in on-target mode. Can be configured to collect $2\times10^{20}$ POT in beam-dump mode after on-target run, with expected sensitivity an order of magnitude better than MiniBooNE.  Upgrades to BNB in 2016 will enable simultaneous on/off-target running.  Significantly improved sensitivity can be achieved with reduction in neutrino background rates by replacing neutrino horn with an iron target.

\item {\bf Near Detector at FNAL's LBNF/DUNE}: 120 GeV Protons. 1 MW beam power. Fig.~\ref{fig:LBNFsetup} shows the kind of dual purpose facility that can be built for the LBNF neutrino source. A dipole magnet is used to sweep the charge particles, that decay into neutrinos, into a different direction while the neutral beam particles that can couple to LDM continue in the forward direction. This effectively decouples the two beams and produces the most physics reach for both neutrino oscillations and LDM searches. To leverage the investment in the LBNF/DUNE experiment to run simultaneously in beam dump and neutrino mode could be cost effective but will require more funding and design work in the next few years. However, this would provide the ultimate proton beam-dump search. 
In the absence of a dipole magnet that would sweep the neutrino beam, it is possible to
build an off-axis near-detector: the dark matter beam is rather broad and would reach
the near-detector while the neutrino beam is well collimated reducing the background off-axis
Timeline: $>$ 2020. See Ref.~\cite{Coloma:2015pih} for more details.

\item {\bf SHiP at CERN}: 400 GeV protons at CERN's SPS. Expected to be able to deliver $10^{20}$ POT. A neutrino detector consisting of OPERA-like bricks of laminated lead and emulsions, placed in a magnetic field downstream of the muon shield, will allow to measure and identify charged particles produced in charged current neutrino interactions. It is followed by a tracking system and muon magnetic spectrometer. Timeline: $>$ 2026. See Ref.~\cite{Alekhin:2015byh}.

\end{itemize}

\subsection{Electron Beam-Dump Experiments}

\begin{itemize}
\item {\bf BDX at JLab}: 11 GeV CW (ns-spaced bunches) electron beam. Capable of delivering $10^{22}$ EOT in one year's running to Al dump. BDX detector proposed to be situated 20 m downstream, for which new infrastructure is requested. The detector is composed of 1 m$^3$ of CsI scintillator. Main production mode via $e^- Z\rightarrow e^- Z (A'\rightarrow \chi\bar\chi)$. Most promising detection via $\chi e \rightarrow \chi e$ or $\chi N \rightarrow \chi N$. For Majorana-DM models: production via $e^- Z\rightarrow e^- Z (A'\rightarrow \chi_1 \chi_2)$ and detection via $\chi_1 (e/Z/N) \rightarrow \chi_2 (e/Z/N)$ followed by $\chi_2 \rightarrow \chi_1 e^+ e^-$ inside detector. Another strategy is to use a different detection technique in the same experiment (e.g. calorimetry and gas-based detectors BDX-DRIFT) to have an independent confirmation of any possible finding. Proposal submitted to JLab's 44th PAC. See Ref.~\cite{Izaguirre:2013uxa,Battaglieri:2014qoa}. for more details.
 
\item {\bf BDX-like at SLAC}: 4 GeV LCLS-II beam to BSY beam dump. Possible to upgrade to 8 GeV in future. Unlike JLab, features 1 MHz repetition bunches. Capable of delivering $3\times10^{21}$ EOT in one year's running to dump. New infrastructure to host the detector is not needed. However, the detector would have to contend with two beam pipes for LCLS-II X-ray beam users carrying a few KHz of high energy pulses past the detector, and the backgrounds must be evaluated. Currently under study. Timeline: $\gsim$ 2020.
\end{itemize}

\subsection{Electron Missing Energy and Momentum Experiments}

\begin{itemize}

\item {\bf VEPP-3 at BINP}: 500 MeV, 30 mA positron beam incident on internal Hydrogen target at the Budker Institute of Nuclear Physics (BINP) in Novosibirsk, Russia  \cite{Wojtsekhowski:2012zq}. Signal is the missing mass reconstructed from $e^+ e^- \to \gamma (A^\prime  \to \chi \chi)$ annihilation. Timeline: $\sim$ 2019.

\item {\bf MMAPS at Cornell}: 6 GeV, 2.3 nA positron beam incident on a Be target at 
the Wilson Lab at Cornell University in Ithaca, NY, USA  \cite{cornell}. 
Signal is the missing mass reconstructed from $e^+ e^- \to \gamma (A^\prime  \to \chi \chi)$ 
annihilation.  Timeline: TBD

\begin{figure*}[t]
\includegraphics[width=0.6\textwidth]{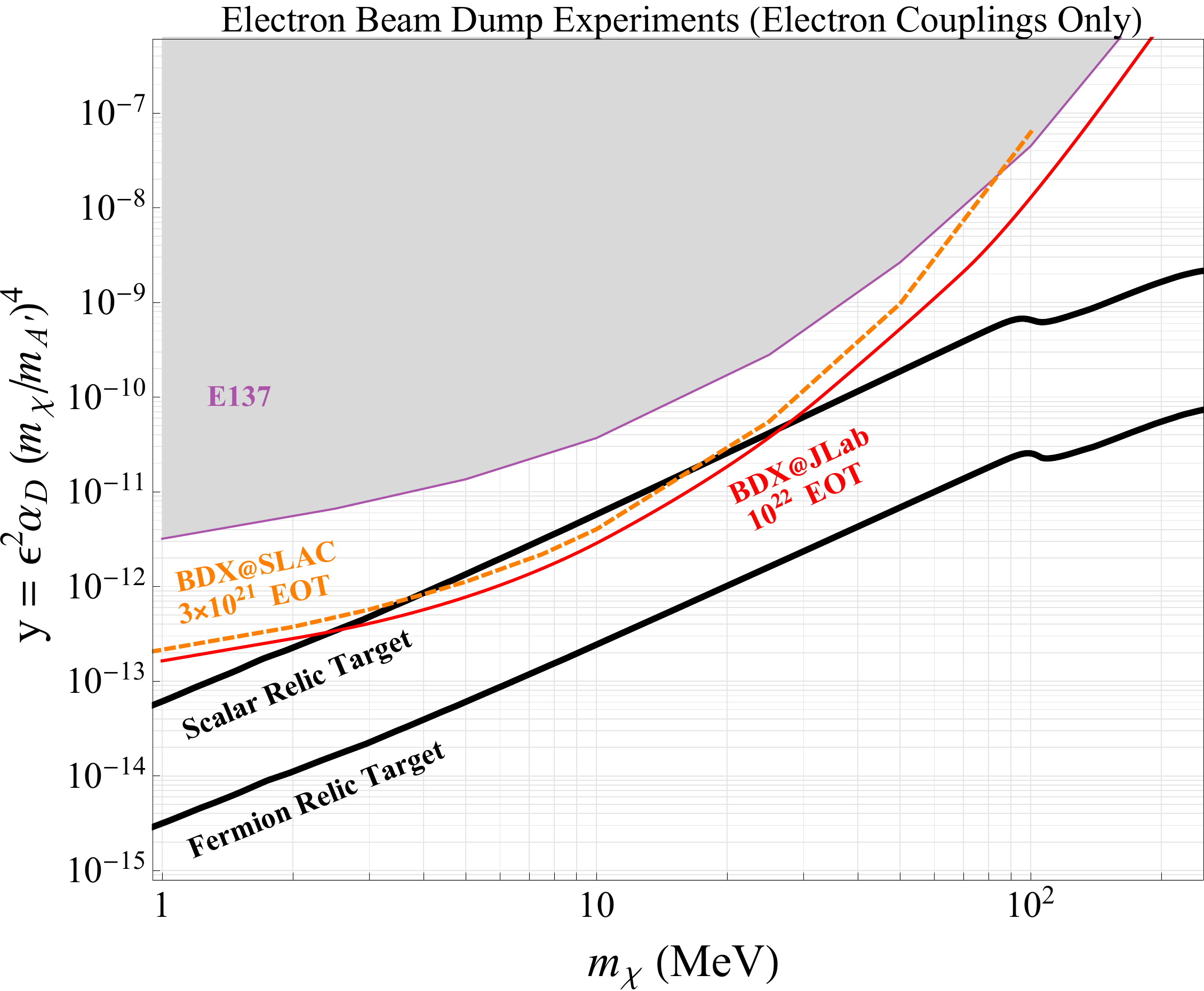}
\caption{\footnotesize  
Same as Fig \ref{fig:MMPlot} but for electron beam-dump experiments.  As in  Fig.~\ref{fig:MMPlot}, here we adopt the conservative prescription $m_{A^\prime} = 3 m_\chi$ and $g_\chi = 0.5$ where
applicable (see \cite{Izaguirre:2015yja} for a discussion). }
 \label{fig:BDPlot}
\end{figure*}

\begin{figure*}[t]
\includegraphics[width=0.6\textwidth]{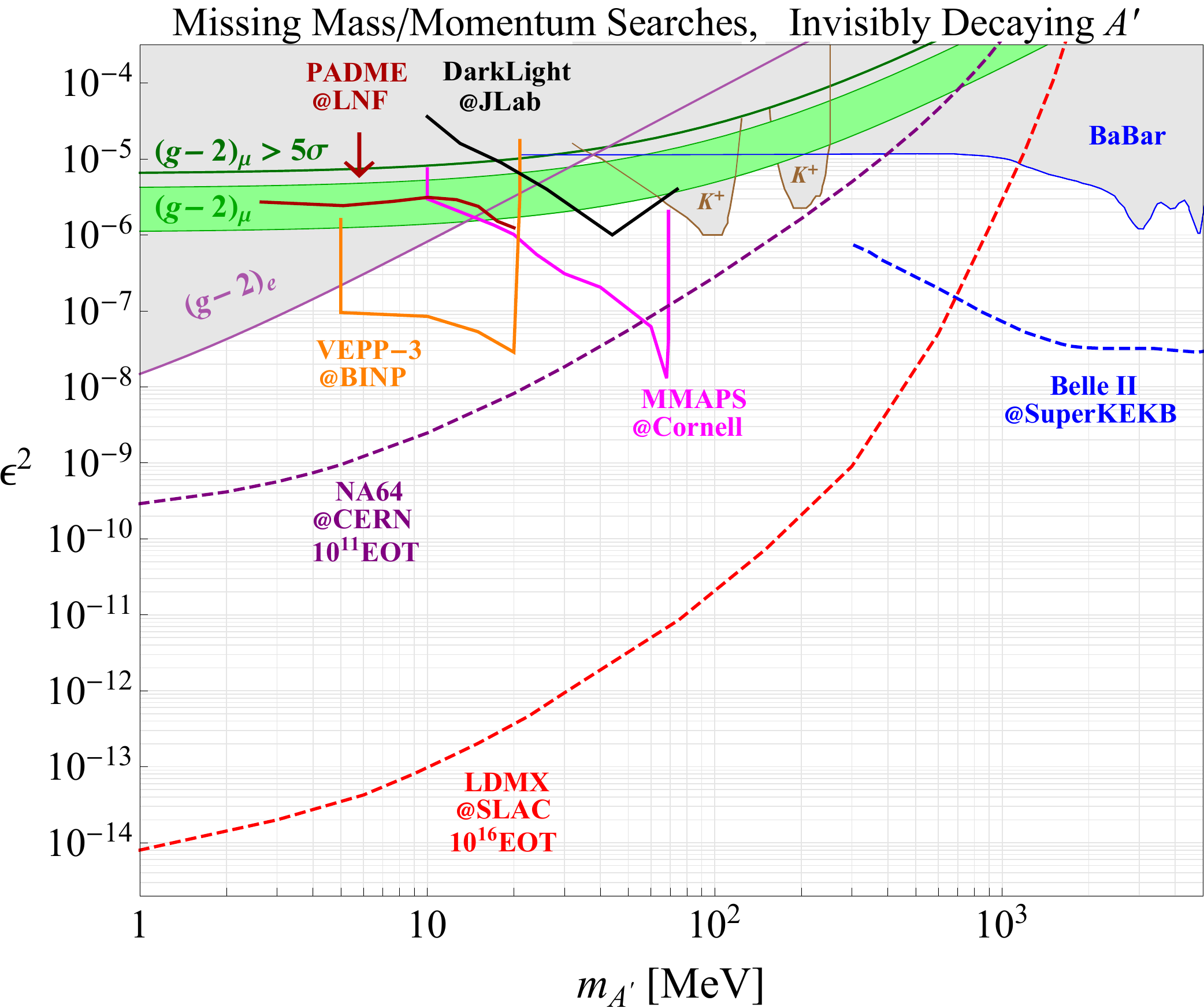}
\caption{\footnotesize   Parameter space for an invisibly decaying dark photon with mass $m_{A'}$ and kinetic
mixing parameter $\epsilon$ with no assumptions about its decay products so long as they are invisible on relevant experimental length scales. 
The shaded regions are model independent constraints  from $(g-2)_\mu$ \cite{Pospelov:2008zw},  $(g-2)_e$ \cite{Giudice:2012ms},  \babar~\cite{Aubert:2008as}, E787/E949 \cite{Adler:2004hp,Artamonov:2009sz}, and the green band represents
the parameter space for which $A'$ resolves the  $(g-2)_\mu$ anomaly \cite{Pospelov:2008zw}. Curves corresponding to unshaded regions represent
 projections for future dedicated searches for invisibly decaying $A'$ using missing mass/momentum techniques also shown in Fig. \ref{fig:MMPlot}.   }
 \label{fig:TraditionalPlot}
\end{figure*}

\item {\bf PADME at LNF}: 550 MeV  positron beam incident on a diamond target at the INFN Laboratory in Frascati, Italy \cite{Raggi:2015gza}. Signal is the missing mass reconstructed from $e^+ e^- \to \gamma (A^\prime  \to \chi \chi)$ annihilation.  
 Timeline: $\sim 10^{13} e^+$ on target by the end of 2018. 

\item {\bf DarkLight at JLab}: 100 MeV Low Energy Recirculating Facility (LERF), formerly the Free Electron Laser, at Jefferson Lab \cite{Balewski:2014pxa}.  Beam current of order 5 mA impinges on a windowless gas target of molecular hydrogen. The complete final state including scattered electron, recoil proton, and $e^+e^-$ pair will be detected.  By reconstructing full final state kinematics, a missing mass
experiment sensitive to  $e p \to e p (A^\prime \to \chi \chi)$ is also possible. Timeline: a phase-I experiment has been funded and is currently taking data. The complete phase-II experiment is under final design and could run within two years after phase-I is completed.

\begin{figure*}[t]
\includegraphics[width=0.57\textwidth]{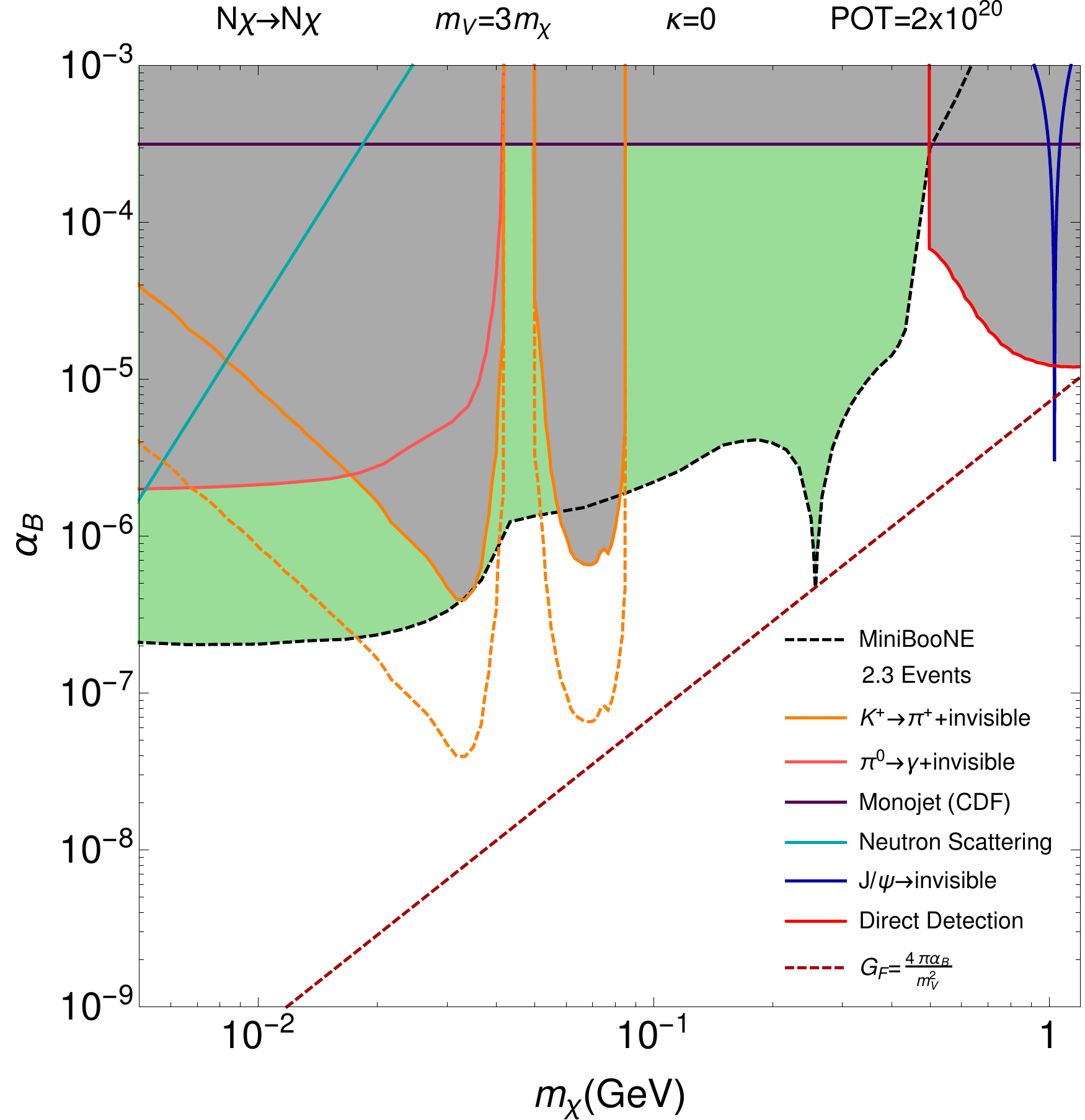}
\caption{\footnotesize  MiniBooNE yield projection for a leptophobic vector force coupled to dark matter \cite{Batell:2014yra}. The 
projections are compared against  the weak scale benchmark for which $4\pi \alpha_B/mV^2 = G_F$, where $\alpha_B$ is the 
dark coupling and $m_V$ is the mediator mass. } 
 \label{fig:protonplot}
\end{figure*}

%\begin{figure*}[t]
%\includegraphics[width=0.6\textwidth]{DMAElectroProtonBeamDump.pdf}
%\caption{\footnotesize  
%Same as Fig \ref{fig:MMPlot} but for electron beam dump experiments. Here BDX\@JLab and \@SLAC assume $10^{22}$ and $2\times 10^{21}$ EOT, respectively and the MiniBooNE curves assume $2\times 10^{20}$ POT.  }
% \label{fig:BDPlotEP}
%\end{figure*}

\item {\bf NA64 at CERN}: 100 GeV secondary electrons at CERN's SPS. Range of currents: $(1-5)\times 10^5 e^-/s$. Pulse structure: 2-4 spills of 4.8 s per minute. Production mode via $eZ\rightarrow e Z (A'\rightarrow \chi\bar\chi)$. The detector consists of a tracker combined with a bending magnet that acts as a spectrometer in order to identify the momentum of the incoming particles. The synchrotron radiation produced by the incoming particles is detected by a BGO detector placed 12 m downstream the magnet which is used to suppress hadron contamination. The particles are dumped in the ECAL where bremsstrahlung photons can be produced. The ECAL is followed by a VETO and a highly hermetic HCAL. The signal is defined as energy deposition in the ECAL below a given energy threshold and no energy deposition in the VETO or the HCAL. Timeline: approved for 2 weeks test beam, and 4 weeks physics run in 2016. Expected to collect $10^{10}$ EOT.

\item {\bf LDMX at SLAC}: 4 GeV electrons from proposed DASEL beamline. Possible to upgrade to 8 GeV in future. LDMX uses missing momentum to detect $eZ\rightarrow e Z (A'\rightarrow \chi\bar\chi)$ reactions by measuring the momentum and energy of the soft outgoing electron, as well as relying on excellent hermiticity to infer the missing energy carried by the DM particles escaping. A beam of $10-1000$ MHz electrons is used, with the momentum of each incoming electron precisely measured in an analyzing magnet before reaching a thin high$-Z$ target. The target is followed by tracking layers to measure the momentum of the outgoing electron, as well as an electromagnetic and hadronic calorimeter optimized to veto SM reactions which can mimic the signal, largely dominated by rare photo-nuclear reactions. LDMX is designed to accumulate enough statistics to probe the thermal relic target over most of the MeV-GeV mass range, to measure potential backgrounds in-situ, and to reject backgrounds using precisely measured final state kinematics. Timeline: $>$ 2020. See Refs.~\cite{Izaguirre:2014bca,Izaguirre:2015yja} for more details.
\end{itemize}

\subsection{Colliders}

\begin{figure*}[t]
\includegraphics[width=0.6\textwidth]{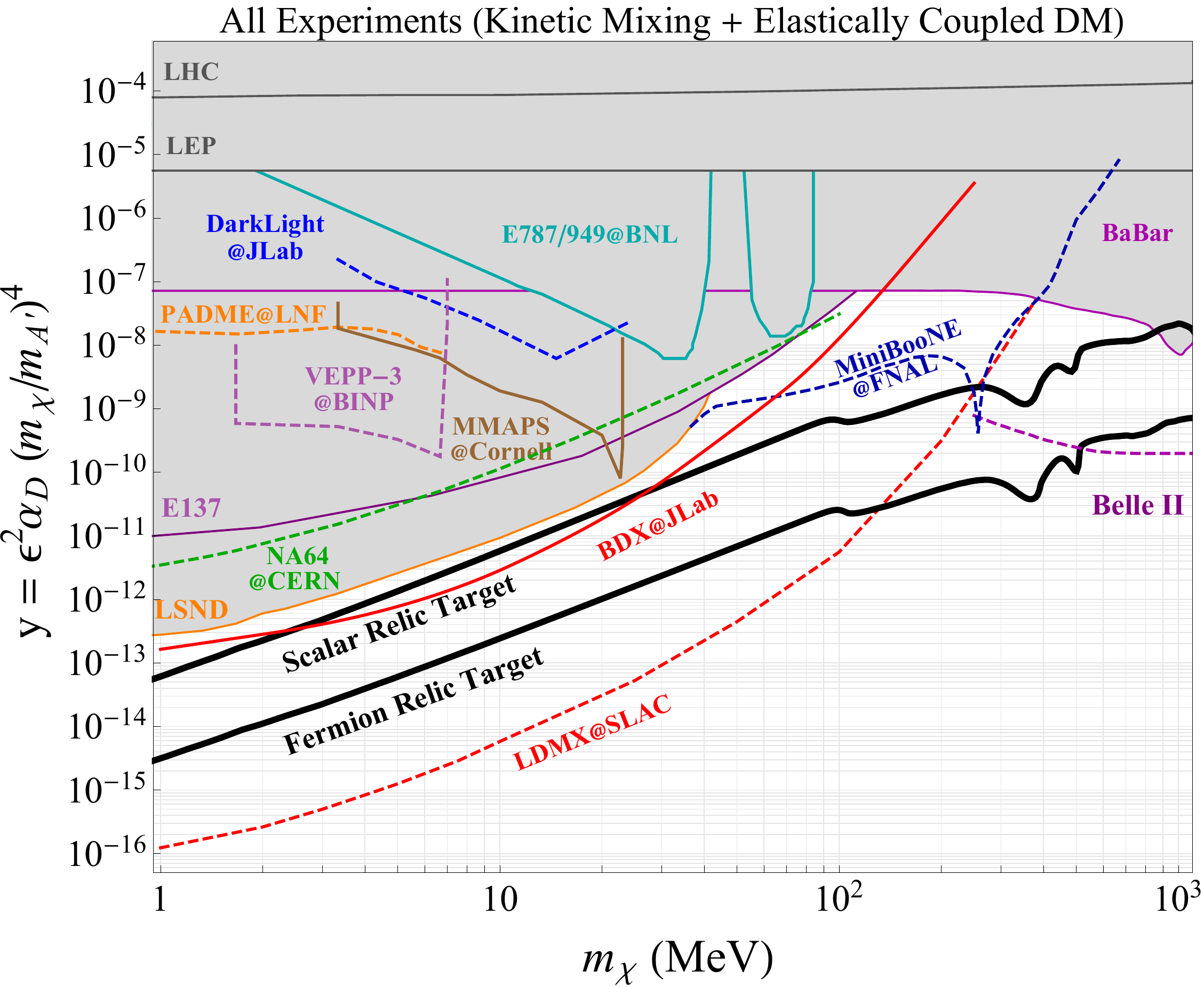}
\caption{\footnotesize  
Combined projections and constraints from Figs.~\ref{fig:MMPlot} and \ref{fig:BDPlot}, %and \ref{fig:BDPlotEP} 
encapsulating direct production LDM constraints in the
context of a kinetically mixed dark photon coupled to an LDM state that scatters elastically (or nearly elastically) at beam-dump, missing energy, and missing momentum experiments. As in  Figs.~\ref{fig:MMPlot} and \ref{fig:BDPlot}, here we adopt the conservative prescription $m_{A^\prime} = 3 m_\chi$ and $g_\chi = 0.5$ where
applicable (see \cite{Izaguirre:2015yja} for a discussion).}
 \label{fig:everything}
\end{figure*}

\begin{itemize}
\item {\bf Belle-II at SuperKEKB}. $e^+ e^-$ asymmetric collider at the $\Upsilon$ resonance $\approx 10.3$ GeV. Expected to achieve integrated luminosities of 
$50\,\abinv$ ($\sim$ 100 times greater than \babar), it can look for DM through the reaction $e^+ e^-\rightarrow \gamma (A'\rightarrow \chi\bar\chi)$. Its signature is a monoenergetic photon. Expected to start collecting data in 2018. 
See Ref.~\cite{TheBelle:2015mwa,Essig:2013vha} for more details.

\item {\bf LHC at CERN}. Analogously to a B-factory, at the LHC, one can produce DM by looking for $pp\rightarrow j A^\prime$ or $\gamma A^\prime$ followed
by $A'\rightarrow \chi\bar\chi$. In this case, however, one does not look for a bump, but instead for missing energy.  For more details, see Ref.~\cite{Abdallah:2015ter}. Additionally, new striking signatures may result in models where the DM is a ``pseudo-Dirac" fermion whose Weyl components are split in
mass. In those models, the $A'$ decays via $A'\rightarrow \chi_1 \chi_2$, with $\chi_2$ possibly giving striking signatures through its decay into $\chi_1 (A^{'(*)}\rightarrow \ell \ell)$. See, for instance, Refs.\cite{Bai:2011jg,Weiner:2012cb,Autran:2015mfa,Bai:2015nfa,Izaguirre:2015zva}.

\end{itemize}

\subsection{Projections}

In Fig. \ref{fig:MMPlot} we show yield projections and sensitivity estimates for various proposed LDM experiments based on missing mass/momentum techniques. Plotted against relevant constraints based on comparable $A^\prime \to $ invisible searches on the  $y$ vs. $m_{A^\prime}$ as described above.  In Figs. \ref{fig:BDPlot} and \ref{fig:protonplot} we show
the analogous figure for electron and proton beam dump experiments assuming only electron and nucleon couplings respectively. In Fig \ref{fig:everything} we show the combined projections of various techniques plotted in the $y$ vs. $m_{A'}$ parameter space for a kinetically mixed dark photon. Finally, in Fig. \ref{fig:TraditionalPlot} we show the projections for the same missing mass/momentum searches presented in Fig. \ref{fig:MMPlot}, but in the $\epsilon^2$ vs. $m_{A'}$ parameter space  without making any assumption about the identity of the $A'$ decay products so long as they are invisible on the length scales probed by each projection and constraint. 
\medskip

\subsection{Outlook}

At the leading edge of Intensity and Energy Frontier science is the simple possibility that DM resides in a Dark Sector. Accelerated progress testing this idea in the next 5-10 years is possible thanks to a healthy interplay between maturing technologies and theoretical input. 
Several approaches have been summarized in this section, and the prospects for experimentally addressing the scientific priorities defined in this report are excellent. 

Comparing the approaches, the question of prioritization was discussed during the DMA WG session, and it was decided that no {\it one} experiment can furnish a robust probe of the important dark matter scenarios that merit study.  Strengths and weaknesses however can be identified. Ignoring the dependence on DM and mediator masses, and focusing on the scenario of the vector portal for concreteness, the experiments that generally will be able to probe the smallest DM-SM interaction couplings are those relying on missing energy, or missing mass --- namely, the signal yield in those experiments goes like $\epsilon^2$. Beam-dump experiments, in turn, scale as $\epsilon^4 \alpha_D$, where $\alpha_D\equiv \frac{g_D^2}{4\pi}$. Thus, {\it within} the vector portal, where the $A'$ couples to all charged SM-fermions democratically, experiments like NA64 and LDMX have the potential to probe the most parameter space for DM and mediator masses below a GeV. However, one must think more generally than about just kinetic mixing. For instance, there are models where the mediator couples preferentially to protons \cite{Dobrescu:2014fca,Batell:2014yra,Coloma:2015pih}, a possibility best tested with future proton beam-dump experiments. Similarly, models where the DM is part of a sector where there are heavier but very short-lived (on collider scales) excited states --- such as the Majorana-like DM scenario with very large mass splittings --- are a potential blind spot of experiments like NA64 and LDMX, but a strength of experiments like BDX, or of any of the future proton beam-dump experiments \cite{Izaguirre:2014dua}. Finally, for DM and mediator masses above a GeV, Belle-II and the LHC will have stronger sensitivity.  Thus, to achieve maximum coverage of the best motivated theoretical benchmarks, a combination of these techniques is required. 

%\bibliographystyle{apsrevM}
%\bibliography{DMAReport}

\newpage

\section{Direct Detection of sub-GeV Dark Matter}
\vspace{-0.3cm}
\begin{flushleft}
\textit{Conveners:~Jeremy Mardon, Matt Pyle.~Organizer Contact: Rouven Essig}
\end{flushleft}

%%%%%%%%%%%%%%%%%%%%%%%%%%%%%%%%%%%%
\subsection{Introduction}

Among the most important experimental approaches to detecting dark matter (DM) particles in the laboratory are direct detection experiments~\cite{Goodman:1984dc}.  This program has over the last few decades focused on searches for Weakly Interacting Massive Particles (WIMPs), with masses above $\sim\!10$~GeV.  
The focus has been on searching for {\it nuclear recoils} induced by WIMPs in the Milky-Way DM halo 
scattering off nuclei in various target materials.  These detectors are placed deep underground and carefully shielded to reduce cosmic-ray 
and radioactive backgrounds.  They employ sensitive equipment to measure recoils of $\mathcal{O}$(keV) and more, and state-of-the-art target-material fabrication and purification to reduce contaminants that could mimic a DM signal.  
This program is well-established, important, and has a clear path forward~\cite{Cushman:2013zza}.  
Nevertheless, as emphasized in this report and elsewhere, the search for DM must be dramatically expanded to capture the range of 
possible candidates.  
In particular, in recent years it has been shown that many production mechanisms that explain the observed DM relic abundance, point to light, sub-GeV DM, below the reach of standard direct detection techniques~\cite{Boehm:2003hm, Boehm:2003ha,Hooper:2008im,Feng:2008ya, Kaplan:2009ag,Morrissey:2009ur,Hall:2009bx,Essig:2011nj,Chu:2011be,Lin:2011gj,Hochberg:2014dra,Hochberg:2014kqa,Kuflik:2015isi,Essig:2015cda}.  
It is thus crucial  to develop direct-detection strategies for such low DM masses.  

The traditional search for nuclear recoils rapidly loses sensitivity for DM masses, $m_\chi$, below a few GeV, since 
the energy of the recoiling nucleus is $E_{\rm NR} \le 2 \mu^2_{\chi, N} v_\chi^2/m_N$, where $\mu_{\chi, N}$ is the reduced 
mass of the DM and a nucleus of mass $m_N$, and $v_\chi\sim 10^{-3}c$ is the DM velocity.  
For a silicon nucleus and $m_\chi\sim \!100$~MeV, we have $E_{\rm NR}\sim \!1$~eV. 
Such low energies are very challenging to measure. The lowest nuclear-recoil threshold is set by cryogenic calorimeters, with CRESST recently achieving a $\sim$\,300\,eV threshold~\cite{Angloher:2015ewa} and SuperCDMS having achieved 315~eV without amplification in their recent R\&D detectors, with the goal of achieving 70~eV.
New experimental concepts are thus required to probe sub-GeV DM. 

In this section, we first review several models that can be probed with direct detection experiments sensitive to sub-GeV DM. 
We then discuss the experimental challenges for detecting sub-GeV DM in general terms, before reviewing various  
proposals for detection.  
Many suggested techniques benefit from the on-going R\&D that is already taking place in the design of the next-generation searches for standard WIMPs and for DM candidates in the mass range from $1-10$~GeV. Other techniques require a separate, dedicated effort to demonstrate their feasibility. 
In all cases, understanding or controlling backgrounds will be crucial to making a discovery.  
Given the relatively small scales of these experiments, even a modest investment of funds can enable the necessary R\&D that 
could allow for significant progress over the next few years.  

%%%%%%%%%%%%%%%%%%%%%%%%%%%%%%%%%%%%
\subsection{Theory Motivation}\label{subsec:DD-models}

There are many well-motivated production mechanisms for DM, including several that suggest DM in the keV--GeV~mass range, which we 
will briefly review. 
We then provide several concrete experimental targets for direct-detection experiments, 
assuming that DM is coupled to a dark photon (see also Sec.~II.B). 
The lower limit on thermally produced DM is $\simeq \! 3.3$~keV, which comes from constraints on small-scale structure formation as 
measured by Lyman-$\alpha$ forest data~\cite{Viel:2013apy}.  It is therefore important to probe for DM down to such masses.   
Of course, such limits strongly depend on the DM production mechanism, and viable sub-keV DM scenarios exist, as we mention below. 
\begin{itemize}[leftmargin=*]
\item {\bf Freeze-out DM:}  
WIMPs at the Weak-scale are motivated by the ``WIMP'' miracle: if the strength of a WIMPs interaction with ordinary matter is similar to the 
Weak force, then generically they can obtain the correct relic abundance from thermal $2\to 2$ freeze-out.  Just like most solutions to the 
Higgs hierarchy problem, this suggests new physics at the Weak scale.  However, if DM is part of a dark sector with new mediators, 
it is simple for DM with vastly different masses (including sub-GeV) to obtain the correct relic abundance from thermal 
freeze-out~\cite{Boehm:2003hm, Boehm:2003ha,Hooper:2008im,Feng:2008ya,Zurek:2008qg}. 
\item {\bf Asymmetric DM:}  
Baryons are known to have a relic abundance that is determined by an initial asymmetry between Standard Model (SM) particles and 
antiparticles (e.g.~baryons and anti-baryons).  
It is quite plausible that a similar asymmetry exists among DM particles/antiparticles, which determines the final relic abundance.  
Moreover, it is plausible that the DM asymmetry is linked to the baryon asymmetry.  
The predicted DM mass strongly depends on the realization of this scenario, although in many cases one finds the mass to be around the GeV scale. Since the precise mechanism is unknown, 
a much larger mass range $\sim$keV--GeV should be probed, see e.g.~\cite{Kaplan:2009ag,Falkowski:2011xh}.  
\item {\bf Freeze-in DM:}  
DM that is not in thermal equilibrium with the SM, but nevertheless has a small interaction with ordinary matter, can be produced via "freeze-in" in which the SM particles slowly annihilate or decay into DM, see e.g.~\cite{Hall:2009bx}.  
This irreducible source for DM production can fully account for the observed relic abundance, depending on the couplings and DM mass. In some light DM scenarios, this interesting region of parameter space can be probed in the near future with the technologies discussed below~\cite{Essig:2011nj,Essig:2015cda,Essig:2016crl}.  
% via the This again underscores the fact that DM can lie in a much wider mass range than that probed by typical WIMP experiments.  
%% 
\item
{\bf  Strongly Interacting Massive Particle (SIMP) ($3\to 2$ freeze-out):}
DM can be thermally coupled to the SM, but its final relic abundance is set by the freeze-out of 
number-changing $3\to 2$ processes among strongly-interacting-massive-particles (SIMPs) in a dark 
sector~\cite{Hochberg:2014dra,Hochberg:2014kqa}.  
In this scenario, DM naturally has sub-GeV masses.  Moreover, the required thermal equilibrium is only maintained for sizable 
DM-SM interactions, which could lead to interesting direct-detection signal (see example in Fig.~\ref{fig:DD-models}).  
\item
{\bf Elastically Decoupling Relic (ELDER):}
A variation of the SIMP scenario is that the final DM relic abundance is determined not by the $3\to 2$ process, but by the cross-section 
of its elastic scattering on SM particles~\cite{Kuflik:2015isi}.  
The predicted DM masses are a few to a few hundred MeV.  
As for the SIMP, a signal is again possible at direct-detection experiments (see example in Fig.~\ref{fig:DD-models}~\cite{ELDER}). 

\item
{\bf Bosonic DM from misalignment mechanism, decay of cosmological defects etc.:}
Pseudoscalar, vector, or scalar particles can be much lighter than 1~keV and also serve as interesting DM candidates through a variety of 
production mechanisms.  
This includes the QCD axion, axion-like particles, and the dark photon itself.  
Such particles do not scatter off bound electrons, but instead are directly {\it absorbed} by them. 
This allows direct-detection experiments sensitive to low-energy electron recoils to probe not only DM as light as 
keV-to-MeV that {\it scatters} off electrons, but also the absorption of bosonic DM as light as meV-to-keV.  
These particles could also be produced and emitted from the Sun.  
See e.g.~\cite{Dimopoulos:1985tm,Avignone:1986vm,Pospelov:2008jk,Derevianko:2010kz,Nelson:2011sf,Arias:2012az,Graham:2015rva,An:2013yfc,An:2013yua,Redondo:2008aa,Aalseth:2008rx,Ahmed:2009ht,Armengaud:2013rta,Aprile:2014eoa,Yoon:2016ogs,Hochberg:2016ajh,Bloch:2016sjj,Hochberg:2016sqx}.  
%%
%\item 
%{\bf Warm DM mass limit of $\sim$keV:}
%DM that is in thermal equilibrium with the SM in the early Universe can wash out small-scale structure if its mass is low enough.  
%The current lower bound, from Lyman-$\alpha$ forest measurements~\cite{Viel:2013apy}, is $\simeq \! 3.3$~keV for thermally produced DM.  
%It is therefore important to probe for DM down to such masses. 
\end{itemize}

While the above summary motivates searches for sub-GeV DM in general, we now describe a specific model (DM coupled to a dark photon),  
which provides several simple and concrete experimental targets.  
More theoretical work is needed to find additional DM models and map out their parameter space.  
\begin{center} 
{\bf Concrete example: DM, $\chi$, coupled to a dark photon, $A'$}  
\end{center}

A simple, predictive scenario is if DM is a Dirac fermion or a complex scalar 
that is charged under a broken Abelian gauge group, 
$U(1)_D$, whose mediator is the $A'$, with a mass $m_{A'}$ and kinetic-mixing parameter, $\epsilon$.  
The DM can scatter off ordinary matter (electrons and nuclei) 
by exchanging an $A'$.  
It is useful to parameterize the full DM-electron scattering cross section in terms of a reference cross section, $\overline{\sigma}_e$, and 
a DM form factor, $F_{\rm DM}$, which captures the momentum-dependence of the interaction (see~\cite{Essig:2011nj} for details). 
The reference cross section is evaluated at the typical momentum transfer of DM-electron scattering, which for bound electrons in 
atoms/semiconductors is $\sim\alpha m_e\sim \!4$~keV.  
Depending on $m_{A'}$, we find 
\bea \label{eq:sigmae}
\overline\sigma_e 
\simeq
\begin{cases}
\frac{16 \pi \mu_{\chi e}^2 \alpha \alpha_D \epsilon^2}{m_{A'}^4}\,, & m_{A'} \gg \alpha m_e \\
\frac{16 \pi \mu_{\chi e}^2 \alpha \alpha_D \epsilon^2}{(\alpha \, m_e)^4}\,, & m_{A'} \ll \alpha m_e
\end{cases}\,,
{\rm and}~~~ 
 F_{DM}(q) 
\simeq
\begin{cases}
1\,, & m_{A'} \gg \alpha m_e \\
\frac{\alpha^2 m_e^2}{q^2}\,, & m_{A'} \ll \alpha m_e\,,
\end{cases}
\eea
where $\mu_{\chi e}$ is the DM-electron reduced mass, $m_e$ ($m_\chi$) the electron (DM) mass, $\alpha$ is the fine-structure constant, 
and $\alpha_D = g_D^2/4\pi$, where $g_D$ is the $U(1)_D$ gauge coupling. 
For $m_{A'}\gg \alpha m_e\sim 4$~keV, the interaction is momentum independent, $F_{\rm DM}=1$, while for 
$m_{A'}\ll \alpha m_e$, the interaction increases as the momentum transfer is lowered.  
In the latter scenario, the rates are significantly enhanced for low momentum-transfer events (relevant for direct detection) 
compared to high momentum-transfer events (relevant for collider or fixed-target searches).  
This simple setup provides interesting experimental targets for a wide range of parameters 
(see Sec.~II.B~and~\cite{Essig:2015cda} for more details): 
\begin{itemize}[leftmargin=*]
\item[~{\bf (1)}] {\bf Freeze-out DM target:} 
If $\chi $ is a complex scalar and $m_{A'} > 2 m_\chi$, the thermal relic abundance of $\chi$ is set by the 
freeze-out of the process 
$\chi \chi^*\to A'^* \to {\rm SM}$, which scales as 
$\langle\sigma v\rangle \propto \alpha \alpha_D \epsilon^2/m_{A'}^4$, i.e.~similar to $\overline\sigma_e$ defined above.  
As in Sec.~IV, we can set e.g.~$m_{A'} = 3m_\chi$ to fix the value of $\overline\sigma_e$ as a function of $m_\chi$ 
(see thick blue line, labeled ``freeze-out'', in Fig.~\ref{fig:DD-models} (left)).   This relation between $\overline\sigma_e$ and $m_\chi$ may be easily altered if additional (hidden) DM annihilation channels exist. 
The magenta shaded region 
shows a constraint from XENON10~\cite{Essig:2012yx}, reviewed below.  
Beam-dump, collider, and traditional direct-detection constraints searching for elastic nuclear recoils are shown in gray~\cite{Boehm:2013jpa,Nollett:2013pwa,Ade:2015xua,Akerib:2013tjd,Agnese:2015nto,Angloher:2015ewa,deNiverville:2011it,Batell:2009di,Kahn:2014sra,Bjorken:1988as,Batell:2014mga,Aubert:2008as,Essig:2013vha} (where relevant for mapping these constraints onto 
the $\overline\sigma_e-m_\chi$ plane, we set $\alpha_D=0.5$~\cite{Davoudiasl:2015hxa} 
and do not include self-interaction constraints~\cite{Randall:2007ph,Tulin:2013teo}; the latter are somewhat uncertain, but would 
only affect constraints for $m_\chi\lesssim 20$~MeV). 
Note that the parameter $y$ defined in Sec.~IV is 
related to $\overline\sigma_e$ as $\overline\sigma_e  = (16 \pi\mu_{\chi e}^2/m_\chi^4) \times y$.
\item[~{\bf (2)}] {\bf Asymmetric DM target:} 
A simple variation of the previous scenario is if $\chi$ is a Dirac fermion instead of a complex scalar 
(also with $m_{A'} > 2 m_\chi$).  Here an initial asymmetry in $\chi$ versus $\bar\chi$ must set the final relic abundance~\cite{Kaplan:2009ag}.  
The same annihilation process as above of DM to the SM is now $s$-wave (as opposed to $p$-wave) 
and we have to ensure that the symmetric 
$\chi\bar\chi$ component is small enough to avoid CMB and 
gamma-ray bounds~\cite{Madhavacheril:2013cna,Essig:2013goa}.  
This sets a lower bound on the annihilation rate and thus also on $\overline\sigma_e$~\cite{Lin:2011gj}, 
which is shown by the thick green line in Fig.~\ref{fig:DD-models} ({\it left}) for $m_{A'}=3m_\chi$.   As above, additional DM annihilation channels may ameliorate the constraint. 
In this model, all the white space above the green line provides an important target.  
\item[~{\bf (3)}] {\bf Freeze-in DM target:} 
For $m_{A'} \ll \alpha m_e$, a non-zero DM abundance can be generated from freeze-in~\cite{Hall:2009bx,Essig:2011nj,Chu:2011be}, 
which again fixes $\overline\sigma_e$ and provides another important experimental target, as indicated by the 
thick blue line in Fig.~\ref{fig:DD-models} ({\it right}).  
The magenta shaded region
again shows a constraint from XENON10~\cite{Essig:2012yx}, reviewed below.  
Since the $A'$ is light, constraints applicable to millicharged particles also apply here (even though the DM is not millicharged). These 
are shown in gray~\cite{Davidson:2000hf}, together with a direct detection bound from traditional elastic nuclear recoil 
searches~\cite{Agnese:2013jaa,Akerib:2013tjd}.  
\end{itemize}
As indicated above, some model adjustments, e.g.~the existence of other low-mass dark-sector states or taking $m_{A'} \lesssim m_\chi$, could 
open up more/different parameter space.  Moreover, DM could constitute a sub-dominant component.  
This makes it desirable to probe to even lower sensitivities than those indicated by these important targets.  

We note that the ``freeze-out'' and ``asymmetric'' targets above are discoverable 
both with direct detection experiments and with beam-dump / collider experiments.  
If the DM scattering is inelastic from one state, $\chi_1$, to another, $\chi_2$ (with $m_{\chi_2}-m_{\chi_1}\gtrsim 10$'s of eV), 
direct-detection probes have very little sensitivity, far weaker than accelerator-based probes, since the DM in our halo is non-relativistic. 
However, direct detection experiments are far superior to accelerator-based probes for models that have a very light mediator 
($\ll $~few keV) as in the ``freeze-in'' target above, or for bosonic DM with masses below $\sim \!1$~keV.  
We also note that several materials and techniques discussed below cannot probe well the various DM-$A'$ scenarios.  
For example, in-medium effects greatly suppress the DM-electron scattering rates in superconductors~\cite{Hochberg:2015fth}, and 
the rates for $\sim$keV DM scattering off helium is also suppressed due to the small momentum transfers and the fact that 
the $A'$ couples to electric charge~\cite{Schutz:2016tid}.

%%%%%%%%%%%%%%%%%%%%%%%%%%%%%%%%%%%%
\subsection{General Goals and Challenges}

In general, the main goals and challenges in direct detection of sub-GeV DM are:

\begin{itemize}[leftmargin=*]
\item 
\emph{Finding detectable low-threshold processes}: 
Sub-GeV DM has much less energy to deposit than Weak-scale DM (${\sim\!1\,\text{eV} \times (m_{\rm DM}/\text{MeV})}$). 
Detectable, low-threshold processes must thus be identified that could be triggered by light DM. 
Since some amplification mechanism is often required, the type of process that can be considered is restricted. 
\item
\emph{Backgrounds}:
Backgrounds can mimic the DM signal.  
These may be totally unrelated to the dominant backgrounds in conventional nuclear recoil searches.  
Indeed, often the ``traditional'' backgrounds like cosmogenics, Compton scattering, or neutrons can be controlled and made 
negligible. Instead, understanding and controlling detector-specific backgrounds, often created by the amplification mechanism, 
may present the dominant experimental design challenge. 
\item
\emph{Scaling up exposure}:
Some new experimental concepts may initially only be possible with very small targets, 
and new challenges need to be confronted when scaling them up. 
\item
\emph{Signal discrimination \& Background model}:
Since discovery is the primary goal of any direct detection experiment (as opposed to setting new limits), it is essential to be able to distinguish real DM scattering events from backgrounds. This may be on an event-by-event basis (such as in many of the existing nuclear-recoil DM searches) or on a statistical basis over many events (for example by annual modulation or directional sensitivity).
\item
\emph{Improved material fabrication}:
Some new ideas require specific target materials with, for example, unprecedented levels of purity or structural coherence. 
This may require advances in the technology for fabricating these materials. 
\end{itemize}

\begin{figure}[t]
\includegraphics[width=0.8\textwidth]{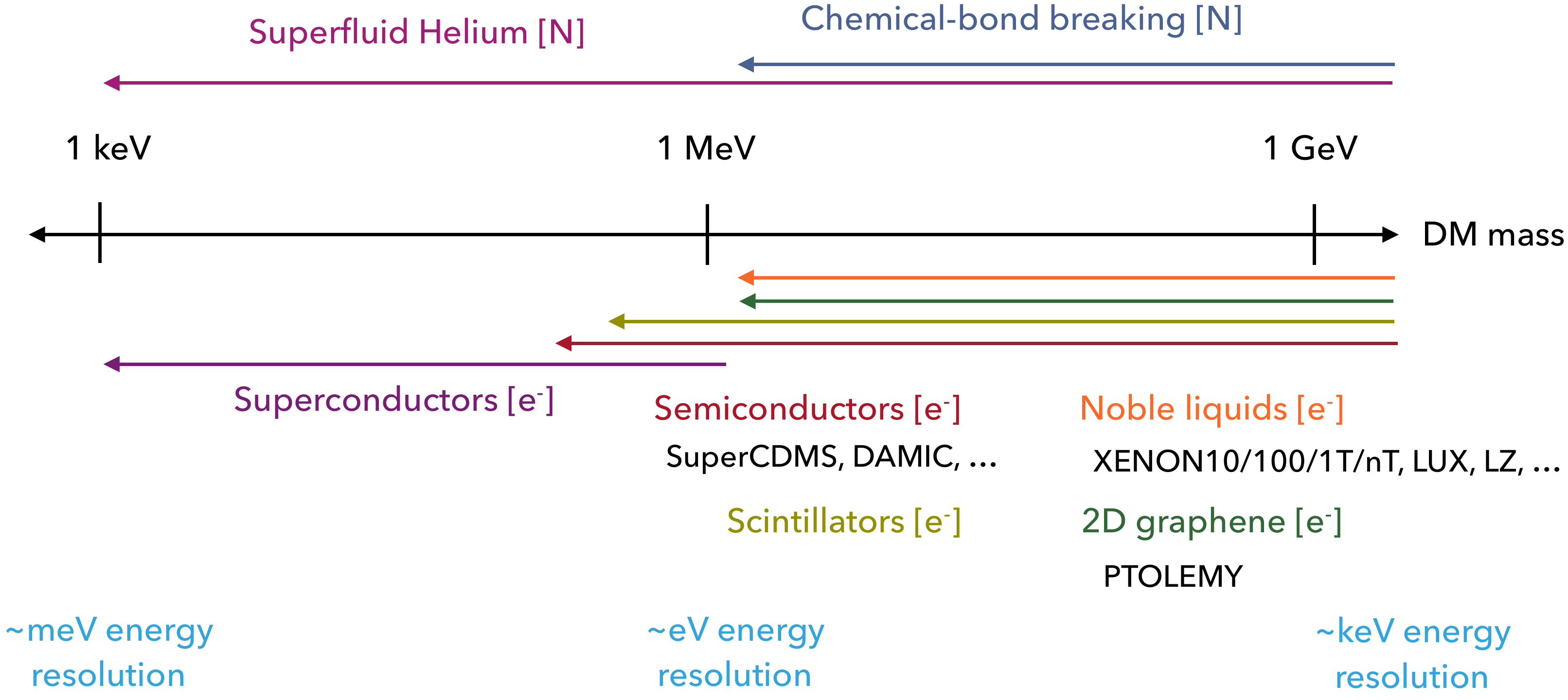}
\caption{\footnotesize 
Materials that could be used to probe sub-GeV DM, down to keV masses, by {\it scattering} off electrons [$e^-$] or nuclei [$N$].  
Certain DM candidates, which can instead be {\it absorbed} by bound electrons in these materials, could be probed down to 
meV masses (not shown).  Adapted from~\cite{dmlandscape}. 
\label{fig:DMlandscape}
}
\end{figure}
\begin{figure}[t]
\includegraphics[width=0.48\textwidth]{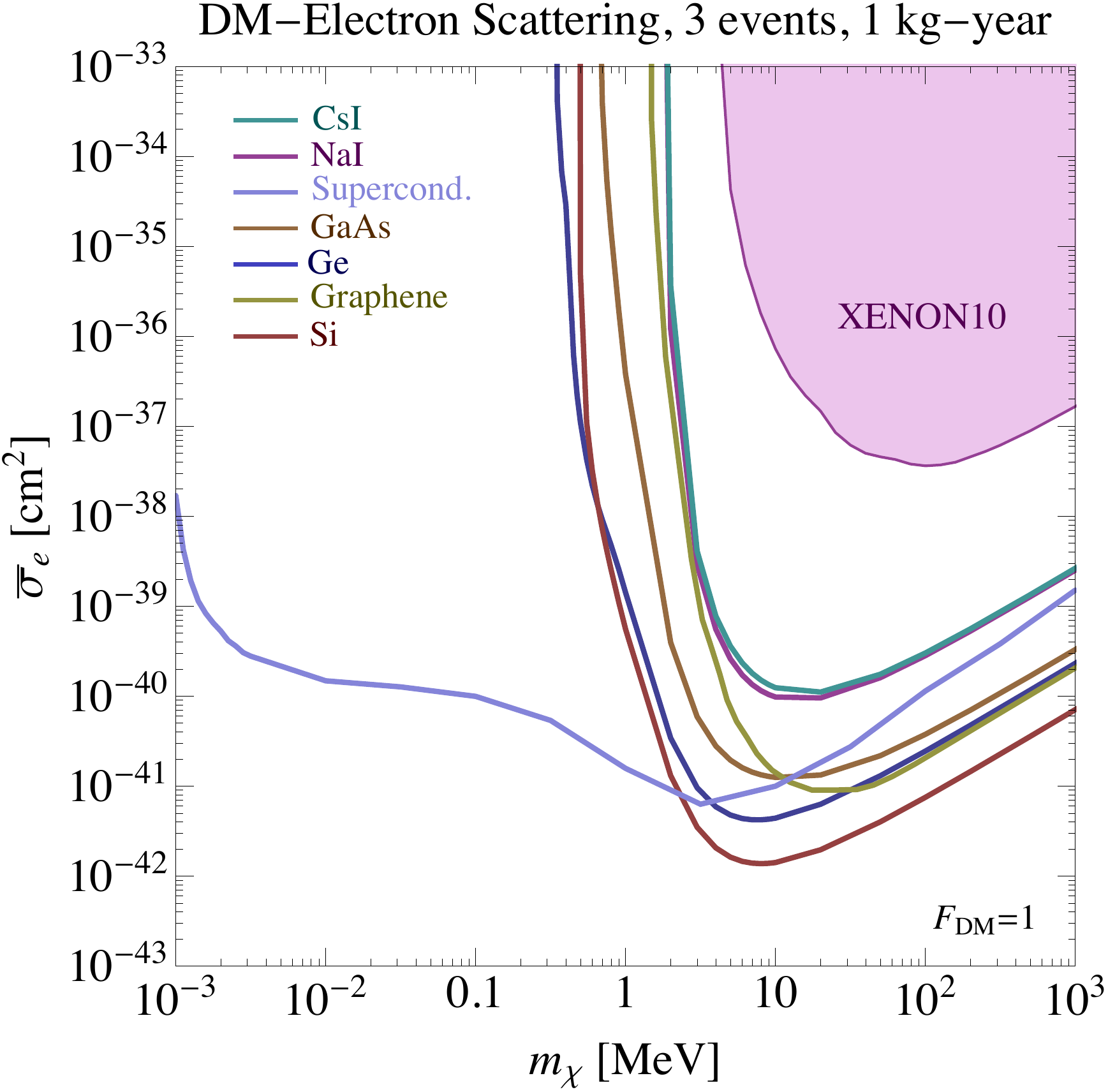}
~
\includegraphics[width=0.48\textwidth]{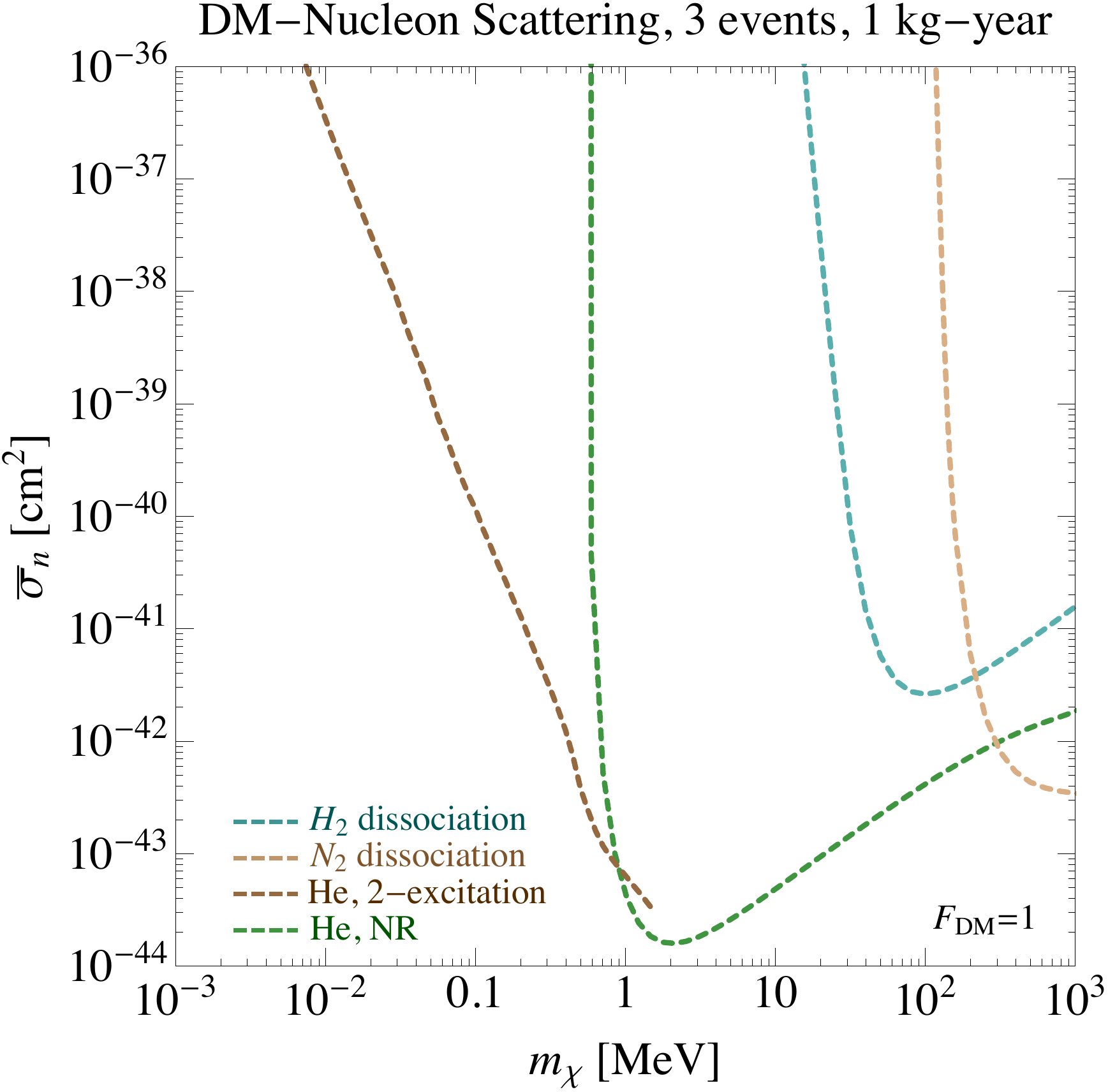}
\caption{\footnotesize 
Projections for DM-electron ({\bf left}) and DM-nucleon ({\bf right}) cross section needed in order to produce 3 events in various materials, and for an exposure of 1~kg-year.  
For the DM-electron scattering case, the materials include 
semiconductors (Ge, Si), scintillators (GaAs, CsI, NaI), 2D materials (graphene), and superconductors.  For DM-nucleon scattering we show projections for 
2-excitation processes in helium (He, 2-excitation), ordinary nuclear recoils in helium (He, NR), 
and dissociation of either H$_2$ or N$_2$ molecules.  
The technology needed to detect these events differs for various elements; challenges and rough potential timescales are summarized   
in Table~\ref{tab:DDtimescales} and the text.  
The magenta shaded region  
shows a constraint on DM-electron scattering from~\cite{Essig:2012yx}, using XENON10 
data~\cite{Angle:2011th}. 
Both plots assume the fundamental interaction between the DM and the electron/nucleus is momentum-independent.  No specific model is assumed here.  
\label{fig:DD-models-generic}
}
\end{figure}
%%

%%%%%%%%%%%%%%%%%%%%%%%%%%%%%%%%%%%%
\subsection{Overview of Strategies and Target Materials}

While searching for (elastic) nuclear recoils rapidly loses sensitivity for DM below a few GeV, a fruitful strategy is to search for 
DM scattering off bound {\it electrons} (instead of a nucleus)~\cite{Essig:2011nj}.  
This allows all of the available DM kinetic energy to be transferred, so that for a bound electron with a binding energy $\Delta E_B$, 
one can in principle probe masses of 
\begin{equation}
m_\chi \gtrsim 250~{\rm keV} \times \frac{\Delta E_B}{1~{\rm eV}}\,.
\end{equation}
The signal depends on the material, but consists of one or more electrons (possibly amplified by an electric field) in 
noble liquids~\cite{Essig:2011nj,Essig:2012yx}, 
semiconductors~\cite{Essig:2011nj,Graham:2012su,Lee:2015qva,Essig:2015cda}, 
superconductors~\cite{Hochberg:2015pha,Hochberg:2015fth}, 
 graphene~\cite{Hochberg:2016ntt}, 
or one or more photons in scintillators~\cite{Essig:2011nj,Derenzo:2016fse}. 

Another strategy to probe below the GeV-scale is to search for DM scattering off nuclei using {\it inelastic} processes.  
The breaking of chemical bonds in molecules or crystals could produce measurable signals for few-MeV DM 
masses~\cite{Essig:2011nj,Essig:2016crl}, while multi-phonon processes in superfluid helium or insulating crystals 
could provide sensitivity to keV DM masses~\cite{Schutz:2016tid}. 
Photon emission in the nuclear recoil could also probe below the GeV-scale~\cite{Kouvaris:2016afs}. 

\begin{figure}[t]
\includegraphics[width=0.48\textwidth]{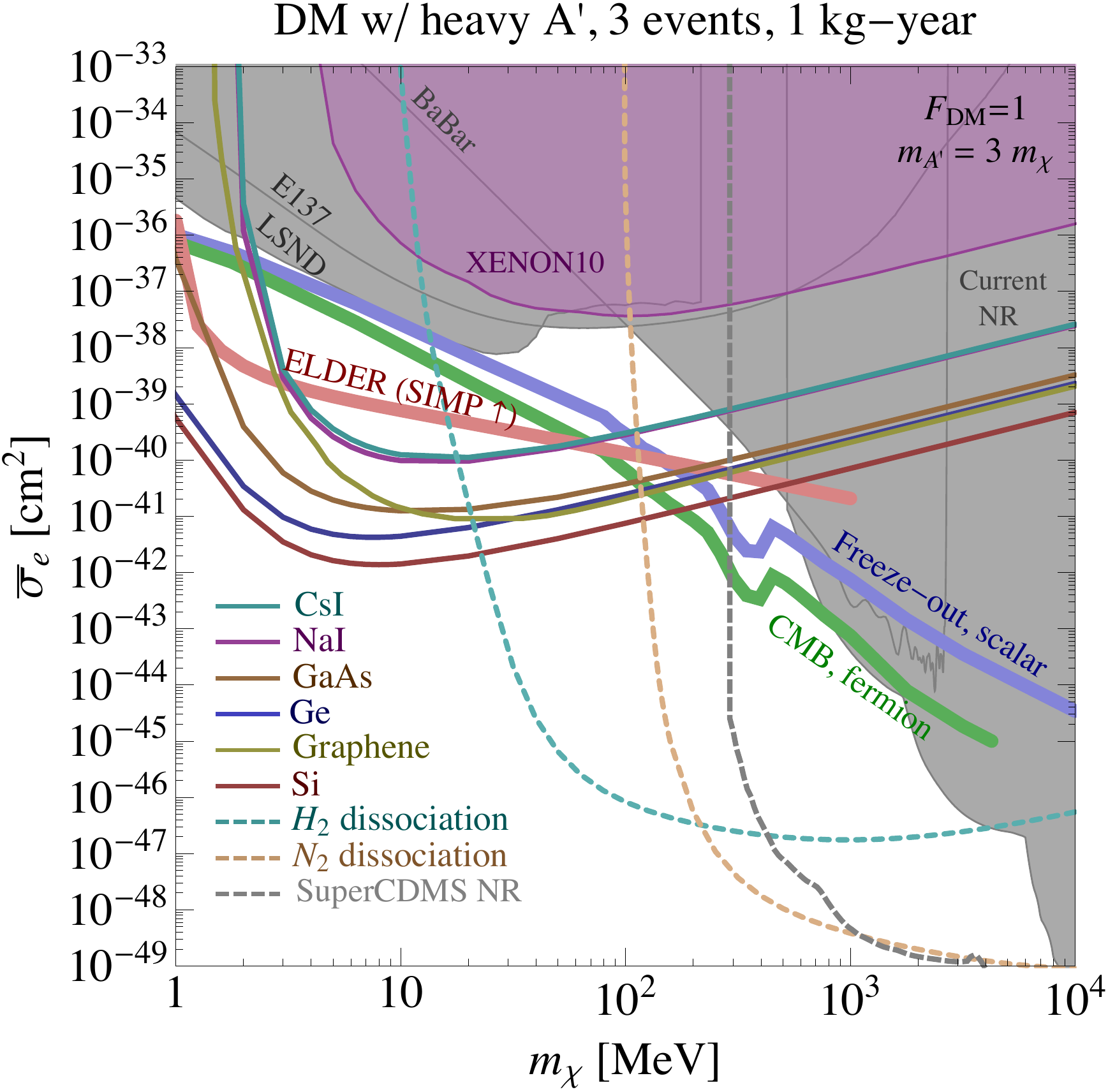}
~
\includegraphics[width=0.48\textwidth]{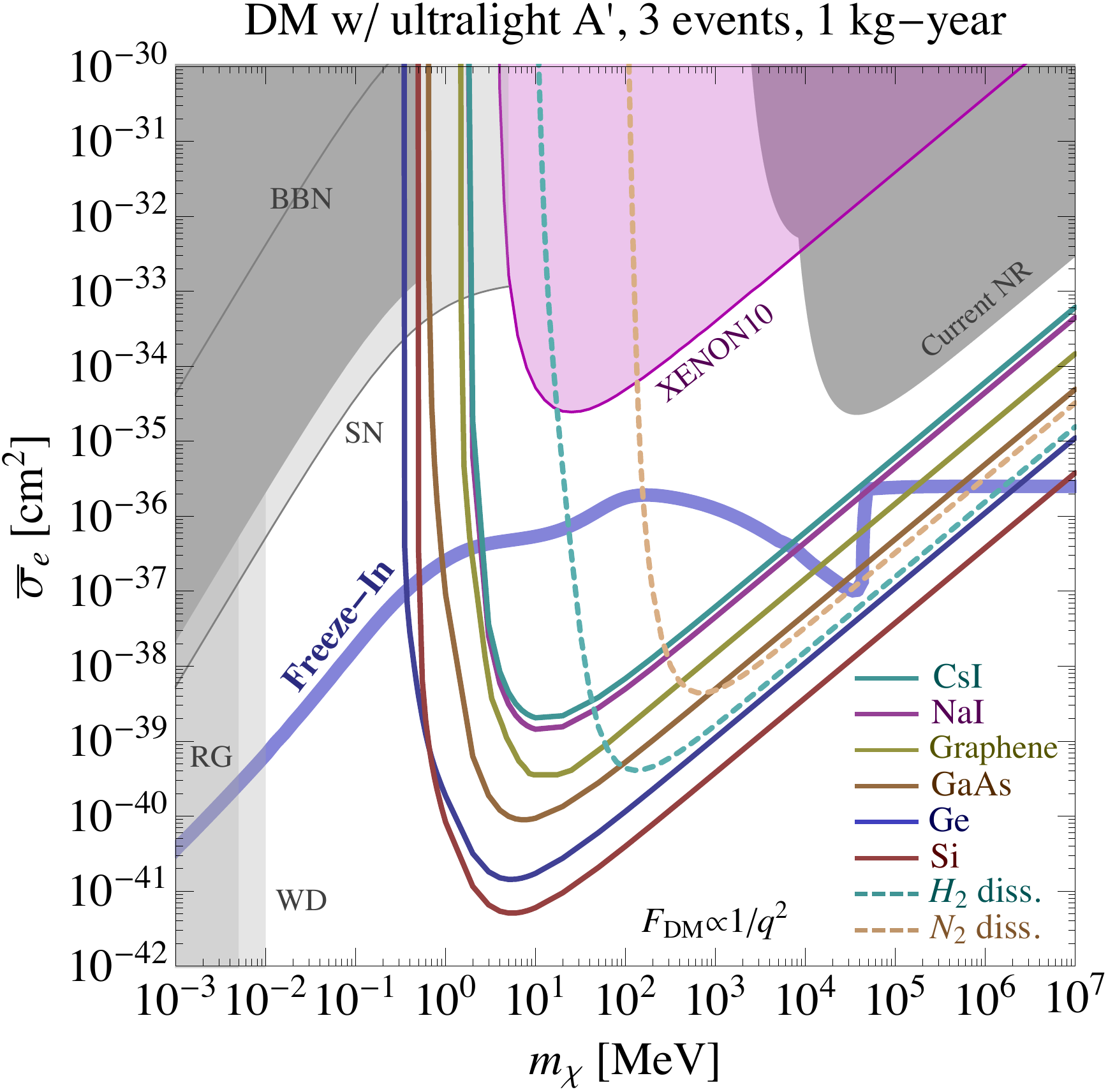}
\caption{\footnotesize 
DM scattering rates in various materials, overlaid on the parameter space for DM, $\chi$, interacting with a 
dark photon, $A'$. 
Solid (dashed) colored curves show cross section needed for DM scattering off electrons (nuclei) to produce 3 events, 
assuming an exposure of 1~kg-year 
(superconductors, known to have a large optical response for a dark-photon mediator, are not included~\cite{Hochberg:2015fth}).  
The magenta shaded region shows a constraint on DM-electron scattering from~\cite{Essig:2012yx}, using XENON10 
data~\cite{Angle:2011th}. 
{\bf Left:} 
Here $m_{A'}=3 m_\chi$, so $F_{\rm DM}=1$, see Eq.~(\ref{eq:sigmae}).  
Complex scalar DM obtains the observed DM relic density along the thick blue line (``Freeze-out, scalar''). 
For a Dirac fermion as DM, the abundance is determined by an initial asymmetry; the region above 
the thick green line is allowed (``CMB, fermion''). 
An elastically decoupling relic (ELDER) lies along the thick orange line~\cite{ELDER}, while a SIMP lies above 
it~\cite{Hochberg:2014dra,Hochberg:2014kqa}.  
Shaded gray regions show bounds from beam-dump, collider, and direct-detection searches for elastic nuclear recoils
(see also Sec.~IV).  
{\bf Right:} 
Here $m_{A'} \ll $~few keV (the precise value is irrelevant), so 
$F_{\rm DM}\propto 1/q^2$, see Eq.~(\ref{eq:sigmae}).  
Freeze-in produces the correct DM abundance along the thick blue curve.  
Shaded gray regions show bounds from direct-detection searches for elastic nuclear recoils, as well as stellar and BBN constraints. 
%See Sec.~\ref{subsec:DD-models} for more details.  
\label{fig:DD-models}
}
\end{figure}

The strategy to search for recoiling {\it electrons} has been {\it proven} to probe DM as light 
as a few MeV in {\it existing} two-phase xenon-based time projection chambers (TPC) (XENON10 and XENON100), 
which have been able to set the only direct-detection constraints on (scattering) sub-GeV DM so far~\cite{Essig:2012yx, futureXe100}. 
Future improvements utilizing this method will depend critically on reducing a detector-specific (spurious) background (a ``dark count''). 
To gain sensitivity to even lower DM masses and/or scattering cross sections many orders of magnitude smaller, new 
experimental strategies, improved technologies, and other target materials (as mentioned in the previous two paragraphs) are needed. 
Fig.~\ref{fig:DMlandscape} summarizes the mass range that can in principle be probed by these target materials.  
Fig.~\ref{fig:DD-models-generic} shows the required cross section for DM to produce 3 events when it scatters off electrons or 
nuclei in various materials, for an exposure of 1~kg-year.    
Fig.~\ref{fig:DD-models} shows the DM-electron scattering rates overlaid for several concrete models 
(DM coupled to a dark photon, ELDER, and SIMP, as discussed in Sec.~\ref{subsec:DD-models}).  
Only those materials are shown that are known not to have a large suppression for this model, 
due e.g.~to in-medium effects as in superconductors. 
We stress that some of the recently-suggested experimental strategies require significant (and possibly long-term) R\&D.   Table~\ref{tab:DDtimescales} summarizes the needed technologies and challenges.  
We now review various material choices and strategies.

\begin{itemize}[leftmargin=*]
\item {\bf Noble liquids} 

Two-phase (liquid and gas) xenon-based time projection chamber (TPC) detectors, whose primary purpose is to search for elastic nuclear recoils from $m_\chi >\! 4$~GeV~\cite{Akerib:2015rjg}, also provide excellent sensitivity to electron recoils induced by sub-GeV DM searches~\cite{Essig:2011nj}.    
An electron in the outer shell of a xenon atom has $\Delta E_B \sim \mathcal{O}(12.1~{\rm eV})$, so it can be 
ionized by DM with $m_\chi \gtrsim$~a few MeV.  With large E-fields, these ionized electrons are drifted through the liquid xenon and extracted into xenon gas, where the larger electron drift velocity means that a measurable number of scintillation photons are produced via scattering off xenon-gas atoms (``S2'' signal).  By searching for this S2 signal 
world-leading sensitivity to DM down to a few MeV masses is obtained~\cite{Essig:2012yx} using published XENON10 data~\cite{Angle:2011th}. 

 The dominant background for these searches was not the expected radiogenic electron recoils in the outermost shielding layer of xenon 
but rather a spurious electron dark count rate.  A significant but not complete fraction of this rate has been determined to be due to ionized 
electrons, originally created by highly ionizing background events outside of the DM scattering region of interest, that become trapped 
at the liquid-gas interface and are released spontaneously at a later time. R\&D into minimizing this source is ongoing in all xenon TPC collaborations. For example, LZ is attempting to increase the extraction electric field at the liquid-gas interface to improve the electron transport efficiency from the $\sim$65\% found in LUX~\cite{Akerib:2015rjg} to 97.5\%. Such efforts, however, may increase dark leakage current from electrons tunneling out of the electrode grids and thus all aspects of the two-phase TPCs must be carefully optimized.

\begin{table}[t!]
\begin{footnotesize}
\begin{center}
\begin{tabular}{|c|c|c|c|c|}
\hline
  Material   				& $m_{\rm DM, th}$ (theoretical) 	& Technology 						& Challenges  		& (Optimistic) Timescale  		\\ \hline \hline
Noble liquids  		& \multirow{2}{*}{few MeV} 					& \multirow{2}{*}{two-phase TPC}					& \multirow{2}{*}{dark counts} 		& \multirow{2}{*}{existing} 			\\ 
(Xe, Ar) 		&  					& 		&  		&  			\\ \hline 
Semiconductors 	& \multirow{2}{*}{$\sim 0.1-1$~MeV} 			& CCDs \& 			 	& \multirow{2}{*}{dark counts (?)} 	& \multirow{2}{*}{$\sim 1-2$ years}	\\ 
(Si, Ge) 			& 						 			& Calorimeter			 	&  						&  		\\ \hline
Scintillators 		& \multirow{2}{*}{$\sim 0.5-1$~MeV} 		& Calorimeter: 	&  sensitivity \& 	& \multirow{2}{*}{$\lesssim 5$ years} \\ 
(GaAs, NaI, CsI) 	&  		 			& $\sigma_E \sim 0.2$~eV	&  afterglow (?)		& 							 \\ \hline
Superconductors 	& \multirow{2}{*}{$\sim 1$~keV}	& Calorimeter: 			 	& sensitivity \&  			& \multirow{2}{*}{$\sim 10-15$ years} 	\\ 
(Al) 				& 						& $\sigma_E \sim 1$~meV 	& unknown backgrounds 			& 							 \\ \hline
\hline
\hline
Superfluid He & \multirow{2}{*}{$\sim 1$~MeV}  & Calorimeter:  			&  sensitivity \& & \multirow{2}{*}{$\lesssim 5$ years} \\ 
(NR) 			&  						   & $\sigma_E \sim 1   $~eV 	&  unknown backgrounds &  \\ \hline
\multirow{2}{*}{Bond Breaking}  & \multirow{2}{*}{$\sim \!$ few MeV} 				& \multirow{2}{*}{color centers}& sensitivity \& & \multirow{2}{*}{$\lesssim 5$~years}  	\\ 
		&  				& 							& unknown backgrounds 			& \\ \hline
Superfluid He & \multirow{2}{*}{$\sim 1$~keV} 	& Calorimeter  		& sensitivity \&  			& \multirow{2}{*}{$\sim 5-10$~years}  		\\ 
(2-excitation) &  				& $\sigma_E \sim 10$~meV 	& unknown backgrounds 			& \\ \hline
2D-targets 		& \multirow{2}{*}{few MeV}	& based on  	
														& low exposure, 	& \multirow{2}{*}{$\sim 5-10$ years}	\\ 
(graphene) 		& 		&  							PTOLEMY			& unknown backgrounds	& 	\\ \hline
\end{tabular}
\caption{\label{tab:DDtimescales}
\footnotesize 
Material, theoretical mass threshold, required technology to achieve lowest mass threshold, potential or known challenges, 
and optimistic timescales.  
All materials and techniques, besides two-phase TPCs in noble liquids, still need to be demonstrated to have sensitivity to sub-GeV DM. 
Most timescales are thus only illustrative of the time needed to study their feasibility. 
Materials/techniques sensitive to DM-electron (DM-nucleus) interactions are at the top (bottom) of the table.
}
\end{center}
\end{footnotesize}
\end{table}

\item {\bf Semiconductors} 

Due to having an electronic bandgap an order of magnitude smaller than those found in xenon and other insulators, semiconductors are conceptually sensitive to electronic recoil-dark matter scattering signatures from dark matter with an order of magnitude small mass (sub-MeV) and with smaller cross sections than those accessible to xenon~\cite{Essig:2011nj,Graham:2012su,Lee:2015qva,Essig:2015cda}. Furthermore, semiconductor detection technology is incredibly common, and can thus be relatively easily repurposed for light mass dark matter searches.

DAMIC, for example, repurposes CCD technology and has set the best non-xenon target constraint on sub-GeV DM~\cite{Essig:2015cda} with their initial prototype experiment~\cite{Barreto:2011zu}. R\&D for their next generation experiment has also proven quite successful. They have decreased their ionization measurement noise ($\sigma$) from 2 e$^{-}$ down to 0.2 e$^{-}$ and believe that another factor of 2 improvement is possible \cite{Fernandez_12_SkipperCCD}. Their new 1mm thick CCDs also have a dark count rate $<$ 10$^{-3}$ e$^{-}$/pixel/d, two orders of magnitude less than in their original prototype detector \cite{Aguilar_16_DAMIC}. Finally, by upgrading their external shielding and through better material selection they have achieved a radioactive background of 5 ${\rm \frac{evt}{keV kg d}}$, a reduction of two orders of magnitude from their initial prototype experiment. After accounting for these improvements, it's fully expected that their upcoming experiment will probe currently unexplored light mass parameter space. Subsequent R\&D will primarily focus on backgrounds that are expected to limit their sensitivity in this upcoming run. Specifically, it's expected that random coincidence of 2 dark count events within a single pixel during a single measurement period will limit their sensitivity to 1 and 2 $e^{-}/h^{+}$ producing recoils, while for higher energies recoils their sensitivity will be limited by radiogenic backgrounds. 

Just as with noble liquid detectors, SuperCDMS~\cite{CDMSliteRun2_2015}, a world leading cryogenic semiconductor calorimeter-based experiment for $1-4$~GeV DM, can also be repurposed to search for electronic recoils from DM. To gain sensitivity to such a small ionization signal created by either sub-keV nuclear 
recoils or few-eV electron recoils, the planned SuperCDMS SNOLAB detector drifts the electron/hole pairs through the semiconducting crystal with 
an external $E$-field.  This converts their electrostatic potential energy into athermal phonon energy, which is collected and measured with 
sensors fabricated onto the surface of the detector.  The potential disadvantage of using this phonon amplification technique is that the semi-conducting crystal is fundamentally out of 
equilibrium, and thus any tunneling or IR-photon excited process could produce a dark current and limit the sensitivity in a similar way 
to that seen in xenon TPCs. Over the next two years, SuperCDMS plans to study a variety of insulating surface layers and to improve their phonon-sensor sensitivity such that $E$-fields as low as 10~V/cm can be used to drift the ionization (two orders of magnitude less than the $E$-fields used in xenon TPCs).

\item {\bf Scintillators} 

Another strategy to search for electron recoils from sub-GeV DM is to use scintillators~\cite{Essig:2011nj,Derenzo:2016fse}, 
e.g.~direct band gap semiconductors like GaAs or insulators like NaI and CsI.  
Their band gaps are $\Delta E_B \sim \mathcal{O}(1-5~{\rm eV})$, so that similar DM masses can be probed as with Ge or Si.  
However, since no external $E$-fields are required to measure scintillation photons produced by the excitation and subsequent de-excitation  of the recoiling electron, no dark-count rate exists, in potential contrast to semiconductor detectors discussed above. 

New large-area photon detectors are needed to measure single- or few-photon events with negligible 
dark counts (i.e.~no PMTs). 
One promising pathway is to build scintillation detectors from microwave kinetic inductance detector (mKID) and transition 
edge sensor (TES) superconducting technology 
(which should be dark-count free). In particular, it is hoped that the TES-based athermal phonon technology 
from SuperCDMS could be repurposed within the next two years.
An additional potential advantage over traditional SuperCDMS technology is that the separation of the active absorber volume from the sensor 
means that cosmogenic backgrounds like $^{3}$H can be drastically reduced, because the crystals can be stored underground 
for their entire lifetime. 

In order for this approach to be feasible, the scintillators must be demonstrated to have a scintillation photon production efficiency that is 
$\mathcal{O}(1)$.  Moreover, any ``afterglow'' --- phosphorescence induced from a previous interaction and arising due to long-lived excited states on crystal impurities --- must be negligible.  If a crystal cannot be found with an afterglow rate  $\lesssim 10$~events/kg/yr, then optical filters or photon sensors with excellent energy resolution would be needed to distinguish DM-induced photons from the lower-energy afterglow photons. 

\item {\bf Superconductors} 

Superconductors, like aluminum, have much lower thresholds to generate a signal ($\mathcal{O}$(1~meV)) 
that is determined by the binding energy of a Cooper pair.  
This could allow for sensitivity to DM masses as light as a keV~\cite{Hochberg:2015pha}, the warm DM mass limit.  
Ultrasensitive phonon detectors requiring significant R\&D over the next decade could achieve sensitivity to such low 
thresholds~\cite{Hochberg:2015fth}.   

\item {\bf Two-dimensional targets} 

Two-dimensional targets, like sheets of graphene, have similar thresholds to semiconductor and scintillator targets.  
Moreover, they could provide a means to distinguish signal from background by allowing for directional sensitivity 
to the recoiling electrons~\cite{Hochberg:2016ntt}. 
One challenge is to obtain $\mathcal{O}$(1~kg) target masses, but this may be possible with a setup based on the proposed PTOLEMY 
experiment~\cite{Betts:2013uya}.  

Single-wall carbon nanotubes (CNT) are wrapped graphene sheets, with variable lengths 
and diameters. The possibility of using CNTs as directional detectors of WIMPs as
light as $\sim 1$~GeV, by exploiting their ion-channeling properties, has been considered in~\cite{Capparelli:2014lua,Cavoto:2016lqo}, 
Studies are underway to see if coaxial electric fields allow the detection of much softer ion recoils, which would probe 
lighter WIMP masses. 

\item {\bf Chemical Bond Breaking} 

DM scattering off nuclei could break apart chemical bonds~\cite{Essig:2016crl}. This includes the dissociation / excitation of 
molecules and the creation of defects in a lattice. With thresholds for dissociation of a few--10s of eV, such an experiment could probe the nuclear couplings of DM particles as light as a few MeV. 
For the case of excitation, thresholds are much lower while detection and background reduction may be more challenging.

%The choice of target material (target-nucleus mass and binding energy) will determine the scattering rate and DM mass threshold. 
%Heavy nuclei allow for signal enhancement due to a coherence effect, while light nuclei reduce the mass threshold. 
%Targets can be chosen to consist of two or more types of nuclei, thus gaining sensitivity to a large range of DM parameters.
%
Since dissociated final states are often extremely long-lived, the measured signal can potentially be amplified. Creation of lattice defects~\cite{futureCC} is particularly appealing, since amplification can be achieved by exciting sites (creating defects) within the lattice with photons, and measuring the fluorescent change. Background discrimination is possible, since high- or low-energy events and nuclear- or electron-recoils create different signals. Moreover, the binding potential for nucleons in a lattice is not spherically symmetric, so that the interaction rates have a daily modulation, helping with background reduction.  R\&D is underway to develop this possibility.

\item {\bf Superfluid Helium} 

Superfluid $^{4}$He has long-lived vibrational excitations, with energies $\sim$~1~meV, and long-lived electronic excitations, with 
energy near $\sim$18 eV (ionization, EUV singlet scintillation photons, and metastable triplet excimers).  A high detection efficiency 
for these varied excitations would enable a detector sensitive to low-energy electronic and nuclear recoils from DM scattering. 

Scintillation light from both singlet and triplet states could be collected and measured with cryogenic large-area single-photon detectors similar to those proposed for crystal scintillator DM experiments. The ionization could be measured using a standard two-phase TPC configuration. However, the mobility of an e$^{-}$ in He is much less than seen in other noble liquids. Consequently, very high E-fields would be necessary, which may lead to a significant dark-count rate. Finally, rotons and other athermal phonon vibrations naturally eject He atoms when they hit a liquid/vacuum interface. HERON showed that these ejected atoms could then be adsorbed onto the surface of a large area calorimeter.  Importantly, this process has natural amplification, since the attraction potential of the He atom to the surface is an order of magnitude larger than vibrational energies themselves.

In~\cite{Schutz:2016tid}, it was proposed to use off-shell elastic scattering processes, whereby DM could produce two nearly back-to-back 
helium jets that carry large fractions of the DM kinetic energy.  If detection is feasible, ultra sensitive superfluid He detectors could be sensitive 
down to the keV warm DM limit. 
The use of superconducting calorimeters (see above) has also been proposed to probe down to the warm DM limit.  Compared to 
liquid He, the calorimeter sensitivity requirements are over an order of magnitude more severe due to the lack of amplification. 
However, the scattering rates themselves are tree level.

\end{itemize}

\subsection{Distinguishing Signal from Background}

Several suggestions exist to distinguish signals from backgrounds.  
First, the DM-induced electron recoil spectrum is expected to look rather distinct from that of 
background events. 
Nevertheless, several additional handles exist.  
The classic approach is to search for an annual modulation of the event rate induced by the motion 
of the Earth around the Sun~\cite{Drukier:1986tm}.  
Since most processes under consideration are inelastic (e.g., scattering off a bound electron), the expected modulation amplitude is 
larger than that for elastic nuclear recoils~\cite{TuckerSmith:2001hy,Essig:2015cda}.  
This amplitude also has a distinct dependence on threshold energy, and is unlikely to be mimicked by a modulating background. 
Another possibility in the future may be to use 2D materials (like graphene) to detect the directionality of the 
recoiling-electron event~\cite{Hochberg:2016ntt}.  
Gravitational focusing of DM by the Sun also induces interesting modulation effects~\cite{Lee:2015qva}. 
Also, the inherent asymmetries in crystals could be used to search for a daily modulation of the event rate~\cite{Essig:2011nj, futureCC}. 
Another technique to confirm a DM signal would be to use multiple targets and multiple experimental techniques; dark counts, for 
example, would be very unlikely to mimic the expected ratio of DM scattering rates between different targets. 

\subsection{Outlook}

Direct-detection searches for keV-to-GeV mass DM play an essential role in probing various types of dark sectors, which 
is often complementary to accelerator-based probes. 
While xenon-based experiments have demonstrated direct-detection sensitivity to DM down to masses of a few MeV and 
further research is needed before their ultimate sensitivity is known, the large dark-count rate makes it highly desirable 
to use other techniques and materials to probe this mass range, as well as to probe lower DM masses. 
Over the next few years, the potential of semiconductors will be better quantified.  In particular, the dark-count rates in semiconductors 
must be quantified to determine if they provide the best path forward.  
Large-area single-photon sensitive detectors must be developed for superfluid He and scintillating-crystals experiments, and the next 
few years may see significant progress in this arena.  Chemical bond breaking techniques, and in particular, detectors based on the detection of color-centers, may also allow for significant progress in the next few years.   
Other techniques, including the use of 2D targets, and superconductors will continue to be developed on a 
longer time-scale.
The multitude of ideas suggests a healthy field from which much progress can be expected over the next 5--10 years.  
The relatively small scale of most of these experiments means that even a modest investment of funds can enable the necessary R\&D that 
could allow for significant progress over the next few years.

\newpage

\section{Rich Dark Sectors}
\vspace{-0.3cm}
\begin{flushleft}
\textit{Convenors:~Bertrand Echenard, David~E.~Morrissey, Brian Shuve.~Organizer Contact: Matt Graham}
\end{flushleft}

\subsection{Introduction}

The physics of the Standard Model~(SM) is complex:~there are multiple non-gravitational forces, mass scales originate both from dimensionful parameters of the theory and via dimensional transmutation, and the origins of flavor and $CP$ violation are unknown. Furthermore, the SM is just the tip of the iceberg:~it comprises only 15\% of the matter in the universe, and identifying the composition and nature of the remaining dark matter~(DM) is one of the major outstanding problems in particle physics.  While DM could be just a single new particle, it may also arise as part of an entire dark sector containing many new particles and forces.  There is no reason for such a dark sector to be substantially simpler than the visible matter of the SM.

When DM is part of a larger dark sector, its experimental signatures can differ substantially from minimal DM scenarios, such as those of a Weakly Interacting Massive Particle~(WIMP). The complementarity of different experiments in probing the DM parameter space, as well as the range of parameters motivated by cosmology and astrophysics, are also more complicated in DM models with dark sectors. While minimal models offer simple motivations for many DM searches, it is possible that some striking signatures of DM physics would be missed with the current program that focuses primarily on WIMPs. Indeed, much of the rich phenomenology observed in the SM is independent of the physics establishing the cosmological abundance of protons and electrons. Given the lack of any positive detection of DM to date, it is imperative to broaden our search strategies to allow a discovery over as wide a range of theories as possible. Conversely, should a DM signal be observed, a thorough exploration of the dark sector would be required.

Dark sectors with a characteristic mass below the electroweak scale have been the subject of recent interest~\cite{Essig:2013lka}. This possibility was motivated by the potential for qualitatively new signals from low-mass hidden sectors at the LHC~\cite{Strassler:2006im}, and putative astrophysical observations of DM annihilation that suggested the presence of a new, light gauge boson that mediates interactions between the SM and DM~\cite{ArkaniHamed:2008qn,Pospelov:2008jd}, typically referred to as a ``dark photon''. New experimental programs dedicated to the discovery of dark photons in their visible and invisible decays were established soon after~\cite{Bjorken:2009mm}, and such experiments are the focus of earlier sections in this report. 

Even in such a simple scenario, however, many questions arise. What is the origin of the mass of the dark gauge boson? Does it have its own Higgs mechanism, and if so, does this induce a new hierarchy problem? Are there any states in the dark sector beyond the DM particle and the dark photon, and how do these states influence the cosmology and phenomenology? Is there strong dynamics in the hidden sector?  Are the couplings of the new dark states different from those assumed in the minimal dark photon model, and is there a way to ensure a comprehensive experimental program for discovering all of the dark sector states? These questions, among others, motivate the study of so-called ``rich dark sectors''.  The goal of this working group is to identify the challenges and opportunities in expanding our sensitivity to a wide range of dark sector models beyond just the minimal dark sector scenario.

\subsection{Exploring Rich Dark Sectors}

A major challenge in studying rich dark sectors is the large number of possible particles and couplings. In this workshop, we focused on three broad categories to provide an organizational framework for discussing experimental signatures and identifying areas for future development. The categories are:~\emph{dark-sector masses and naturalness}, which focused on models explaining the origin of dark-sector masses or their naturalness, and potential connections with the SM Higgs mechanism; \emph{non-minimal dark matter}, which highlighted theories where DM has non-standard interactions and examined the consequent astrophysical, cosmological, and experimental implications; and \emph{exotic dark sectors}, which studied an array of models extending the minimal dark photon paradigm.

\subsubsection{Dark-Sector Masses and Naturalness}

Dark sectors with massive vector, fermion, and scalar fermion degrees of freedom na\"ively raise many of the same questions as the SM:~what sets the masses of the dark states, and is the hierarchy between the dark sector masses and the Planck scale natural? For dark sectors with masses at or below the weak scale, it is reasonable to consider whether the dark masses are set by an independent Higgs mechanism, a Stueckelberg mechanism if the gauge symmetry is Abelian, or some extended symmetry breaking that is shared amongst the visible and dark sectors. In most models accounting for dark sector masses, new states are required beyond a single DM candidate or a dark force mediator, and this leads to new experimental constraints and possibilities.

One of the simplest mechanisms of dark mass generation is a dark Higgs field. The corresponding dark Higgs boson $h^\prime$ can be produced in the dark Higgs-strahlung process, ${A'}^*\rightarrow A' h'$, where $A'$ is the massive dark gauge boson connected to the SM through the vector portal~\cite{Batell:2009yf,Essig:2009nc}. Such a process can occur wherever dark gauge bosons are produced, such as in $B$ factories, proton colliders, and fixed-target experiments. The experimental signals of this process depend on how the dark Higgs $h'$ decays.  When $m_{h'} \gtrsim m_{A'}$, the dominant decay is $h'\rightarrow A'A^{'(*)}$, which can give rise to multi-lepton (or multi-pion) final states~\cite{Batell:2009yf}. Such signatures have been the subject of searches at \babar\ and Belle and are nearly background free~\cite{Lees:2012ra,TheBelle:2015mwa}, suggesting that Belle II should have very good sensitivity. If $m_{h'} < m_{A'}$,  the $h'$ decays through a loop of $A'$ vectors to a pair of SM states~\cite{Batell:2009yf}, or by its mixing with the SM Higgs~\cite{Clarke:2013aya}. For both such $h'$ decay modes, the decay rate receives a strong suppression, from two factors of the kinetic mixing in the former and the small SM Yukawa couplings in the latter, often leading to displaced or invisible decays~\cite{Batell:2009yf}.

Dark sectors with fundamental scalars can suffer from a hierarchy problem, and models which alleviate their fine tuning generally predict additional states connected with the dark Higgs sector. As a concrete example, supersymmetric theories with a dark photon and Higgs predict additional dark gaugino and Higgsino states~\cite{ArkaniHamed:2008qp,Cheung:2009qd,Morrissey:2009ur}. Searches at ATLAS and CMS for dark photons produced in SM Higgs decay are sensitive to some of these scenarios by looking for lepton jets, highly collimated collections of leptons~\cite{Aad:2014yea,Khachatryan:2015wka}. New dark supersymmetric (SUSY) states can also be produced in dark photon decays and give rise to additional new signals.  For example, if the dark photon decays to the lightest and next-to-lightest dark neutralinos, $A'\to \chi_1^x\chi_2^x$, the heavier dark neutralino $\chi_2^x$ often has a long lifetime and can be seen in fixed-target/beam-dump experiments, as well as in displaced vertex searches at $B$ factories and hadron colliders~\cite{Morrissey:2014yma}.  Similar signatures arise in simplified models of inelastic DM~\cite{TuckerSmith:2001hy,Bai:2011jg,Izaguirre:2015zva}. Such signatures are not completely covered in existing experimental analyses.

Even more exotic signatures are possible when the dark photon is part of an extended non-Abelian gauge sector~\cite{Barello:2015bhq}, or if  the  breaking of the gauge symmetry is due to strong dynamics~\cite{Alves:2009nf,Essig:2009nc,Harigaya:2016rwr}. In the latter scenarios, the mass scale is natural in the same manner as the proton mass in Quantum Chromodynamics~(QCD). The resulting spectrum of states is very rich and is similar to that found in Hidden Valley models~\cite{Strassler:2006im}. Many Hidden Valley signatures have been proposed over the years, including long-lived decays~\cite{Strassler:2006ri,Strassler:2006qa,Han:2007ae}, emerging jets~\cite{Schwaller:2015gea}, and semi-visible jets~\cite{Cohen:2015toa}. Many of these possibilities have not yet been directly studied at low-energy experiments.

\subsubsection{Non-Minimal Dark Matter}

Cosmological and astrophysical observations of DM can be largely explained by a cold (non-relativistic), collisionless DM particle~\cite{Ade:2015xua}.  However, a variety of studies of small-scale structure over the past two decades have uncovered conflicts between the predictions of cold dark matter~(CDM) simulations and the observed DM halo profiles and distributions~\cite{Moore:1994yx,Flores:1994gz,BoylanKolchin:2011de}.  Recent simulations of structure formation that include the effects of baryonic feedback on DM halos give better agreement with observations, but it is an open topic of debate whether this can account for all of the discrepancies~\cite{Governato:2012fa,Onorbe:2015ija,Oman:2015xda,2015arXiv151108741P}.  If the disagreements between simulations and observations persist, this may give evidence for new physics beyond collisionless DM, such as DM with self interactions.

The cross sections needed to resolve the small-scale structure anomalies ($\sigma/m_{\rm DM}\sim 1\,\mathrm{cm}^2/\mathrm{g}$) are much larger than a typical electroweak scattering rate~\cite{Spergel:1999mh,Rocha:2012jg,Vogelsberger:2012ku}, and can be challenging to realize in models of DM that are consistent with other constraints on interacting DM such as from the Bullet Cluster~\cite{Randall:2007ph,Buckley:2009in,Peter:2012jh}. In models of self-interacting dark matter~(SIDM), where DM interacts with itself via a light mediator, the velocity dependence of the cross section can render SIDM consistent with all current constraints~\cite{Tulin:2013teo}, and observations of small-scale structure may even allow for the determination of the DM and mediator masses without any direct interactions with SM fields~\cite{Kaplinghat:2015aga}.

The SIDM paradigm provides an additional astrophysical motivation for light mediators within the dark sector. However, the coupling of the mediator to SM fields can be much smaller in these models than is accessible in standard dark photon or Higgs searches. Such tiny couplings may be tested by precision probes of Big Bang Nucleosynthesis~(BBN) and the Cosmic Microwave Background~(CMB), as late decays of the mediator can disrupt the standard cosmology~\cite{Fradette:2014sza,Berger:2016vxi}.

Large dark scattering rates mediated by new light dark forces can also lead to the formation of DM bound states~\cite{Alves:2009nf,Feng:2009mn,Kaplan:2009de,An:2015pva}. Such bound states can then decay via annihilation of the constituent DM particles, giving an alternative mechanism for DM indirect detection. Indeed, the formation of bound states can dominate over the Sommerfeld enhancement that is typically considered in the annihilation of interacting DM~\cite{An:2016gad}. Bound states of DM can also be produced in $B$ factories and high-energy colliders, leading to striking decays to multi-mediator final states~\cite{An:2015pva,Tsai:2015ugz}. If the dark sector contains multiple bound states, then the spectrum of such states can be studied at colliders. Even if the bound states are invisible, the photon energy in monophoton searches traces the invariant mass of the invisible states, allowing for spectroscopy to be performed at lepton colliders with a monophoton trigger \cite{Hochberg:2015vrg}. Experimental searches for the full range of these signals remain to be performed at both low- and high-energy colliders.

Finally, non-minimal DM scenarios can dramatically change the nature of DM signals in various experiments. For example, the presence of strong DM self-interactions can lead to a radically different evolution of the DM abundance due to cannibalization and $3\rightarrow2$ scattering~\cite{Carlson:1992fn}, such as in SIMP~\cite{Hochberg:2014dra} or ELDER~\cite{Kuflik:2015isi} models. These mechanisms are typically realized for DM candidates that are qualitatively different from conventional WIMPs.  There are also non-thermal scenarios that, once again, break the relation between couplings and masses found in WIMP models~\cite{Gelmini:2006pq}. An intriguing example of a non-thermally established DM abundance are models of asymmetric DM, where the DM abundance arises from a dark matter-antimatter asymmetry~\cite{Kaplan:2009ag}. If this asymmetry is connected with baryon- or lepton-number-violation in the visible sector, then DM can lead to dramatic signatures such as induced nucleon decay~\cite{Shelton:2010ta,Davoudiasl:2011fj}. The halo structure of DM can also be different in extended dark sectors relative to collisionless WIMP models~\cite{Fan:2013yva,Fan:2013tia,Foot:2014mia}, modifying predictions for direct- and indirect-detection experiments. Non-minimal DM models clearly motivate a wide array of non-traditional searches for dark-sector states and forces.

\subsubsection{Exotic Dark Sectors}

Exotic dark sectors encompass any model that extends beyond the minimal vector, Higgs, or neutrino portals.  Such sectors frequently lead to (optimal) search strategies that are not included in the current dark-sector discovery program.  These theories can be motivated by various anomalies such as the excess in muon $(g\!-\!2)_{\mu}$~\cite{Blum:2013xva}, the puzzle of the proton charge radius~\cite{Pohl:2010zza}, as well as hints of new particles in beryllium nuclear transitions~\cite{Krasznahorkay:2015iga}.
% and at high-energy colliders~\cite{atlas750,CMS:2015dxe}. 
Exotic models are also motivated by many of the mass and DM considerations discussed above. An overarching theme of the workshop discussions was that the related ``exotic signatures'' do not necessarily arise from complicated or contrived hidden sectors. For instance, a single new light mediator with large couplings to DM can lead to the formation of DM bound states that qualitatively change the phenomenology of the hidden sector~\cite{Alves:2009nf,Feng:2009mn,Kaplan:2009de,An:2015pva}. 

In other examples, new light scalars or vectors could be hidden from existing experiments if their couplings differ from the predictions of the minimal dark-photon or dark-Higgs scenarios. For example, if SM fields are charged directly under the gauge interaction of a new hidden sector, the couplings can be very different from a kinetically mixed dark photon:~$B\!-\!L$ interactions would lead to new couplings between neutrinos, leptons, and nucleons~\cite{Heeck:2014zfa}, a $B$ gauge interaction would be more difficult to detect but can modify rare meson decays~\cite{Tulin:2014tya}, while an $L_\mu\!-\!L_\tau$ gauge interaction would decouple the new gauge boson from the initial states of low-energy electron accelerator experiments~\cite{Altmannshofer:2014cfa}. Similarly, a scalar coupling via mixing with an extended electroweak symmetry breaking sector can modify the mixing of the dark photon with the Z boson~\cite{Davoudiasl:2012ag}, or give scalars with enhanced couplings to heavy-flavor leptons and suppressed couplings to electrons and quarks~\cite{Chen:2015vqy,Batell:2016ove}.  Finally, a protophobic force could account for the recently hypothesized new particle produced in $^8{\rm Be}^*$ decays while remaining consistent with dark-photon bounds~\cite{Feng:2016jff}.

It is challenging to comprehensively explore exotic dark signatures with existing or new experiments. The Search for Hidden Particles~(SHiP) Experiment, a proposal to build a particle detector 50~m downstream from the beam dump of the Super Proton Synchrotron~(SPS) at CERN~\cite{Anelli:2015pba,Alekhin:2015byh}, could provide sensitivity to a wide variety of models with long-lived particles. Since most models connecting dark forces to the SM will induce couplings to nucleons via higher-order processes, SHiP would be an excellent model-independent probe of exotic new physics. In background-limited environments such as the LHC or $B$ factories, it is possible to design targeted searches for well-motivated exotic signatures that give excellent sensitivity. Existing experiments such as SeaQuest can also be re-purposed with electron and pion reconstruction abilities to enhance the prospects for exotic signature discoveries~\cite{gardner:2015wea}. Finally, dedicated experiments may be needed to probe new particles that would not register in any other experiment; a recent example is the MilliQan proposal at the LHC to search for new milli-charged particles~\cite{Haas:2014dda}. Opportunities to implement relatively low cost, parasitic experiments at existing facilities should be explored further.

\subsection{Synthesis and Summary}

While each discussion focused on different aspects of rich dark sectors, several common themes emerged from the presentations, questions, and conversations. It is impossible to design experimental and theoretical programs that are able to probe every possible DM model; however, the proposals below are meant to expand sensitivity to as many rich dark sector scenarios as possible. While this should not come at the expense of searches for the canonical kinetically-mixed dark photon models, small variations in experimental or analysis strategies could broaden the spectrum of final states that can be discovered at each experiment.

\benum

\item  {\bf Existing or planned experiments can be sensitive to rich dark sectors.}\\
It is not necessary to reinvent every search and experiment! For example, in dark photon scenarios with a large dark gauge coupling leading to the formation of DM bound states, existing searches for visibly-decaying dark photons can also be sensitive to invisibly decaying dark photons followed by bound-state decay~\cite{An:2015pva}. Similarly, constraints on kinetically-mixed dark photons can easily be mapped into the parameter spaces of new gauge interactions like $B\!-\!L$ or $L_e\!-\!L_\mu$.   Existing experiments designed for other purposes could also have sensitivity to new dark states. For example, neutrino beam experiments are sensitive to $L_\mu\!-\!L_\tau$ interactions via the neutrino trident process~\cite{Altmannshofer:2014cfa}, while neutrino detectors could serve as ``direct detection'' experiments for boosted DM coming from the Galactic Center or the Sun~\cite{Schuster:2009au,Agashe:2014yua,Berger:2014sqa}. 

\begin{figure}[ttt]
\centering
\includegraphics[width=0.47\textwidth]{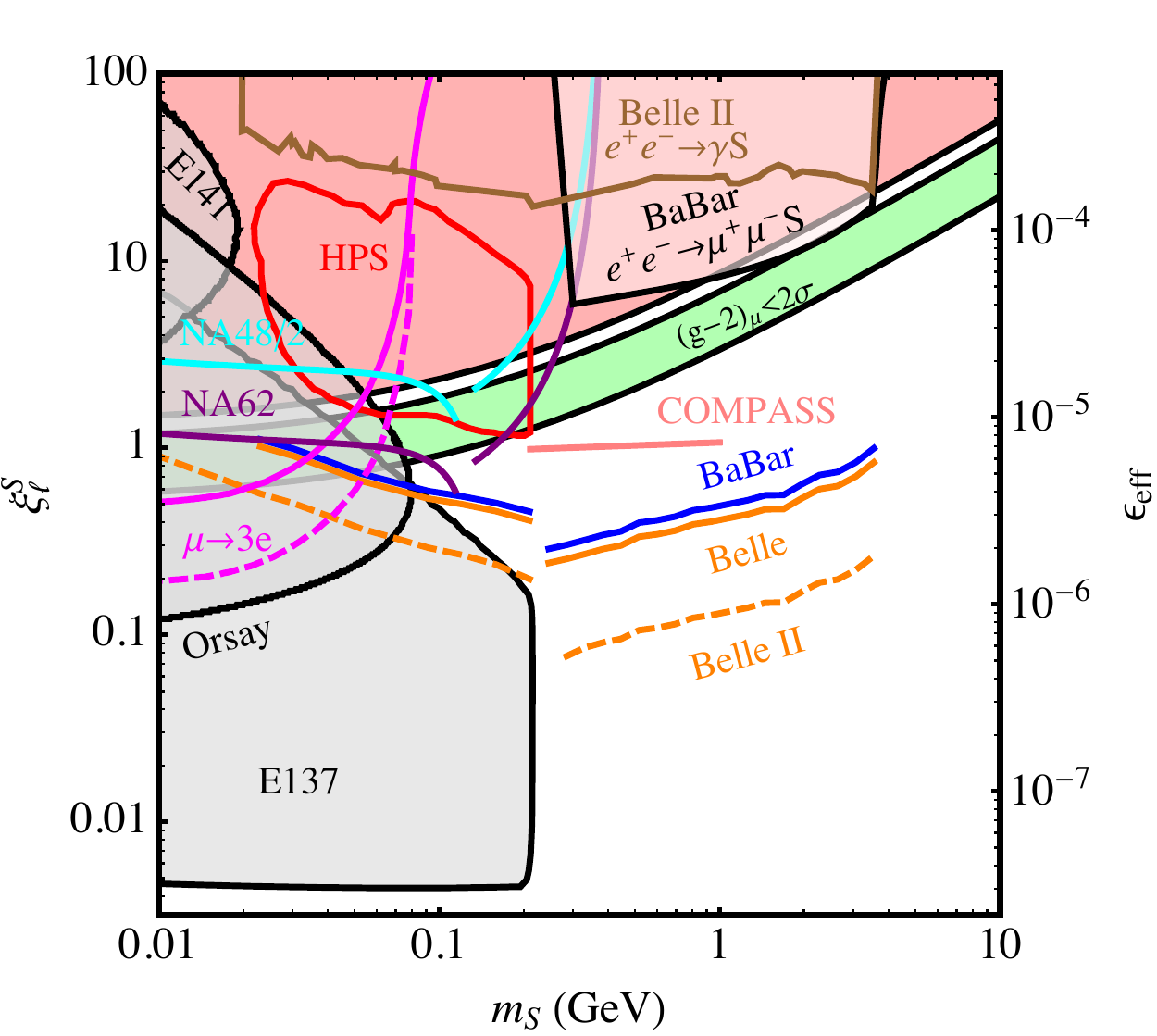}~~~~~
\includegraphics[width=0.47\textwidth]{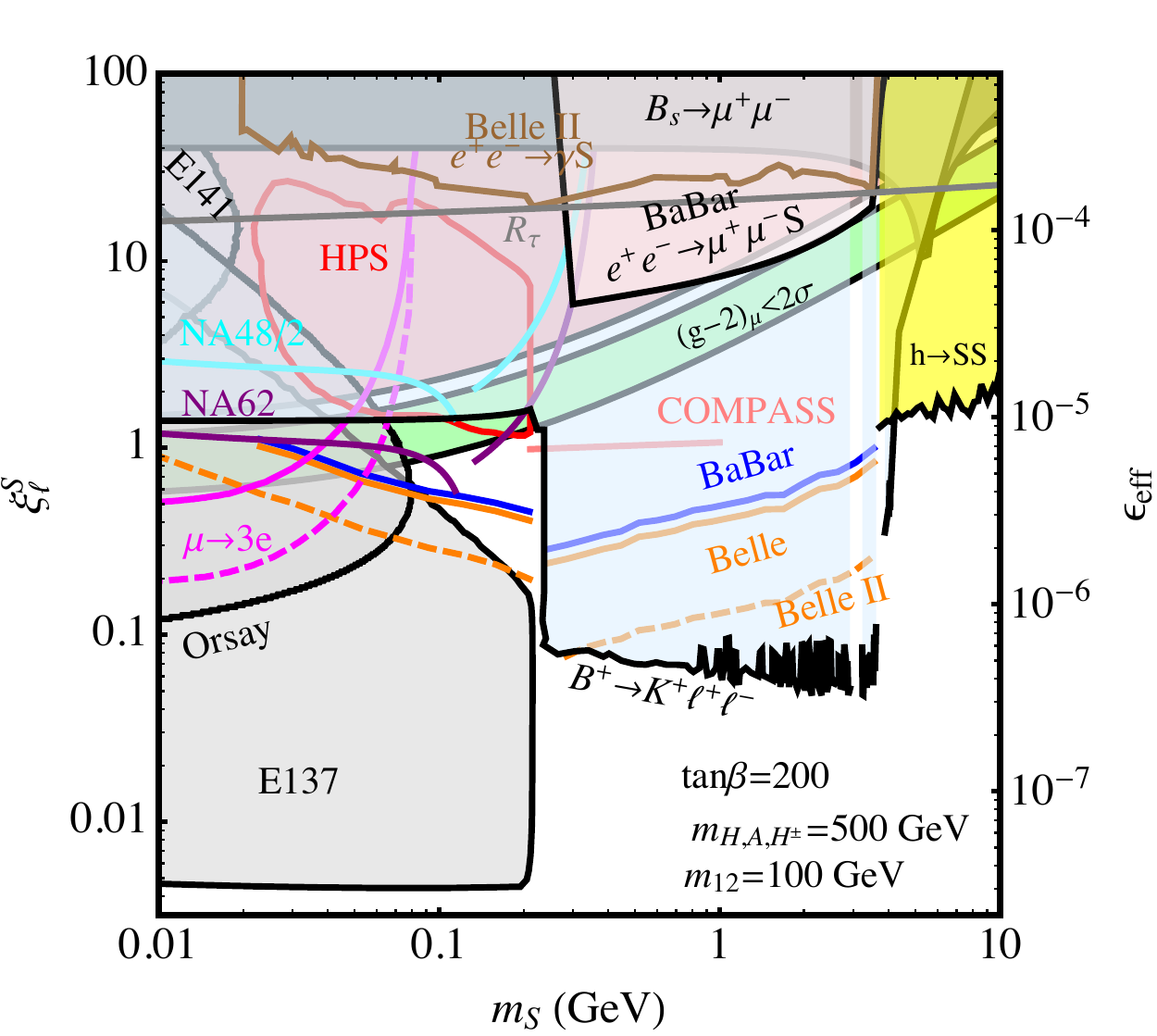}
\caption{Constraints on a leptophilic scalar $S$ from Ref.~\cite{Batell:2016ove}; the scalar coupling to each lepton $\ell$ is equal to $\xi_\ell^S$ times the SM Yukawa coupling for $\ell$.  \emph{Left:}~Model-independent limits and projections assuming only the coupling of $S$ to leptons. \emph{Right:}~Limits and projections for a UV-complete model that leads to additional couplings between $S$ and $b$ quarks (see Ref.~\cite{Batell:2016ove} for details).
\label{fig:leptohiggs}}
\end{figure}

To facilitate the reinterpretation of existing results, experiments should provide as much information as possible about the specific couplings that are driving their limits (or discoveries). Such results can be made more transparent by, for example, reporting constraints from dark photon searches as a function of individual couplings to $e$, $\mu$, $q$ (or products thereof) in addition to the $m_{A'}-\epsilon$ parameter space. To illustrate how experimental sensitivities can change in an exotic dark sector, we show in Fig.~\ref{fig:leptohiggs} the constraints on a boson that couples mass-proportionally to leptons~\cite{Batell:2016ove}; the constraints are very different than for a kinetically mixed dark photon (see Sec.~\ref{sec:VDP}). Also, experiments should consider how minor additions or modifications to their trigger and reconstruction strategies could be used to ensure sensitivity to harder-to-reach dark sector scenarios. In conjunction with these experimental efforts, theorists should develop a set of simplified models that capture a broad range of behaviors of rich dark sectors to facilitate such planning.  

\item {\bf Dedicated rich dark sector searches are also needed.}\\ 
Many signals predicted by rich dark sectors do not arise in more minimal models. However, the number of possible signals is nearly limitless. Theoretical guidance will be important to select the most promising search directions, but it is also essential that experimental proposals be designed to be sensitive to as wide a range of models as possible. As an example, consider searches for six-lepton final states at $B$-factories~\cite{Lees:2012ra,TheBelle:2015mwa}.  These can arise both by Higgs-strahlung, $e^+ e^-\rightarrow A' h', h'\rightarrow A'A'$ with $A'\rightarrow \ell^+ \ell^-$~\cite{Batell:2009yf}, as well as from the direct production and decays of dark bound states, $\Upsilon_{\rm D}\rightarrow A'A'A'$, $A'\rightarrow \ell^+ \ell^-$~\cite{An:2015pva}.  By expanding the selection requirements of these searches to include events where the total final-state energy does not reconstruct the beam energy, they could also be sensitive to more general dark sector processes with some invisible particles in the final state.  

A useful guide for organizing dark-sector experimental channels can be found in SUSY multi-lepton searches at the LHC \cite{Aad:2014nua,Chatrchyan:2014aea}.  These searches are organized in signal regions of different lepton flavors, associated jet multiplicities, $b$-jets, and so forth; although theory models rarely predict signatures that populate a single bin, the background estimates and event counts in each bin give sensitivity to many models beyond those officially studied in the analysis. Where appropriate, organizing searches at low-energy experiments in a similar fashion  could greatly enhance their ranges of applicability.

\item {\bf Communication between theorists and experimentalists is key.}\\ 
In many cases, experiments are capable of performing various searches without knowing whether any theoretical motivations exist, while theorists are not necessarily aware of all of the opportunities and limitations of each experiment. This problem will only intensify with a proliferation of new experiments and renewed theoretical interest in dark sectors. An online repository or forum that collects experimental and theoretical ideas, classified by signature, could be a helpful resource to promote more fruitful interactions between theorists and experimentalists, as well as to reduce redundancy in efforts among researchers in the field. 

For such an effort to be successful, theorists would have to provide Monte Carlo~(MC) tools and events for use by experimentalists. In return, experimentalists should work to report results in as model-independent a manner as possible. For example, approximate publicly available parameterizations of reconstruction efficiencies, detector geometries, and fiducial cross section limits have been very useful in broadening the applicability of searches for new physics at high-energy colliders\footnote{This is in the form both of published efficiency information, as well as cards for fast detector simulations such as \texttt{Delphes}~\cite{deFavereau:2013fsa} that are validated with experimental data.}. Given the nature of low-energy experiments, it may not be possible to provide some or all of this information for a given experiment, but a description that allows even a rough estimate of sensitivity can be helpful to theorists for determining whether an experiment could have any sensitivity to a given model.

\begin{figure}[ttt]
\centering
\includegraphics[width=0.7\textwidth]{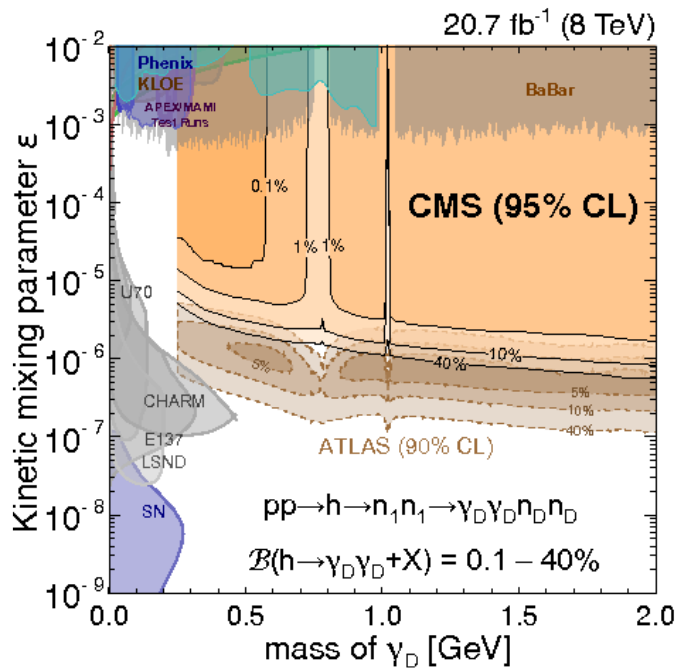}
${}$\vspace{-3cm}${}$
\caption{Constraints on a dark-sector SUSY model with a dark photon and multiple dark states (combination  from Ref.~\cite{Khachatryan:2015wka}). ATLAS and CMS constrain production of the dark matter through the Higgs portal and subsequent decays, while other experiments directly constrain the dark photons.
\label{fig:darkphoton_ATLAS_CMS}}
\end{figure}

\item {\bf Complementarity is complicated.}\\ 
One of the most appealing aspects of accelerator-based studies of DM is their complementarity to cosmological, astrophysical, and direct-detection probes of DM. However, the crossing symmetry relating these various search strategies applies strictly only to minimal DM candidates. In rich dark sectors, the connections between different experiments might be broken. In light of the complexity of possible signatures, it is imperative that experimental results be stated as precisely as possible (both by theorists and experimentalists), with all caveats clearly identified. For example, explanations for the muon $(g\!-\!2)$ anomaly or the putative $^8\mathrm{Be}$ signal based on minimal dark photons are ruled out by direct dark photon searches, but allowed in variants of the minimal model with a slightly more complicated coupling structure. In a concrete model, it is possible to apply experimental searches for different signatures to the same parameter space:~for example, in a model with a dark photon and multiple dark states, it is possible in certain limits to combine experimental results from LHC lepton-jet searches with canonical dark photon searches.  An example of this is given in Fig.~\ref{fig:darkphoton_ATLAS_CMS}, that shows the current sensitivity of the ATLAS and CMS experiments to a specific scenario in which dark photons are produced in the decays of the SM Higgs boson.  The LHC searches cover a large region of the model space that is inaccessible to lower-energy probes.

With extended rich dark sectors, a whole host of new constraints can also arise that are absent in minimal models. New forces coupled to neutrinos could face formidable bounds from neutrino scattering constraints, while sufficiently long-lived particles can unacceptably modify the predictions from various cosmological epochs. Since every ground-based, astrophysical, and cosmological search for new physics could, in principle, have sensitivity to dark sector models, a repository as proposed above would be useful in ensuring that all proposed models are consistent with existing constraints.

\item {\bf New tools may be needed.}\\ 
Given the plethora of existing bounds on dark sectors, as well as the challenge of MC simulation for some low-energy experiments, developing a framework for automating Monte Carlo~(MC) event generation and limit setting would be invaluable. Tools such as \texttt{MadDM} exist that take generic models in \texttt{UFO} format and compute various experimental rates and properties of interest~\cite{Backovic:2013dpa,Backovic:2015cra}. However, many of these tools are very WIMP focused, and an extension to low-mass DM would be needed. A framework which allows  users to write  modules to automate the computation of relic abundances of light thermal DM, non-thermal DM, asymmetric DM, and other non-WIMP scenarios, as well as a whole host of cosmological and astrophysical bounds, would be desirable; this would minimize redundant computations and ensure more accurate automated results across a broad range of rich dark sector scenarios.

\eenum

\subsection*{Appendix: Summary of Current Experiments}\label{rds:appendix}

\begin{itemize}
\item \textbf{$B$- and $\phi$-Factories}\\
In the past two decades, high-luminosity $e^+e^-$ collider experiments such as \babar\, Belle, and KLOE were conducted with collisions tuned to the energies of various hadronic resonances. The low center-of-mass energies and large data sets give these experiments excellent sensitivity to low-mass dark sectors. KLOE II is now taking data and Belle II will come online in a few years to further improve sensitivity.

Complementing the searches for standard visibly and invisibly decaying dark photons, \babar\ and Belle have searches for dark Higgs production via Higgs-strahlung, $e^+e^-\rightarrow A'h',\,h'\rightarrow A'A'$ \cite{Lees:2012ra,TheBelle:2015mwa}, while KLOE searched for the same process with invisible dark Higgs decays \cite{Babusci:2015zda}. \babar\ additionally studied a scenario where the dark photon is embedded in a dark non-Abelian gauge group via the reaction $e^+e^-\rightarrow {A'}^*\rightarrow W' W'\rightarrow 4\ell$ \cite{Aubert:2009af}. Finally, \babar\ recently performed a model-independent search for a boson coupled exclusively to muons via $e^+e^-\rightarrow \mu^+\mu^-Z',\,Z'\rightarrow\mu^+\mu^-$ \cite{TheBABAR:2016rlg}, providing a test of $L_\mu-L_\tau$ gauge bosons and leptophilic scalar scenarios above the dimuon threshold. \babar\ is currently searching for a dark scalar boson $S$ in $e^+e^- \rightarrow \tau^+ \tau^- S, S \rightarrow \mu^+\mu^-,e^+e^-$, as well as self-interacting dark matter in 6 lepton final states.

\item  \textbf{High-Energy Colliders}\\
Low-mass dark sectors can be challenging to discover at high energy colliders because the characteristic energy of dark-sector processes is small compared with the collision energy. If, however, dark-sector states are predominantly produced through the decays of heavy particles such as gauge or Higgs bosons, as is the case in many rich dark sectors, then high-energy colliders provide the best sensitivity to dark-sector physics. The new dark particles are typically produced in the boosted regime, leading to distinctive signatures. ATLAS, CMS, and LHCb each have growing search programs for RDS physics.

ATLAS results include searches for pairs of both displaced and prompt lepton jets \cite{Aad:2014yea,Aad:2015sms}, which are interpreted in terms of a model with Higgs decays into a hidden sector; searches for new low-mass, long-lived, hadronically decaying particles predicted in various Hidden Valley and other dark-sector models \cite{Aad:2015uaa,Aad:2015asa}; searches for new dilepton resonances produced in Higgs decays \cite{Aad:2015oqa,Aad:2015sva}; and searches for new low-mass diphoton resonances \cite{Aad:2015bua} or hidden-sector particles decaying to photons and dark matter \cite{ATLAS-CONF-2015-001}.

CMS results include a search for pairs of muon jets \cite{Khachatryan:2015wka}; searches for new low-mass, long-lived, leptonically or hadronically decaying particles \cite{CMS:2014hka,CMS:2014wda};  searches for Higgs decays to new dark-sector particles, including dilepton and $b$-quark resonances \cite{Khachatryan:2015nba,CMS-PAS-HIG-14-041,CMS-PAS-HIG-15-011};  and a search for exotic Higgs decays to low-mass dark matter and photons \cite{Khachatryan:2015vta}.

As a forward detector dedicated to the study of heavy-flavor physics, LHCb has a relatively high trigger efficiency and acceptance for low-mass particles, and its excellent vertexing capabilities allow it to perform sensitive searches for long-lived particles. However, LHCb has a much lower integrated luminosity than ATLAS or CMS. LHCb results include a search for new low-mass, long-lived, hadronically decaying particles \cite{Aaij:2014nma}; a search for new leptonic resonances in $B\rightarrow K^*\mu^+\mu^-$ \cite{aaij:2015tna}; and searches for new lepton-number-violating sterile neutrinos \cite{Aaij:2012zr,Aaij:2014aba}.

MilliQan is a proposal to search for milli-charged particles produced in $pp$ collisions at the LHC with an increased sensitivity in the 1--100 \gev\ mass range compared to current bounds~\cite{Haas:2014dda,Ball:2016zrp}.

\item \textbf{Beam Dump Experiments}\\
Rich dark sectors frequently contain new light, long-lived states.  In beam dump experiments, such particles can be created from collisions in the primary target and detected through their scattering or decays in the downstream detector.  Signals from scattering are typically very similar to those discussed in Section~\ref{sec:DMA}, but a new possibility in RDS models is the decay of long-lived states in the beam dump detector.  This was used in Refs.~\cite{Schuster:2009au,Batell:2009di,Morrissey:2014yma} to derive limits on RDS theories with a dark photon that decays primarily to long-lived dark Higgs scalars or fermions using data from CHARM~\cite{Bergsma:1985qz}, E137~\cite{Bjorken:1988as}, with future improvements expected from BDX~\cite{Battaglieri:2014qoa} and SHiP~\cite{Alekhin:2015byh}.  Beam dumps can also be used to probe dark sectors with light vectors that couple to visible matter more generally than just gauge kinetic mixing.  For example, the observed rates of neutrino trident scattering in CHARM-II~\cite{Geiregat:1990gz} and CCFR~\cite{Mishra:1991bv} were used in Ref.~\cite{Altmannshofer:2014pba} to place bounds on a new $L_{\mu}\!-\!L_{\tau}$ force.
\item \textbf{Nuclear Decays}
Detailed measurements of nuclear transitions offer a further probe into light dark sectors. A recent example is the distribution of opening angles between $e^+e^-$ pairs emitted in the decay of an $18\,\mev$ excited state of $^8$Be down to the ground state~\cite{Krasznahorkay:2015iga}. The observed distribution shows an excess at large angles, and can explained by the existence of a protophobic dark force~\cite{Feng:2016jff}.

\end{itemize}

%\bibliographystyle{JHEP}
%\bibliography{RDS}

%\end{document}

\newpage

\section{Conclusions}

While the presence of Dark Matter on galactic and extragalactic scales is firmly established from astrophysical observations, its microscopic nature remains a mystery. DM cannot consist of any of the known particles of the Standard Model, and thus represents the first convincing experimental indication of the existence of physics beyond the SM. A number of theoretical ideas for extending the SM to accommodate and explain DM have been proposed. Many of these ideas incorporate the DM particle as one of the states in a new ``Dark Sector", a set of new particles that are not charged directly under any of the SM forces (weak, strong, and electromagnetic). In this report, we summarized 
the current status and near-future prospects for experimental searches for this type of DM particle, other states in the associated Dark Sectors, and the ``mediator" force carriers that are responsible for interactions between the Dark Sector and the SM. We identified a number of theoretically motivated milestones in the parameter space of this class of models. We then described experimental strategies that may reach sufficient sensitivity to probe these milestones, as well as a number of proposed experiments that aim to implement these strategies in the coming years. A broad experimental program, encompassing both DM direct detection experiments and accelerator-based searches for mediator force carriers as well as the DM particles themselves, has a tremendous potential to discover the new physics of the Dark Sector, revolutionizing our understanding of both particle physics and cosmology. We advocate that such a program be pursued with vigor and determination.

\newpage

\subsection*{Acknowledgments}

We thank the SLAC National Accelerator Laboratory for hosting the Dark Sectors 2016 workshop and for their hospitality.  
R.E.~is supported by the DoE Early Career research program DESC0008061 and through a Sloan Foundation Research Fellowship. B.E. is supported by the U.S. Department of Energy (DoE) under grant DE-FG02-92ER40701 and DE-SC0011925. 
This work is partly supported by the project PGR-226 of the Italian Ministry of Foreign Affairs and International Cooperation (MAECI), CUP I86D16000060005.

\newpage

%\input{acknowledgments.tex}

%\newpage

\bibliography{fullbib_latest}

%\bibliographystyle{apsrevM}
%\bibliography{WorkshopReport.bib}

\end{document}